\pdfoutput=1
\documentclass[a4paper,12pt]{article}

\usepackage{amsfonts}
\usepackage{mathrsfs}
\usepackage{amssymb}
\usepackage{amsmath}
\usepackage{shuffle}
\usepackage{framed} 
\usepackage{subfigure}
\usepackage[medium]{titlesec}
\usepackage{bm}
\usepackage{cite}

\usepackage[normalem]{ulem}
\usepackage{extarrows}
\usepackage{slashed}
\usepackage{isodateo}
\usepackage{graphicx}
\usepackage{array}
\usepackage[dvipsnames]{xcolor}
\usepackage[bookmarksnumbered=true,bookmarksopen=true]{hyperref}
 \hypersetup{colorlinks,%
             linkcolor=NavyBlue, %
             citecolor=PineGreen, %
             urlcolor=PineGreen}
\usepackage[hmargin=.7in,vmargin=1.1in]{geometry}
\usepackage{indentfirst}
\usepackage{booktabs}

\usepackage{bbm}

\newcommand{\FR}[2]{\displaystyle\frac{\,{#1}\,}{#2}}
\newcommand{\fr}[2]{\mbox{$\frac{\,{#1}\,}{#2}$}}
\newcommand{\n}{\nonumber}

\graphicspath{{fig/}}

\def\bge{\begin{equation}}
\def\ede{\end{equation}}
\def\bga{\begin{aligned}}
\def\eda{\end{aligned}}
\def\bgb{\begin{bmatrix}}
\def\edb{\end{bmatrix}}
\def\bgp{\begin{pmatrix}}
\def\edp{\end{pmatrix}}
\def\bgm{\begin{matrix}}
\def\edm{\end{matrix}}
\def\bgs{\begin{subequations}}
\def\eds{\end{subequations}}
\newcommand{\order}[1]{\mathcal{O}({#1})}
\def\di{{\mathrm{d}}}

\def\mb{\mathbf}

\def\pd{\partial}

\def\la{\langle}\def\ra{\rangle}

\def\to{\rightarrow}

\def\ii{\mathrm{i}}

\def\ga{\gamma}
\def\de{\delta}
\def\ep{\epsilon}

\def\lam{\lambda}

\def\E{\mathcal{E}}

\def\aa{\mathsf{a}}
\def\bb{\mathsf{b}}
\def\cc{\mathsf{c}}

\newcommand{\ft}[1]{\big[{#1}\big]}
\newcommand{\ei}[1]{\big\{{#1}\big\}}
\newcommand{\ef}[1]{\big\la{#1}\big\ra}

\DeclareFontFamily{U}{bigshuffle}{}
\DeclareFontShape{U}{bigshuffle}{m}{n}{
  <5-8> s*[1.7] shuffle7
  <8->  s*[1.7] shuffle10
}{}
\DeclareSymbolFont{BigShuffle}{U}{bigshuffle}{m}{n}
\DeclareMathSymbol\bigshuffle{\mathop}{BigShuffle}{"001}
\DeclareMathSymbol\bigcshuffle{\mathop}{BigShuffle}{"002}

\usepackage{mdframed}

\newmdenv[skipabove=0pt,%
          skipbelow=5pt,%
          leftmargin=0pt,%
          rightmargin=0pt,%
          innertopmargin=-5pt,%
          innerbottommargin=7pt,%
          innerleftmargin=2pt,%
          innerrightmargin=2pt,%
          splittopskip=0pt,%
          splitbottomskip=0pt,%
          linewidth=0pt,%
          nobreak=true]%
          {keyeqn2}

\newmdenv[backgroundcolor=gray!15,%
          skipabove=0pt,%
          skipbelow=5pt,%
          leftmargin=0pt,%
          rightmargin=0pt,%
          innertopmargin=-5pt,%
          innerbottommargin=7pt,%
          innerleftmargin=2pt,%
          innerrightmargin=2pt,%
          splittopskip=0pt,%
          splitbottomskip=0pt,%
          linewidth=0pt,%
          nobreak=true]%
          {keyeqn}

\makeatletter
\@ifpackageloaded{hyperref}%
  {\newcommand{\mylabel}[2]
    {\protected@write\@auxout{}{\string\newlabel{#1}{{#2}{\thepage}%
      {\@currentlabelname}{\@currentHref}{}}}}}%
  {\newcommand{\mylabel}[2]
    {\protected@write\@auxout{}{\string\newlabel{#1}{{#2}{\thepage}}}}}
\makeatother

\usepackage{titlesec}          
\titleformat{\section}
{\normalfont\fontsize{15}{20}\bfseries}{\thesection}{1em}{}

\newcommand{\wt}[1]{\mkern 2mu \widetilde{\mkern -2mu #1 \mkern -2mu}\mkern 2mu}
\newcommand{\wh}[1]{\mkern 2mu \widehat{\mkern-2mu#1\mkern-2mu}\mkern 2mu}

\newcommand{\fnemail}[1]{\footnote{Email: \href{mailto:#1}{\nolinkurl{#1}}}}

\begin{document}

\title{\Large\textbf{Cosmological Amplitudes in Power-Law FRW Universe\\[2mm]}}

\author{Bingchu Fan\fnemail{fbc23@mails.tsinghua.edu.cn}~~~~~ and ~~~~~Zhong-Zhi Xianyu\fnemail{zxianyu@tsinghua.edu.cn}\\[5mm]
\normalsize{\emph{Department of Physics, Tsinghua University, Beijing 100084, China}}}

\date{}
\maketitle

\vspace{20mm}

\begin{abstract}
\vspace{10mm}

The correlators of large-scale fluctuations belong to the most important observables in modern cosmology. Recently, there have been considerable efforts in analytically understanding the cosmological correlators and the related wavefunction coefficients, which we collectively call cosmological amplitudes. In this work, we provide a set of simple rules to directly write down analytical answers for arbitrary tree-level amplitudes of conformal scalars with time-dependent interactions in power-law FRW universe. With the recently proposed family-tree decomposition method, we identify an over-complete set of multivariate hypergeometric functions, called family trees, to which all tree-level conformal scalar amplitudes can be easily reduced. Our method yields series expansions and monodromies of family trees in various kinematic limits, together with a large number of functional identities. The family trees are in a sense generalizations of polylogarithms and do reduce to polylogarithmic expressions for the cubic coupling in inflationary limit. We further show that all family trees can be decomposed into linear chains by taking shuffle products of all subfamilies, with which we find simple connection between bulk time integrals and boundary energy integrals. 
\end{abstract}

\newpage
\tableofcontents

\newpage

\section{Introduction}
\label{sec_intro}

Modern cosmology begins with the spectacular observation of redshift-distance relation which revealed that our universe is expanding rather than stationary. Throughout the century-long explorations, we know nowadays that the expansion of our universe is a dynamical process, determined by its energy content which has been successively dominated by the radiation, the cold matter, and the dark energy. We also know that our universe is approximately homogeneous and isotropic at large scales, and has a vanishing spatial curvature. Technically, such a universe is described by the famous Friedmann-Robertson-Walker (FRW) metric in general relativity. 

It is remarkable that we can gather so much information about the evolution history and physical properties of our universe. This is largely achieved by delicate measurements of cosmic perturbations, which are deviations of energy distribution and spacetime geometry from a uniform background. Given the horizon problem of the thermal big-bang cosmology, it is generally believed that these large-scale fluctuations are generated during a primordial phase in the cosmic history, when the universe evolves entirely different from the current thermal big-bang phase, an era called the primordial universe. 

Many possibilities for primordial cosmology have been proposed and studied, and the most prevailing scenario is the cosmic inflation, according to which the large-scale perturbations were generated during an exponentially fast expanding phase \cite{Achucarro:2022qrl}. In the meantime, alternative scenarios have also been actively studied \cite{Khoury:2001wf,Lehners:2007ac,Gasperini:1992em,Gasperini:2002bn,Wands:1998yp,Finelli:2001sr,Brandenberger:1988aj,Nayeri:2005ck}. Different primordial cosmologies could lead to qualitatively different correlation patterns of large-scale perturbations \cite{Chen:2011zf,Chen_2011fingerprints,Chen:2014joa,Chen:2014cwa,Chen:2015lza,Chen:2018cgg}. Therefore, measuring large-scale perturbations can be a useful way to probe the primordial universe. Moreover, the primordial universe was likely evolving at an extremely high energy scale, and therefore, we can also study high-energy particle physics through measuring large-scale fluctuations. This program, dubbed cosmological collider physics,  has been actively explored in recent years\cite{Chen:2009we,Chen:2009zp,Baumann:2011nk,Chen:2012ge,Pi:2012gf,Noumi:2012vr,Gong:2013sma,Arkani-Hamed:2015bza,Chen:2015lza,Chen:2016nrs,Chen:2016uwp,Chen:2016hrz,Lee:2016vti,An:2017hlx,An:2017rwo,Iyer:2017qzw,Kumar:2017ecc,Chen:2017ryl,Tong:2018tqf,Chen:2018sce,Chen:2018xck,Chen:2018cgg,Chua:2018dqh,Wu:2018lmx,Saito:2018omt,Li:2019ves,Lu:2019tjj,Liu:2019fag,Hook:2019zxa,Hook:2019vcn,Kumar:2018jxz,Kumar:2019ebj,Alexander:2019vtb,Wang:2019gbi,Wang:2019gok,Wang:2020uic,Li:2020xwr,Wang:2020ioa,Fan:2020xgh,Aoki:2020zbj,Bodas:2020yho,Maru:2021ezc,Lu:2021gso,Sou:2021juh,Lu:2021wxu,Pinol:2021aun,Cui:2021iie,Tong:2022cdz,Reece:2022soh,Chen:2022vzh,Niu:2022quw,Niu:2022fki,Aoki:2023tjm,Chen:2023txq,Chakraborty:2023qbp,Tong:2023krn,Jazayeri:2023xcj,Jazayeri:2023kji,Aoki:2023dsl,McCulloch:2024hiz,Craig:2024qgy,Meerburg:2016zdz,MoradinezhadDizgah:2017szk,MoradinezhadDizgah:2018ssw,Kogai:2020vzz}. 

Technically, we use the $n$-point $(n\geq 2)$ correlation functions of large-scale fluctuations to bridge the observational data with the predictions made by all kinds of primordial models, similar to how we make use of $S$-matrix to study particle physics with collider experiments. Thus, it is of central importance to study these correlation functions, which are called cosmological correlators. Typically, primordial perturbations are generated by quantum fluctuations of some fields living in the primordial universe. Thus, the study of cosmological correlators requires a good understanding of quantum field theory amplitudes in curved spacetime. This is a fast developing direction and has gained lots of attentions from both cosmology and amplitude communities \cite{Arkani-Hamed:2018kmz,Baumann:2019oyu,Baumann:2020dch,Pajer:2020wnj,Hillman:2021bnk,Baumann:2021fxj,Hogervorst:2021uvp,Pimentel:2022fsc,Jazayeri:2022kjy,Wang:2022eop,Baumann:2022jpr,Goodhew:2020hob,Goodhew:2021oqg,Melville:2021lst,Meltzer:2021zin,DiPietro:2021sjt,Tong:2021wai,Salcedo:2022aal,Agui-Salcedo:2023wlq,Sleight:2019hfp,Sleight:2019mgd,Sleight:2020obc,Sleight:2021plv,Jazayeri:2021fvk,Premkumar:2021mlz,Qin:2022lva,Qin:2022fbv,Xianyu:2022jwk,Qin:2023ejc,Qin:2023bjk,Qin:2023nhv,Xianyu:2023ytd,Loparco:2023rug,Maldacena:2011nz,Baumann:2017jvh,Bonifacio:2022vwa,Lee:2022fgr,Cabass:2022rhr,Cabass:2022oap,Lee:2023jby,Arkani-Hamed:2017fdk,Arkani-Hamed:2018bjr,Hillman:2019wgh,Gomez:2021qfd,Gomez:2021ujt,Wang:2021qez,Werth:2023pfl,Stefanyszyn:2023qov,DuasoPueyo:2023viy,Pinol:2023oux,Chen:2023iix,Werth:2024aui,Arkani-Hamed:2023bsv,Arkani-Hamed:2023kig,Benincasa:2024leu,Benincasa:2024lxe}.

An interesting class of cosmological correlators are generated by a conformal scalar field with self-interactions. By a conformal scalar, we mean a massless scalar field $\phi_c$ which nonminimally coupled to the Ricci scalar of the background metric via $-\fr12\xi R\phi_c^2$ with the coupling tuned to the conformal value $\xi=(d-1)/(4d)$, where the spatial dimension $d=3$ for our universe. The nice feature of a conformal scalar is that its quadratic action in arbitrary spatially-flat FRW universe can be brought to that of a massless scalar in Minkowski spacetime. Thus, its correlation function is much simpler than more general cases. 

Given the simplicity of the conformal scalar, we can study their correlation functions with arbitrary (and possibly time dependent) self-interactions. It is important that we allow for conformally non-invariant self-couplings, for otherwise the whole correlator would be identical to the corresponding flat space result up to overall rescaling. Thus, a conformal scalar with nonconformal self-interactions makes an interesting prototypical model generating relatively simple wavefunction coefficients or correlation functions, which can be taken as a starting point for studying more complicated and more realistic correlation functions. For simplicity, in this work we collectively call cosmological wavefunction coefficients and correlation functions the ``\emph{cosmological amplitudes}.'' Furthermore, by a slight abuse of terminology, we call the amplitudes generated by the aforementioned self-interacting conformal scalar ``\emph{conformal amplitudes}.''  

The conformal amplitudes in inflationary universe or more general power-law FRW universes are receiving increasing attentions recently \cite{Arkani-Hamed:2017fdk,Arkani-Hamed:2018bjr,Hillman:2019wgh,Arkani-Hamed:2023bsv,Arkani-Hamed:2023kig,Benincasa:2024leu,Benincasa:2024lxe}.  It is known that the conformal amplitudes in power-law FRW universe can be expressed as a twisted energy integral, whose integrand can be built from the corresponding flat-space amplitudes, ``twisted'' by powers of time introduced by the spacetime evolution and/or time-dependent self-couplings\footnote{\cite{Arkani-Hamed:2023bsv,Arkani-Hamed:2023kig} only considered the case where all ``twist factors'' share the same power. In this work, we allow for the possibility that different vertices have different powers, since they may correspond to different self-interactions in a general tree graph.}. The energy integrands, namely the flat-space amplitudes, are (sums of) simple fractions so that one can find a recursive rule to directly build them up \cite{Arkani-Hamed:2017fdk}. On the other hand, the full integrated result, namely the conformal amplitudes in FRW background, can be significantly more complicated, and a complete analytical answer was not known. Recently, it has been shown that one can derive a set of differential equations satisfied by these amplitudes \cite{Arkani-Hamed:2023bsv,Arkani-Hamed:2023kig}. The differential equations are constructed via a rather sophisticated algorithm. In principle, one can get analytical answer by solving these equations, as shown in \cite{Arkani-Hamed:2023kig} for simple examples. However, with the increasing number of vertices, the differential equations quickly become very complicated that solving them directly seems rather impractical. Thus, while it is nice to have differential equations governing the change of correlators with external kinematics, explicit analytical expressions for these correlators are certainly desirable as well. 

In this work, we introduce a simple algorithm for directly writing down the analytical answer for arbitrary tree-level conformal amplitudes in power-law FRW universe, including both the wavefunction coefficients and correlators. Our method works directly in the bulk and is based on simple manipulations of (nested) bulk time integrals. A key ingredient of our algorithm is the family-tree decomposition technique introduced in a previous work \cite{Xianyu:2023ytd}. With this method, we can decompose arbitrary tree-level amplitudes into a sum of a few terms, each of which is a product of several nested time integrals with fixed partial orders. A tree graph with a partial order can be naturally thought of as a (maternal) family tree. Thus we call each of such partially ordered time integrals a \emph{family tree}, and therefore we call our algorithm ``family-tree decomposition.'' 

In short, our algorithm consists of three steps: First, we pick up an arbitrary vertex from the tree graph and call it the ``earliest site.'' For a tree graph, the designation of the earliest site uniquely fixes the partial order for the graph. Second, we write down an expression for the partially ordered graph in terms of kinematic variables and decorate every bulk line by a binary variable taking values from $\pm 1$. Third, we take appropriate ``cuts'' of the graph, and each cut yields a new expression based on the original expression in the second step. Then, the final result is simply the sum of all possible cuts and all values of binary decorations. We design two types of cuts, one called ``wavefunction cut'' which is used for computing wavefunction coefficients, and the other called ``correlator cut'' used for computing correlators. There are tiny yet important differences between these two types of cuts which are responsible for the different final expressions for the wavefunction coefficients and the correlators. As is clear, every bulk line can be cut or uncut. Together with all possible binary decorations, we see that the final expression for a tree graph with $V$ vertices has $4^{V-1}$ distinct terms, since the number of bulk lines is simply $V-1$ for a tree graph.

A main reason that we decompose a tree graph into sum and products of family trees is that explicit analytical answers for these family trees are available in terms of multivariate hypergeometric series \cite{Xianyu:2023ytd}. The expansion variables of this hypergeometric series are directly related to the partial order of the graph. Thus, once we fix a partial order for the graph and make the corresponding family tree decomposition, we will immediately get the analytical answer for the original graph in terms of hypergeometric series. Moreover, since the partial order of the graph can be chosen at will as we shall show, we can freely choose the expansion variables for the hypergeometric series. This is important because it means that we are able to do analytic continuation for series solutions beyond their domain of convergence. In this work, we will provide more series expansions for the family trees in various kinematic limits, thus enrich our analytical understanding of these functions. 

With the family-tree decomposition, it is straightforward to study various analytical properties of conformal amplitudes. There are certainly many interesting topics in this direction, and we consider several of them in this work.

First, as mentioned above, previous works have exploited the energy integral representation of conformal amplitudes, which can be thought of as a boundary representation since the time variables are fully integrated out. On the other hand, our family-tree decomposition directly works in the bulk spacetime. Thus it would be interesting to see how our family-tree decomposition is connected to the energy integrals. We shall show that this connection is very straightforward for graphs with a total order rather than a partial order. Meanwhile, it is easy to see that any tree graphs with a partial order can be further decomposed into a sum of graphs with total orders. Technically, this is achieved by taking recursive shuffle products among all subfamilies in a family tree. This shuffle product can be thought of as ``comparing birthdays of all family members'' in a family tree. Thus we call it the \emph{birthday rule}. With this rule, we find a direct connection between our bulk family-tree integrals and the boundary energy integrals. In particular, it is straightforward to verify that our ``birthday rule'' leads to energy integrands identical to the ones found in the literature. Therefore, our result naturally satisfies the differential equations proposed in \cite{Arkani-Hamed:2023bsv,Arkani-Hamed:2023kig}.

Second, with our explicit expression, it is easy to examine the singularity structure of conformal amplitudes. In particular, it is easy to see that the conformal amplitudes typically possess (either divergent or finite) branch points in the total-energy limit or various partial-energy limits. The energy integral representation also makes it very simple to see how a particular family tree behaves in these kinematic limits. 

Third, there is a special limit of general conformal amplitudes which is of particular interest. That is, we can consider the $\phi^3$ theory for a conformal scalar in a spacetime with exponential inflation. The $\phi^3$ theory, albeit incomplete as a quantum field theory, is an interesting perturbative model. Recent studies have revealed surprising structures of flavored $\phi^3$ amplitudes \cite{Arkani-Hamed:2023lbd,Arkani-Hamed:2023mvg}. In the context of inflation correlators, the $\phi^3$ model is also explored as a simple test ground, from which one may gain useful insights into more general inflation correlators. Previous works have considered simple correlators in this model where the analytical answers are expressible in terms of polylogarithmic functions \cite{Arkani-Hamed:2015bza,Hillman:2019wgh,Arkani-Hamed:2023kig}. With the family-tree decomposition, it is straightforward to recover the known results and find new ones. As we shall see, it is easy to convert our series representation for family tree integrals to polylogarithmic functions in the inflationary limit. From this result, we observe that a tree graph with $V$ vertices can be expressed as polylogarithmic functions with transcendental weights up to $V$. It is also easy to find the symbols for the conformal amplitudes in this limit, which can be useful for simplifying the final expressions.

The rest of the work is organized as follows. 

In Sec.\ \ref{sec_frw}, we review briefly the basic theoretical input of our work, including the power-law FRW cosmology, the conformal scalar, the wavefunction formalism, and the Schwinger-Keldysh formalism. To make connection with primordial universe models, we will only consider horizon-exiting cosmologies. We also use this section to introduce our diagrammatic notations. Experts can skip or just glance through the entire section. 

In Sec.\ \ref{sec_family}, we introduce the most general class of nested time integrals required for the conformal amplitudes at the tree level. We then review the family-tree decomposition for such nested integrals and show explicit analytical expressions for these integrals in terms of multivariate hypergeometric series. We provide two different series representations, one expanded in terms of inverse powers of the earliest-site energy, the other expanded in terms of inverse powers of the total energy. While the former has been given in \cite{Xianyu:2023ytd}, the latter series was only indirectly mentioned in \cite{Xianyu:2023ytd}, and here we provide the full expression.

In Sec.\ \ref{sec_example}, we demonstrate the family-tree decomposition for a few examples of conformal amplitudes with increasing complexity, including the two-site chain, the three-site chain, and the four-site star graphs. In each case, we separately consider the wavefunction coefficient and the correlator. Along the discussion, we will see how a pattern emerges in the final expressions, from which we can observe the rule for directly writing down the final answers for general amplitudes. The examples considered in this section can all be expressed in known special functions in many different forms, which we collect in App.\ \ref{app_examples}.

In Sec.\ \ref{sec_general_rules}, we summarize the rules for directly writing down analytical expressions of general tree-level conformal amplitudes in terms of family-tree integrals. The rules will be stated with yet another explicit example (4-site chain) for both the wavefunction coefficients and the correlators. We also provide direct proofs of these rules based on a simple analysis of bulk propagators. 

In Sec.\ \ref{sec_energyint}, we explore the connection between the bulk time integrals and the ``boundary'' energy integrals. With a few examples, we first show that the relation between a nested time integral and the energy integrand is trivial for a chain diagram. We then show that any (partially ordered) family tree integral can be converted to a sum of (totally ordered) chain diagrams via a ``birthday rule,'' which amounts to taking shuffle products recursively for all subfamilies within a family tree. 

In Sec.\ \ref{sec_singular}, we briefly examine the singularity structure of family tree integrals. Again, with a few examples, we show how the total-energy and partial-energy singularities are distributed among different terms in the family-tree decomposition. 

In Sec.\ \ref{sec_inflation}, we focus on the special case of $\phi^3$ theory in inflation. This amounts to taking a particular limit of our general results. We will show that this limit is a little subtle in that every single term of the family-tree decomposition is divergent in this limit, but the divergences must cancel out among themselves when we combine all family trees into a full amplitude for a given tree graph. As a result, only the finite (non-divergent) part of the family tree integrals make contributions to the full result. We will show that this finite part typically belongs to functions of polylogarithmic type, and our series representation makes it easy to identify the full expression and also the corresponding symbol. 

We conclude the paper by further discussions and outlooks in Sec.\ \ref{sec_conclusion}. We collect some of our notations and conventions in App.\ \ref{app_notations}, and the special functions used in this paper in App.\ \ref{app_specialfunctions}. In App.\ \ref{app_examples} we collect the explicit analytical expressions for the worked out examples in Sec.\ \ref{sec_example}, including the wavefunction coefficients and the correlators for the two-site chain, the three-site chain, and the four-site star, respectively. In App.\ \ref{app_transf} we collect a few transformation-of-variable formulae for hypergeometric functions derivable from our family-tree decomposition. 

\paragraph{Notation and convention} We will give a detailed list of notations used in this work in App.\ \ref{app_notations}, but here we briefly mention a few basic or important ones for readers' convenience. In this work we assume general spacetime dimensions, with the spatial dimension denoted by $d$. The spacetime metric takes the ``mostly plus'' convention $(-,+,\cdots,+)$. We always assume spatially-flat FRW metric with power-law (including exponential) scale factor $a(t)$ where $t$ denotes the comoving time. However, we will use conformal time $\tau$ more frequently, which is related to $t$ by $\di\tau=\di t/a(t)$. We use italic bold letters to denote spatial vectors such as $\bm x$ (spatial comoving coordinates), $\bm k$ (spatial momentum of external lines), and $\bm K$ (spatial momentum of internal lines). The magnitude of a spatial vector will be denoted by the corresponding unbold letter such as $K\equiv|\bm K|$. Upright letters sans serif such as $\aa,\bb$ always denote summation variables taking values from $\pm 1$. Finally, we make heavily use of quantities with subscripts such as $K_\ell$. We stress that this is \emph{not} a spatial component of $\bm K$, but the magnitude of $\bm K$ for the internal line with index $\ell$. Also, we heavily use the shorthand notation $K_{ijk\cdots}\equiv K_i+K_j+K_k+\cdots$. Other special notations used in this work will be explained when we first introduce them.

\section{FRW Cosmology and Conformal Amplitudes}
\label{sec_frw}

In this work, we consider arbitrary tree-level amplitudes, including the wavefunction coefficients and the correlation functions, of a self-interacting conformal scalar field $\phi$ in a power-law and horizon-exiting FRW universe. Below, we review the relevant ingredients in turn. 

\paragraph{Power-law FRW spacetime}
We assume the spacetime to be $(d+1)$-dimensional, with a rigid background described by the spatially flat FRW metric:
\bge
\label{eq_FRWt}
  \di s^2=-\di t^2+a^2(t)\di\bm x^2,
\ede
where $\bm x\in\mathbb{R}^d$, and the scale factor $a(t)$ takes a power-law form:
\bge
  a(t)=\Big(\FR{t}{t_0}\Big)^{p},
  \label{eq_SF_PT}
\ede
where $t_0$ is an arbitrary time at which we normalize the scale factor as $a(t_0)=1$. The power-law form covers many familiar situations when $d=3$ such as Minkowski spacetime $(p=0)$, the matter dominated universe ($p=2/3$), and the radiation dominated universe $(p=1/2)$. The vacuum-energy dominated universe, namely the cosmic inflation, is not directly included in our consideration since it has $a(t)\propto e^{Ht}$. However, we can formally think of the exponential inflation as a limit of power-law cosmology $a(t)\propto t^p$ with $p\to\infty$. This will be made clearer below after we introduce the conformal time.

The conformal time $\tau$ is conventionally introduced such that the FRW metric (\ref{eq_FRWt}) becomes conformally flat. This is achieved by choosing $\di\tau=\di t/a(t)$. Solving $\tau$ and rewriting the scale factor $a=a(\tau)$ as as a function of $\tau$, we have:
\bge
  a(\tau)=\Big(\FR{\tau}{\tau_0}\Big)^{\wt p},~~~~\wt p\equiv \FR{p}{1-p}.
  \label{eq_SF_CT}
\ede 
Here $\tau_0$ is again an arbitrary time scale introduced to normalize the scalar factor. Then, the metric (\ref{eq_FRWt}) becomes:
\bge
\label{eq_metricct}
  \di s^2=a^2(\tau)(-\di\tau^2+\di\bm x^2)=\Big(\FR{\tau}{\tau_0}\Big)^{2\wt p}(-\di\tau^2+\di\bm x^2).
\ede
The familiar examples in $d=3$ mentioned above can be summarized as:
\begin{align}
\label{eq_wtp}
  \wt p=\begin{cases}
  0, &\text{(Minkowski)} \\
  1, &\text{(radiation dominated)}\\
  2, &\text{(matter dominated)}\\
  -1. &\text{(exponential inflation)}
  \end{cases}
\end{align}

The assignment of $p$ or $\wt p$ does not uniquely determine the spacetime evolution, as we have not fixed the sign of $t_0$ or $\tau_0$. Taking matter domination $a(t)=(t/t_0)^{2/3}$ as an example, we can either choose $t_0>0$ or $t_0<0$. The former describes an expanding universe filled with matter, which becomes singular at $t=0$ but can otherwise expand indefinitely to the future infinity. The latter, on the contrary, describes a contracting universe which possess a future singularity at $t=0$, but has a history that can be extrapolated indefinitely to past infinity. 

The contracting scenario with matter domination does not immediately correspond to the universe of our own in its current expanding phase, but it can make an interesting model for the primordial universe, a phase of cosmic history completely different from the current one. To solve the horizon problem of the big-bang cosmology, a key feature of any model of the primordial universe is that a field mode with fixed comoving momentum always exits rather than enters the horizon. With the additional requirement that modes are horizon-exiting, the spacetime evolution is uniquely labeled by the value of $p$.

With the application to primordial universe in mind, in this work, we shall exclusively consider horizon-exiting cosmologies with power-law FRW metric. We note that a horizon-exiting universe may be either expanding or contracting, depending on the value of $p$. Specifically, $0<p<1$ (and thus $\wt p>0$) corresponds to contracting scenarios and $p>1$ ($\wt p<0$) corresponds to expanding scenarios. See \cite{Chen:2011zf,Chen:2014cwa,Chen:2015lza} for more discussions.

\paragraph{Conformal scalar} For a technical reason to be made clear below, we shall focus on the amplitudes of a conformal scalar field $\phi_c$ in this work. By a conformal scalar, we mean a massless scalar field with a particular nonminimal coupling to the Ricci scalar $R$, as shown in the following action: 
\bge
\label{eq_actionconformal}
  S[\phi_c]=-\int\di^{d+1}x\sqrt{-g}\bigg[\FR12(\pd_\mu\phi_c)^2+\FR12 \xi R\phi_c^2+\sum_{n\geq3}\FR{\lam_n}{n!}\phi_c^n\bigg],
\ede
where the nonminimal coupling between the Ricci scalar $R$ and the scalar field $\phi_c$ is introduced with the coupling $\xi\equiv (d-1)/(4d)$. Also, we have included arbitrary non-derivative self-interactions of $\phi_c$ which take polynomial forms. The significance of the choice $\xi=(d-1)/(4d)$ can be seen by substituting the metric (\ref{eq_metricct}) into the action (\ref{eq_actionconformal}), and redefining the scalar field as $\phi_c=a^{(1-d)/2}(\tau)\varphi$. Then, the action (\ref{eq_actionconformal}) becomes:
\bge
\label{eq_actionflat}
  S[\varphi]=-\int\di\tau\di^{d}\bm x\, \bigg[\FR{1}{2}(\pd_\mu\varphi)^2+\sum_{n\geq 3}\FR{\lam_n(-\tau)^{P_n}}{n!}\varphi^n\bigg],~~~~~~P_n\equiv\Big[d+1-\FR{(d-1)}{2}n\Big] \wt p.
\ede
That is, we get a massless scalar field $\varphi$ living effectively in the Minkowski spacetime, but with unusual self-couplings which are power functions of the conformal time $-\tau$. Note that it is $(-\tau)^{P_n}$ appearing in this expression because we are considering horizon-exiting universes with $\tau<0$. Also, we have absorbed the $\tau_0$ factor in (\ref{eq_metricct}) by redefining the coupling constant $\lam_n$. We note that the power $P_n$ in (\ref{eq_actionflat}) depends on both the interaction type $n$ and the power $\wt p$ in the FRW metric. For instance, in 4-dimensional spacetime ($d=3$), we have $P_n=\wt p$ for $\phi^3$ interaction, which is an interesting perturbative model in its own right. On the other hand, for $\phi^4$, we have the power $P_n=0$, being independent of $\wt p$, which simply says that the $\phi^4$ interaction is conformal invariant classically when $d=3$. Thus, we see that the power $P_n$ can vary from vertex to vertex even when we fix the FRW background, because different vertices may correspond to different types of interactions. For this reason, in the rest of the paper, we will treat $P_n$ as a free parameter, and allow for the possibility that different vertices in a graph may take different values of $P_n$.

With all these steps taken, we can compute the correlation functions of $\varphi$ effectively in flat spacetime, and the time-dependent couplings can be handled in a perturbative way as usual. In particular, we can quantize $\varphi$ field as usual, resolving it into linear combinations of creation and annihilation operators $a_{\bm k}$ and $a_{\bm k}^\dag$ in $d$-momentum space:
\bge
  \varphi(\tau,\bm x)=\int\FR{\di^d\bm k}{(2\pi)^d}e^{+\ii\bm k\cdot\bm x}\Big[\varphi(k,\tau)a_{\bm k}+\varphi^*(k,\tau)a_{-\bm k}^\dag\Big].
\ede
Here $\varphi(k,\tau)$ is the mode function, which is obtained by solving the free equation of motion $\pd^2\varphi(\tau,\bm x)=0$. The general solution is a linear combination of plane waves of positive and negative frequencies. Typically, in the general FRW background, there is no unique way to pick up a combination. However, for a conformal scalar $\phi_c$, there is a natural choice which corresponds to the Minkowski vacuum, called the \emph{conformal vacuum} \cite{Birrell:1982ix}. This choice thus requires that we only keep the positive frequency mode in the mode function $\varphi(k,\tau)$. Then, the normalization is further determined, up to an irrelevant phase, by the canonical commutators $[\varphi(\tau,\bm x),\pd_\tau\varphi(\tau,\bm y)]=\ii\de^{(d)}(\bm x-\bm y)$ and $[a_{\bm k},a_{\bm q}^\dag]=(2\pi)^d\de^{(d)}(\bm k-\bm q)$. The result is:
\bge
\label{eq_modefunction}
  \varphi(k,\tau)=\FR{1}{\sqrt{2k}} e^{-\ii k\tau}.
\ede
This will be the mode function we use throughout this work.

With the mode function at hand, we can go on to compute amplitudes formed by the conformal scalar. However, the details will differ depending on which quantities we are going to compute. In this work, we are interested in both the $n$-point wavefunction coefficients $\psi_n$ and the $n$-point correlation functions $\mathcal{T}_n$. Below we briefly introduce the perturbative bulk construction of these two types of amplitudes, respectively.

\paragraph{Wavefunction coefficients} The idea of computing wavefunction coefficients comes from thinking of the whole story as a quantum mechanical problem phrased in Schrödinger picture. Here we give a short review, and more comprehensive introduction to the wavefunction method in cosmology can be found in, e.g., \cite{Goodhew:2020hob}. In short, we define state vectors $|\Psi\ra$ on equal-time slices, and resolve them with the field-value eigenbasis $|\varphi\ra$. So the quantity of interest is the wavefunction $\Psi[\varphi]=\la\varphi|\Psi\ra$. It is well known that the wavefunction $\Psi[\varphi]$ can be computed by a path integral with boundary conditions given:
\bge
\label{eq_pathint}
\Psi[\wh\varphi]=\int_{\varphi(\tau=-\infty)=0}^{\varphi(\tau=\tau_f)=\wh\varphi}\mathcal{D}\varphi\,e^{\ii S[\varphi]},
\ede
where $S[\varphi]$ is just the classical action given in (\ref{eq_actionflat}). Here the boundary condition at the past infinity ($\tau\to-\infty$) is imposed by the conformal-vacuum condition. With the usual $\ii\ep$-prescription, this simply says that the field  itself should vanish at $\tau\to-\infty$.\footnote{The $\ii\ep$-prescription is effectively saying that the field at the past infinity should be exponentially damped rather than exponentially enhanced. } On the other hand, the boundary condition at $\tau=\tau_f$ is fixed by arbitrarily chosen field configuration $\wh\varphi(\bm k)$ on an equal-time slice. --- Here we have switched to the $d$-momentum space to express the field $\varphi(\bm k)$. The wavefunction defined as such has the obvious physical meaning that $|\Psi[\wh\varphi]|^2$ gives the probability of finding the field $\varphi$ at a particular configuration $\wh\varphi$. Thus, if we want to compute an $n$-point correlation function of $\varphi$ at $\tau=0$, all we need to do is to compute the following statistical average:
\begin{align}
\label{eq_ctowf}
  \la\varphi(\bm k_1)\cdots\varphi(\bm k_n)\ra=\int\mathcal{D}\varphi\,\varphi(\bm k_1)\cdots\varphi(\bm k_n)|\Psi[\varphi]|^2.
\end{align}

As usual in quantum field theory, the path integral (\ref{eq_pathint}) is only formally defined and it is usually impossible to finish it in complete form. Therefore we seek for a perturbative expansion. In this work, we shall only consider the \emph{tree-level approximation} to the path integral, which amounts to evaluating the path integral at its saddle point: 
\bge
\label{eq_wfsaddle}
\Psi[\wh\varphi]\simeq e^{\ii S[\varphi_\text{cl}]},
\ede
where $\varphi_\text{cl}$ is the solution to the classical equation of motion $\de S/\de \varphi=0$ with the same boundary conditions that specify the path integral. In general, even the classical solution $\varphi_\text{cl}$ cannot be computed exactly, and thus we seek to compute the saddle-point wavefunction (\ref{eq_wfsaddle}) perturbatively. As usual, we separate the action $S[\varphi]=S_0[\varphi]+S_\text{int}[\varphi]$ into a free part $S_0[\varphi]$ and an interacting part $S_\text{int}[\varphi]$. We then perturbatively solve for $\varphi_\text{cl}$ from the equation $\de S/\de \varphi=0$. Once we get the solution $\varphi_\text{cl}$, we can substitute it back into (\ref{eq_wfsaddle}) and get the wavefunction. 

In practice, it is conventional to parameterize the wavefunction in the following way:
\begin{align}
\label{eq_wfcoefdef}
  \Psi[\wh\varphi]= \exp\bigg\{-\sum_{n\geq 2}\FR{1}{n!}\int\prod_{i=1}^n\bigg[\FR{\di^d\bm k_i}{(2\pi)^d}\wh\varphi(\bm k_i)\bigg]\psi_n(\bm k_1,\cdots,\bm k_n)(2\pi)^d\de^{(d)}(\bm k_1+\cdots+\bm k_n)\bigg\}.
\end{align}
Here we are working in the $d$-momentum space, and $\psi_n(\bm k_1,\cdots,\bm k_n)$ is the $n$-point wavefunction coefficients. It is well known that the free-theory wavefunction with vacuum boundary condition is Gaussian, where we have $\psi_2\neq 0$ and all higher point coefficients $\psi_{n\geq 3}=0$. Thus, in a weakly coupled interacting theory, we can think of interaction effects as slight deviations from a Gaussian wavefunction. The wavefunction coefficients $\psi_\text{cl}$ at the tree level can then be extracted by comparing (\ref{eq_wfsaddle}) with (\ref{eq_wfcoefdef}).

The above procedure can be reduced to a simple set of Feynman rules which allow us to get wavefunction coefficients $\psi_n$ by drawing diagrams. At the tree-level, for a given $\psi_n$, one draws all possible tree diagrams with $n$ boundary points. Then, all lines come in two types: The \emph{bulk propagator} $G(K;\tau_1,\tau_2)$ that connects two bulk time points $\tau_1$, $\tau_2$ with momentum $\bm K$, and the \emph{bulk-to-boundary propagator} $B(k;\tau)$, which connects a bulk point with time $\tau$ to a boundary point at the boundary $\tau=\tau_f$ with momentum $\bm k$. For conformal scalars, all propagators depend on the momentum $\bm k$ only in its magnitude $k=|\bm k|$. Explicitly, the bulk propagator is given by:
\bge
\label{eq_bulkpropWC}
  G(K;\tau_1,\tau_2)=\FR{1}{2K}\Big[e^{-\ii K(\tau_1-\tau_2)}\theta(\tau_1-\tau_2)+e^{+\ii K(\tau_1-\tau_2)}\theta(\tau_2-\tau_1)-e^{+\ii K(\tau_1+\tau_2)}\Big].
\ede
The bulk propagator is a solution to the inhomogeneous equation of motion with a $\de$-function source, and satisfies the boundary condition $G\to 0$ when either of the two time variables goes to either the past infinity or the future infinity. On the other hand, the bulk-to-boundary propagator is given by:
\bge
\label{eq_btobpropWC}
  B(k;\tau)=e^{\ii k\tau}.
\ede
This propagator is a solution to the homogeneous equation of motion, with the boundary conditions that $B\to 0$ when $\tau\to-\infty$ and $B\to 1$ when $\tau\to \tau_f$.

Then, for each $n$-point interaction vertex, one assigns a time variable $\tau_i$, includes the usual factor $\ii \lam_n(-\tau_i)^{P_n}$ coming from (\ref{eq_actionflat}), and then integrates the time variable $\tau_i$ over $(-\infty,\tau_f)$. In practice, we always consider the case $\tau_f\to 0$ so it is typical that we integrate $\tau_i$ over the entire negative real axis. 

\begin{figure}[t]
\centering
\includegraphics[width=0.35\textwidth]{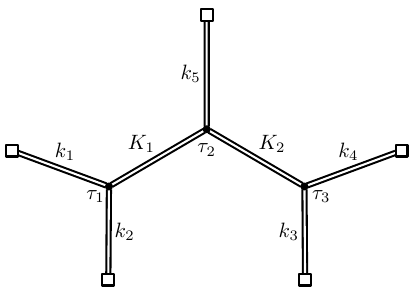}
\hspace{1cm}
\includegraphics[width=0.35\textwidth]{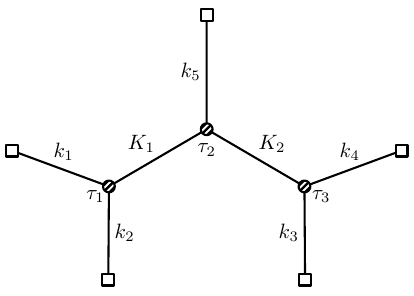}
\caption{A five-point function. The left and right diagrams are for the wavefunction coefficient and the correlator, respectively. (See App.\ \ref{app_notations} for our diagrammatic notations for Feynman diagrams.)  We emphasize that the wavefunction coefficient and the correlation function shown in this figure do not have a direct one-to-one relationship.}
\label{fig_5pt}
\end{figure}

It would be more clear to illustrate this rule with an explicit example, as shown in the left graph of Fig.\;\ref{fig_5pt}. Since we are considering wavefunction coefficients and correlation functions simultaneously in this work, we use two sets of diagrammatic notations for wavefunctions and correlators to highlight their difference.

\begin{align}
\label{eq_5ptWF}
  \psi_5
  =&~\ii^3\int_{-\infty}^0 \di\tau_1\di\tau_2\di\tau_3\,(-\tau_1)^{P_{3}}(-\tau_2)^{P_{3}}(-\tau_3)^{P_{3}} B(k_1;\tau_1)B(k_2;\tau_1)B(k_3;\tau_3)B(k_4;\tau_3)B(k_5;\tau_2)\n\\
  &\times G(K_1;\tau_1,\tau_2)G(K_2;\tau_2,\tau_3)\n\\
  =&~\FR{-\ii}{2K_12K_2}\int_{-\infty}^0 \di\tau_1\di\tau_2\di\tau_3\,(-\tau_1)^{P_{3}}(-\tau_2)^{P_{3}}(-\tau_3)^{P_{3}}e^{\ii(k_{12}\tau_1+k_{34}\tau_3+k_5\tau_2)}\n\\
  &\times\Big[e^{-\ii K_1(\tau_1-\tau_2)}\theta(\tau_1-\tau_2)+e^{+\ii K_1(\tau_1-\tau_2)}\theta(\tau_2-\tau_1)-e^{+\ii K_1(\tau_1+\tau_2)}\Big]\n\\
  &\times\Big[e^{-\ii K_2(\tau_2-\tau_3)}\theta(\tau_2-\tau_3)+e^{+\ii K_2(\tau_2-\tau_3)}\theta(\tau_3-\tau_2)-e^{+\ii K_2(\tau_2+\tau_3)}\Big].
\end{align}

In this work, we only consider tree-level graphs, where the momenta of all bulk propagators are fixed by the external ones, and thus there is no loop momentum integral. It is then clear that, to compute a tree-graph contribution to any wavefunction coefficients, we need to perform an integral over time variables for all vertices, with the integrand being the products of all propagators and vertices involved. A graph of $V$ vertices thus corresponds to a $V$-layer time integral. Thus, contrary to the flat-space Feynman integrals, the level of difficulty for computing wavefunction coefficients increases not only with the number of loops, but also with the number of vertices. In addition, given the time-ordering $\theta$-functions in the bulk propagator (\ref{eq_bulkpropWC}), these $V$-layer integrals can be nested, meaning that the integration variable of one layer can be the integration limits of the inner layer.

In the next section, we will provide a systematic procedure to deal with arbitrarily nested integrals encountered in the computation of wavefunctions. Next, we will turn to the computation of correlation functions. It turns out that, also the correlation function and wavefunction coefficient are conceptually very different objects, their bulk computations  involve time integrals of very similar structure.

\paragraph{Correlation functions} While the tree-level wavefunction coefficients are easily computable from the above perturbative expansion, they are not directly the observables of cosmological interests. The observables are correlation functions.

In principle, once we get a result for the wavefunction $\Psi[\varphi]$ (or equivalently, the wavefunction coefficients $\psi_n$) as above, we can compute the correlation functions of $\varphi$ using (\ref{eq_ctowf}). Expanding the right-hand side of (\ref{eq_ctowf}) in coupling constants, we get a set of equations that relate the correlation functions with the wavefunction coefficients order by order in perturbation theory. 

In practice, so long as the correlation functions are the only concern, one can bypass the wavefunction formalism and develop a perturbation expansion directly for the correlation functions, which is nothing but the well-known Schwinger-Keldysh formalism or in-in formalism (See e.g.\ \cite{Chen:2017ryl} for a review). In this formalism, we compute the correlation function from a closed-path integral:
\begin{align}
  &\la\varphi(\bm k_1)\cdots\varphi(\bm k_n)\ra'(2\pi)^{d}\de^{(d)}(\bm k_1+\cdots+\bm k_n)\n\\
 =&\int\mathcal{D}\varphi_+\mathcal{D}\varphi_-\,\varphi_+(\tau_f,\bm k_1)\cdots\varphi_+(\tau_f,\bm k_n)e^{\ii S[\varphi_+]-\ii S[\varphi_-]}\prod_{\bm k}\de\Big[\varphi_{+}(\tau_f,\bm k)-\varphi_{-}(\tau_f,\bm k)\Big].
 \label{eq_crfuncdef}
\end{align}
We can then compute the correlation function directly by Feynman diagram expansion of the above closed-path integral. Effectively, the phase $e^{\ii S[\varphi_+]-\ii S[\varphi_-]}$ of the path integral tells us that all interaction vertices are doubled, with a plus type and a minus type. A plus-type vertex is associated with a factor of $-\ii\lam_n(-\tau_i)^{P_n}$ and an integration over $\tau_i$; while a minus-type vertex has $+\ii\lam_n(\tau_i)^{P_n}$. Also, the bulk propagators come in four types, depending on the vertex types of its two endpoints: 
\begin{align}
\label{eq_Dpmpm}
  D_{\pm\pm}(k;\tau_1,\tau_2)=&~\FR{1}{2k}\Big[e^{\mp\ii k(\tau_1-\tau_2)}\theta(\tau_1-\tau_2)+e^{\pm\ii k(\tau_1-\tau_2)}\theta(\tau_2-\tau_1)\Big]\\
\label{eq_Dpmmp}
  D_{\pm\mp}(k;\tau_1,\tau_2)=&~\FR{1}{2k}\,e^{\pm\ii k(\tau_1-\tau_2)}.
\end{align}
On the other hand, the bulk-to-boundary propagators come in two types, depending on the type of its bulk endpoint:
\bge
\label{eq_btobpropSK}
  D_{\pm}(k;\tau)=\FR{1}{2k}e^{\pm\ii k\tau}.
\ede
So, once again, the computation of a tree-graph contribution to a correlation function with $V$ vertices involves a $V$-fold nested time integral, with the integrand being the products of all propagators and vertices. The new complication for the correlation function is that one needs to sum over all possible assignment of plus and minus types to the vertices. We again illustrate this rule with the example of a 5-point function, shown in the right diagram of Fig.\;\ref{fig_5pt}.
\begin{align}
\label{eq_5ptC}
  \mathcal{T}_5
  =&\sum_{\aa_1,\aa_2,\aa_3=\pm}\aa_1\aa_2\aa_3(-\ii)^3\int_{-\infty}^0 \di\tau_1\di\tau_2\di\tau_3\,(-\tau_1)^{P_{3}}(-\tau_2)^{P_{3}}(-\tau_3)^{P_{3}}D_{\aa_1}(k_1;\tau_1)D_{\aa_1}(k_2;\tau_1) \n\\
  &\times D_{\aa_3}(k_3;\tau_3)D_{\aa_3}(k_4;\tau_3)D_{\aa_2}(k_5;\tau_2)D_{\aa_1\aa_2}(K_1;\tau_1,\tau_2)D_{\aa_2\aa_3}(K_2;\tau_2,\tau_3).
\end{align}
We see that, at the computation level, the wavefunction coefficients and correlators are quite similar. Both of them involve nested time integrals over arbitrary power functions and exponential functions. In the next section, we explain how to compute these nested time integrals with the method of family-tree decomposition.

\section{Bulk Time Integrals and Family-Tree Decomposition}
\label{sec_family}

In the previous section, we have introduced the perturbative formalism to compute the tree-level wavefunction coefficients and the correlation functions in a FRW universe. Technically, the computation boils down to a $V$-fold time integral, where $V$ is the number of vertices. The integrand consists of simple power functions of time variables coming from the interaction vertices, the exponential functions from various types of propagators, as well as the time-ordering $\theta$ functions. Clearly, the integrals for the wavefunction coefficients and the correlation functions have very similar structure, so that we can treat them in a unified manner. 

\subsection{General bulk tree graphs}

We begin with an arbitrary tree graph $\mathcal{G}(\{\bm k\})$ in the bulk, which can be a graph for either the wavefunction coefficient or the correlator. In general, when a graph is given, its value is a function of all external momenta $\bm k$ subject to $d$-momentum conservation. However, for a given graph, we can use a better description for this dependence. Concretely, we distinguish between the momenta $\{\bm k\}$ of external lines and momenta $\{\bm K\}$ of internal lines, and one can easily see from the previous rules that a tree graph is a function of all $k\equiv|\bm k|$ and all $K\equiv|\bm K|$. We use a widely adopted terminology for the magnitude of a momentum and call it the \emph{energy}. Thus, we can write a tree graph as $\mathcal{G}(\{k\},\{K\})$, where $\{k\}$ denotes the set of all external energies, while $\{K\}$ denotes the set of all internal energies. 

It proves useful to further rescale the wavefunction coefficient $\wt\psi$ and the correlator $\wt{\mathcal{T}}$ in the following way:
\begin{align}
  \label{eq_reswfcoef}
  \psi\big(\{k\};\{K\}\big)\equiv &~\bigg(\prod_{\{K\}}\FR{1}{2K}\bigg)\wt{\psi}\big(\{k\};\{K\}\big), \\
  \label{eq_rescrfunc}
  \mathcal{T}\big(\{k\};\{K\}\big)\equiv &~\bigg(\prod_{\{k\}}\FR{1}{2k}\prod_{\{K\}}\FR{1}{2K}\bigg)\wt{\mathcal{T}}\big(\{k\};\{K\}\big).
\end{align}
That is, we remove all internal-energy factors of $1/(2K)$ from the wavefunction coefficient, which come from the same factor in the bulk propagator (\ref{eq_bulkpropWC}). Also, we remove all energy factors $1/(2k)$ and $1/(2K)$ (both external and internal) from the correlator, which come from the same factor in all types of propagators for computing correlators, namely (\ref{eq_Dpmpm}), (\ref{eq_Dpmmp}), and (\ref{eq_btobpropSK}). 

Thus, to compute a rescaled wavefunction coefficient or a rescaled correlator, we can use the following rescaled propagators:
\begin{align}
  \label{eq_resG}
  \wt G(k;\tau_1,\tau_2)\equiv&~ (2k)G(k;\tau_1,\tau_2);\\
  \label{eq_resD}
  \wt D_{\aa\bb}(k;\tau_1,\tau_2)\equiv&~ (2k)D_{\aa\bb}(k;\tau_1,\tau_2); \\
  \label{eq_resDbtob}
  \wt D_{\aa}(k;\tau)\equiv&~ (2k)D_{\aa}(k;\tau),~~~~~(\aa,\bb=\pm)
\end{align}

At this point, it is useful to note that the bulk-to-boundary propagator $B(k;\tau)$ given in (\ref{eq_btobpropWC}) satisfies an obvious relation:
\bge
  \prod_{i=1}^n B(k_i;\tau)=B\bigg(\sum_{i=1}^n k_i;\tau\bigg).
\ede
So are the \emph{rescaled} bulk-to-boundary propagators $\wt D_\pm(k;\tau)$ in the Schwinger-Keldysh formalism, given in (\ref{eq_resDbtob}). 
Thus, so far as the computation of the bulk time integral is the only concern, the number of bulk-to-boundary propagators attached to a given vertex does not really matter. The above relation tells us we can attach only one bulk-to-boundary propagator to each vertex, with the energy being the sum of all \emph{external} energies at one vertex. Therefore, to fully specify the kinematics for a given graphic contribution to the \emph{rescaled} amplitude (either wavefunction coefficient or correlator) with $V$ vertices and $I$ internal propagators, it is sufficient to assign an \emph{external energy} $E_i$ to each vertex with time variable $\tau_i$ ($i=1,\cdots, V$), and assign an \emph{internal energy} $K_j$ ($j=1,\cdots I$) to each bulk propagator.

\begin{figure}[t]
\centering
\includegraphics[width=0.84\textwidth]{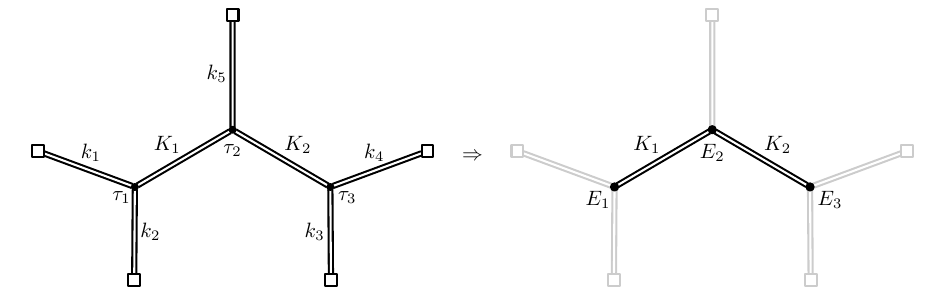} 
\caption{The kinematics of a tree-graph contribution to the rescaled wavefunction coefficient $\wt\psi_5$. The kinematics is fully specified by the three external energies $E_1\equiv k_{12}$, $E_2\equiv k_5$, and $E_3\equiv k_{34}$, together with the two internal energies $K_1$ and $K_2$. Thus, when computing $\wt\psi_5$, all bulk-to-boundary propagators can be neglected.}
\label{fig_5ptRed}
\end{figure}

Diagrammatically, the above observation means that we can remove all external legs (namely, the lines attached to external boxes in Fig.\;\ref{fig_5pt}), and, instead, we attach a \emph{total external energy} variable $E_i$ to the $i$'th bulk vertex ($i=1,\cdots,V$). Then, each vertex (with label $i$) is associated with three variables, including the time $\tau_i$, the external energy $E_i$, and the exponent $q_i$ that appears in the power factor $(-\tau_i)^{q_i-1}$ in the integrand. In addition, when computing correlators, we also need to assign an SK index $\aa_i=\pm$ to the $i$'th vertex. See Fig.\;\ref{fig_5ptRed} for an illustration. Here, for later convenience, we have rewritten the power as $q-1$ in place of $P_n$ defined previously. We will stick to $q$ variable from now on. Using (\ref{eq_actionflat}), we find $q=1+[d+1-(d-1)n/2] \wt p$.

\paragraph{Nested time integrals.}

From the previous section, it is clear that the computation of cosmological amplitudes directly from the bulk is hampered by the difficulty of performing nested time integrals. Fortunately, a recent work \cite{Xianyu:2023ytd} has found a systematic procedure for performing arbitrarily nested time integrals. The integrand can be arbitrary powers and exponential functions, which exactly covers all cases of interest in this work. Also, as shown in \cite{Xianyu:2023ytd}, when used together with the partial Mellin-Barnes representation, this method can also be used to compute arbitrary tree graphs with general massive exchanges.  

The most general time integral arising from the expression of tree-level conformal amplitudes of the last section has the following form:
\begin{align}
\label{eq_NTI}
  \mb{T}^{(q_1\cdots q_N)}_{\mathscr{N}}(\omega_1,\cdots,\omega_N)\equiv (-\ii)^N\int_{-\infty}^0\prod_{\ell=1}^N\Big[\di\tau_{\ell}\,(-\tau_\ell)^{q_\ell-1}e^{\ii \omega_{\ell}\tau_\ell}\Big]\prod_{(i,j)\in\mathscr{N}}\theta(\tau_j-\tau_i).
\end{align}
Here on the left-hand side, we use $\mathscr{N}$ to denote the time order structure of the integral. Thus, on the right-hand side, the products of Heaviside $\theta$ functions are determined by time order structure $\mathscr{N}$.

Here the prefactor $(-\ii)^N$ is inherited from the factor of $-\ii$ at each interaction vertex for computing correlators, and we include it here for later convenience. 

A direct integration of (\ref{eq_NTI}) is clearly difficult. Also, the final answer to this general integral might not correspond to any well-known special functions. In a sense, we may say that the integral (\ref{eq_NTI}) itself defines a large class of special functions worth more thorough studies on their own rights. With this said, as the first step, the best we can do with such integrals is to find their series expansions at points of interest. Also, when a series expansion fails to converge, it would be desirable to find its analytical continuation beyond the region of convergence.

The method of family-tree decomposition is proposed to reach these goals. This method begins with a very simple observation: Although it is impossible to resolve the nested time integrals into factorized time integrals without nesting, it is nevertheless possible to change the direction of the nesting function $\theta(\tau_{1}-\tau_2)$ using the relation $\theta(\tau_1-\tau_2)+\theta(\tau_2-\tau_1)=1$. Thus, we can switch the directions of any number of directional lines as we want, at the expense of producing additional graphs containing less directional lines. 

The family-tree decomposition method makes use of this flexibility of switching directions. According to this method, we decompose any nested time integral into a sum of several ``graphs'', such that, in each graph of this sum, all time variables are either factorized or partially ordered. 

It proves very convenient to adopt the following terminology when studying a partially-ordered graph: For any two vertices $\tau_i$ and $\tau_j$ connected by a directional line $\theta(\tau_j-\tau_i)$, we call $\tau_i$ the \emph{mother} and $\tau_j$ the \emph{daughter}. Then, the structure of partial ordering itself can be described as ``a mother can have many daughters but a daughter can only have one mother.'' Also, with this terminology, a nested time integral with a partial order naturally becomes a maternal family tree, hence the name of family-tree decomposition.

\subsection{Family tree integral}
Once we have decomposed a nested time integral into a sum of products of family trees, we can focus on the computation of a single family tree. Once again, in our terminology, a family tree is a nested time integral with a particular partial order. As shown in \cite{Xianyu:2023ytd}, a partial order allows us to directly write down a series representation for the family tree, expanded as inverse powers of either the earliest energy variable, or the total energy. However, before presenting this expression, let us introduce a convenient notation to keep track of the partial order structure for an arbitrary family tree.

Similar to a general nested time integral, a family tree with $N$ sites is fully specified by the three variables $(\tau_i,\omega_i,q_i)$ at the $i$th vertex ($i=1,\cdots, N$), together with a partial order structure. Thus, without spelling out these variables explicitly every time, we can simply specify a $N$-site family tree by $N$ numbers within a square bracket like $[1\cdots N]$. To manifest the partial order structure, we add additional pairs of parentheses. More explicitly, the rules are the following:
\begin{enumerate}
  \item Every family tree integral is represented by a string of numbers within a pair of square brackets.
  \item The leftmost number denotes the earliest member. 
  \item If a mother has only one daughter, write the daughter number directly to the right of the mother number. 
  \item If a mother has more than one daughter, use a pair of parentheses to enclose the entire sub-family of each of these daughters, and write all these subfamilies (in parentheses) to the right of the mother number. The order of subfamilies at the same level is arbitrary. Here, a sub-family of a site means herself and all her descendants. 
  
\end{enumerate}
We use the following several examples to illustrate this notation:
\begin{align}
  \label{eq_familytree_123def}
  \ft{123}=&~(-\ii)^3\int\prod_{i=1}^3\Big[\di\tau_i(-\tau_i)^{q_i-1}e^{\ii\omega_i\tau_i}\Big]\theta_{32}\theta_{21},\\
  \ft{1(2)(3)(4)}=&~(-\ii)^4\int\prod_{i=1}^4\Big[\di\tau_i(-\tau_i)^{q_i-1}e^{\ii\omega_i\tau_i}\Big]\theta_{41}\theta_{31}\theta_{21},\\
  \ft{1(2)\big(3(4)(5)\big)}=&~(-\ii)^5\int\prod_{i=1}^5\Big[\di\tau_i(-\tau_i)^{q_i-1}e^{\ii\omega_i\tau_i}\Big]\theta_{43}\theta_{53}\theta_{31}\theta_{21}.
\end{align}
Here and in some equations below, we use $\theta_{ij}\equiv \theta(\tau_i-\tau_j)$ for short. Sometimes, we do not want to specify the partial order, but only refer to a family tree with a general partial order. In this case, we write $[\mathscr{P}(1\cdots N)]$, meaning that the $N$ sites here are organized into a particular partial order structure $\mathscr{P}$.

Also, in our definition of family trees, the energy variable of the $i$th site is simply written as $\omega_i$. However, as discussed before, when computing the amplitudes, we need to distinguish between external energies $E_i$ associated to vertices, and the internal energies, which are defined with respect to bulk lines. Therefore, we need to refine our notation here to make this distinction manifest. This can be achieved by adding subscripts to the number for each site. We will elaborate this refined notation in the next section.

Now we introduce series expressions for an arbitrary family tree. As discussed in \cite{Xianyu:2023ytd}, there are many possible representations, expressing the family tree as power series of the inverse earliest energy $1/\omega_1$ or the inverse total energy $1/\omega_{1\cdots N}$. One can even consider a hybrid power series with more than one expansion parameter. The general expression for the earliest energy power series was given in \cite{Xianyu:2023ytd}, and here we directly quote the result and adapt it to our notation:
\begin{keyeqn}
\begin{align}
  \label{eq_family}
   &\ft{\mathscr{P}(\wh{1}2\cdots N)}=\FR{(-\ii)^N}{(\ii \omega_1)^{q_{1\cdots N}}}
   \sum_{n_2,\cdots,n_N=0}^\infty \Gamma(q_{1\cdots N}+n_{2\cdots N})\prod_{j=2}^N\FR{(-\omega_j/\omega_1)^{n_j}}{(\wt{q}_j+\wt{n}_j)n_j!}.
\end{align}
\end{keyeqn}
Here we have added a hat on the variable $\wh 1$ to highlight the fact that we are choosing Site 1 as the earliest site, namely, the ancestor of the whole family.\footnote{Note that our definition $[\mathcal{P}\{\wh 12\cdots N\}]$ differs from the quantity $\mathcal{C}_{q_1\cdots q_N}(\wh E_1,\cdots, E_N)$ used in \cite{Xianyu:2023ytd} by an additional overall factor $(-\ii)^N$.} On the right-hand side, we have introduced $N-1$ summation variables $n_2,\cdots,n_N$ for all but the earliest sites. Also, we use $\wt n_j$ to denote the sum of all $n$ variables of Site $j$ and her descendants. For instance, in $[1(23)(4(5)(67))]$, we have $\wt n_2=n_{23}$ and $\wt n_4=n_{4567}$. The variable $\wt q_j$ is similarly defined. Thus, the above result expresses a family tree as a series of the inverse of the energy variable of the earliest site, namely $1/\omega_1$ in the above equation. This is a Taylor series, except for the overall factor $(\ii\omega_1)^{-q_{1\cdots N}}$ which encodes all the nonanalytic behavior (such as the monodromy) of the family tree in $\omega_1$ as $\omega_1\to \infty$. Thus, we expect that the above series representation is well convergent when $\omega_1$ is much greater than the rest of energy variables. 

There is another series representation for the same object $[\mathscr{P}(\wh{1}2\cdots N)]$, which is written as a power expansion of $\omega_{1\cdots N}^{-1}$ where $\omega_{1\cdots N}\equiv \omega_1+\cdots+\omega_N$ denotes the total energy. This representation was briefly discussed in \cite{Xianyu:2023ytd} but no general expression was given there. Here we provide the general formula:
\begin{keyeqn}
\begin{align}
  \label{eq_family_total}
   \ft{\mathscr{P}(\wh{1}2\cdots N)}
   =&~\FR{-\ii^N}{(\ii \omega_{1\cdots N})^{q_{1\cdots N}}}
   \sum_{n_2,\cdots,n_N=0}^\infty \Gamma(q_{1\cdots N}+n_{2\cdots N}) 
   \prod_{j=2}^N \FR{(-\wt \omega_j/\omega_{1\cdots N})^{n_j}}{(-\wt q_j-\wt n_j)_{1+n_j}} ,
\end{align}
\end{keyeqn}
where we introduce another new variable $\wt\omega_j$ which denotes the sum of all energies of Site $j$ and her descendants. Also, we used the Pochhammer symbol $(z)_n\equiv \Gamma(z+n)/\Gamma(z)$. Although (\ref{eq_family}) and (\ref{eq_family_total}) look different, they are simply different series representations of the same function. 

It is obvious that the family-tree decomposition is highly non-unique. There are many ways to rewrite a nested time integral as sum of partially ordered graphs. This is a key point of the method: As we shall see, different decompositions lead to superficially different expressions, but they all correspond to the same function. In the most general case where only series expansions are possible, different decompositions give the series expansions of the same function but in different kinematic regions. In the event that these series expansions do sum into known special functions, the many ways of family-tree decomposition lead to many transformation-of-variable formulae for these special functions. 

Here we give several examples:
\begin{align}
  &\ft{12}+\ft{21}=\ft{1}\ft{2}.\\
  &\ft{123}+\ft{2(1)(3)}=\ft{1}\ft{23}.\\
  &\ft{123}-\ft{321}=\ft{1}\ft{23}-\ft{21}\ft{3}.
\end{align}
These seeming trivial relations and their higher-site generalizations give a large number of ``transformation of variable'' formulae for various types of hypergeometric functions. We collect some of them in App.\;\ref{app_transf}.
Importantly, when a family tree does not correspond to any well studied special function and only the series representation is available, the formulae as above can be regarded as very convenient ways to take analytical continuation of the series representation beyond its region of convergence. It is quite remarkable that such highly intricate functional relations can be derived through very simple manipulation of the integrand, namely, we only need to switch some time orderings. This is reminiscent of another related area of polylogarithms, where many nontrivial functional relations can be derived by simple manipulations of the integrands that defines the polylogarithmic functions, which are known as symbols. Indeed, as we shall show in Sec.\;\ref{sec_inflation}, our family trees do reduce to polylogarithmic functions when we send all exponent $q$ to zero. Also, the above ``transformation of variable'' formulae for hypergeometric functions reduce to known functional relations among polylogarithmic functions. Thus, the family tree decomposition procedure can be regarded as a kind of symbology for a very large class of multivariate hypergeometric series, worthy of further exploration.

Furthermore, we can make use of the identity $\theta(\tau_i-\tau_j)+\theta(\tau_j-\tau_i)=1$ in another direction: We can insert as many $\theta(\tau_i-\tau_j)+\theta(\tau_j-\tau_i)$ as we want in the time integrand, in such a way that, when combined with original $\theta$ functions in the integrand, the resulting integrand breaks into a sum of many ``family chain''. That is, each term is itself a family tree with a linear chain structure, so that every mother has only one daughter. In this way, we can lift the original partial order structure to total orders. We will discuss this reduction in more detail in Sec.\;\ref{sec_energyint}. 

We will not directly make use of this ``family chain'' decomposition in other places of this work, since we have already got the analytical expression for more general and partially ordered family trees. However, we note that this ``family chain'' reduction shows that the family trees are actually an over-complete set of functions for expressing tree-level conformal amplitudes. Of course, being over-complete is never a drawback given that we have found the analytical expressions for all of them. However, the family chain reduction does show that we can further reduce the cosmological amplitudes into a smaller class of iterated integrals, which starts to look very similar to the role played by the polylogarithmic functions and symbology techniques in the study of $\phi^3$ theory in inflation. Thus, the family-tree or family-chain integrals can be viewed as generalizations of polylogarithmic functions and deserve more detailed studies.

\section{Selected Examples and General Rules}
\label{sec_example}

In this section, we will walk the readers through a few examples, which are simple enough to be computed by brute force, but at the same time complicated enough to reveal the general structure of arbitrary conformal amplitudes, so that readers can see the emergence of a set of general rules for directly writing down answers for more complicated amplitudes. Thus, we think that it would be fun to discuss these examples before presenting the general rules in the next section.

Below, we study three examples, the two-site chain, the three-site chain, and the four-site star. In each case, we will consider both the correlation function and the wavefunction coefficient. It appears that the correlation function is a bit more complicated than the wavefunction. Thus, in each case, we shall study the wavefunction coefficient first, and then study the correlation function. Let us emphasize again that each example studied below does not correspond to a unique amplitude; rather, one can freely associate any number of external lines (bulk-to-boundary propagators) to each site. Thus, one can generate amplitudes of different interactions with different number of external points from a given example. It's just that all these amplitudes boil down mathematically to the same type of integral.

\subsection{Two-site chain} 

The graphs with only one bulk vertex represents a contact graph of arbitrary external legs. This example is in a sense trivial even in general FRW background, since the bulk integral can be directly finished, and the results for the wavefunction coefficient $\wt\psi_\text{1-site}$ and the correlator $\wt{\mathcal{T}}_\text{1-chain}$ are respectively given by:
\begin{align}
  \label{eq_1site_CF}
  \wt\psi_\text{1-site}(E)
  =&~\ii \int_{-\infty}^0\di\tau\,(-\tau)^{q-1} e^{+\ii E\tau}=\FR{\ii\Gamma(q)}{(\ii E)^q}.\\
  \wt{\mathcal{T}}_\text{1-site}(E)
  =&-\ii\int_{-\infty}^0\di\tau\,(-\tau)^{q-1} e^{+\ii E\tau}+\text{c.c.}=\FR{-\ii\Gamma(q)}{(\ii E)^q}+\text{c.c.}.
\end{align}

Thus, the simplest nontrivial example is a two-site chain, which represents an arbitrary tree graph with a single bulk exchange. Below, we consider this example in detail.

\paragraph{Wavefunction coefficient} As mentioned above, the case of wavefunction coefficient is somewhat simpler than the correlator, and so let us begin with the coefficient. According to the general rule presented in Sec.\ \ref{sec_frw} and also in Sec.\ \ref{sec_family}, the bulk integral for the rescaled wavefunction coefficient $\wt\psi_\text{2-chain}$ can be written as:
\begin{align}
\label{eq_2siteWCint}
 & \wt\psi_\text{2-chain}(E_1,E_2,K_1)
 =\ii^2\int_{-\infty}^0\di\tau_1\di\tau_2\,(-\tau_1)^{q_1-1}(-\tau_1)^{q_2-1}B(E_1;\tau_1)B(E_2;\tau_1) \wt G(K_1;\tau_1,\tau_2).
\end{align}
At this point we may insert the explicit expressions for the propagators in (\ref{eq_bulkpropWC}) (\ref{eq_btobpropWC}), and  (\ref{eq_resG}), and compute the integral directly, as the integral in this case is simple enough. However, for later generalizability, we want to make a slight detour, and reorganize the integral with a particular partial-order structure. As we can see, in its original form (\ref{eq_bulkpropWC}), the bulk propagator contains three terms, two of which have time directions opposite to each other. Our strategy is to rewrite this propagator with a fixed time ordering. There are two ways to achieve this:
\begin{align}
\label{eq_Gtau1}
  &\wt G(K_1;\wh\tau_1,\tau_2)=\Big[e^{+\ii K_1(\tau_1-\tau_2)}-e^{-\ii K_1(\tau_1-\tau_2)}\Big]\theta(\tau_2-\tau_1)+e^{-\ii K_1(\tau_1-\tau_2)}-e^{+\ii K_1(\tau_1+\tau_2)};\\
\label{eq_Gtau2}
  &\wt G(K_1;\tau_1,\wh\tau_2)=\Big[e^{-\ii K_1(\tau_1-\tau_2)}-e^{+\ii K_1(\tau_1-\tau_2)}\Big]\theta(\tau_1-\tau_2)+e^{+\ii K_1(\tau_1-\tau_2)}-e^{+\ii K_1(\tau_1+\tau_2)}.
\end{align}
Here we have added a hat to the \emph{earlier} time variable in the argument of $\wt G$ to make clear which of the two possible orderings we are choosing. 
As we shall see, the two different ways of splitting $\wt G$ lead to equivalent yet superficially different results.  

Now, suppose we set $\tau_1$ to be the earlier time variable, and thus substitute (\ref{eq_Gtau1}) into (\ref{eq_2siteWCint}), we get four terms in the integrand:
\begin{align}
  &\wt\psi_\text{2-chain}(E_1,E_2,K_1)
 =\ii^2\int_{-\infty}^0\di\tau_1\di\tau_2\,(-\tau_1)^{q_1-1}(-\tau_1)^{q_2-1}e^{\ii E_1\tau_1+\ii E_2\tau_2}\n\\
  &\times\bigg\{\Big[e^{+\ii K_1(\tau_1-\tau_2)}-e^{-\ii K_1(\tau_1-\tau_2)}\Big]\theta(\tau_2-\tau_1)+e^{-\ii K_1(\tau_1-\tau_2)}-e^{+\ii K_1(\tau_1+\tau_2)}\bigg\}.
\end{align}
Since each term of this integral has the form of a family-tree integral introduced in the previous section, we can directly write down the final answer using our notations for the family-tree integrals, as follows:
\begin{align}
\label{eq_psi2sol1}
  \wt\psi_\text{2-chain}(\wh E_1,E_2,K_1)= \ft{1_1 2_{\bar 1}}-\ft{1_{\bar 1}2_{1}}+\ft{1_{\bar 1}}\ft{2_1}-\ft{1_1}\ft{2_1} .
\end{align}
Here, on the left-hand side, we are putting a hat to $E_1$ which is inherited from our notation for $G(K;\wh\tau_1,\tau_2)$, showing that $E_1$ is the energy variable of the earlier site. On the right-hand side, we added a subscript to each external energy variable to indicate an internal energy to be added, or subtracted if the subscript appeared with a bar. For instance, $1_1$ means $E_1+K_1$ while $2_{\bar 1}$ means $E_2-K_1$.

At this point, let us also introduce the following diagrammatic representation for Eq.\ (\ref{eq_psi2sol1}):
\bge
\parbox{0.8\textwidth}{\includegraphics[width=0.8\textwidth]{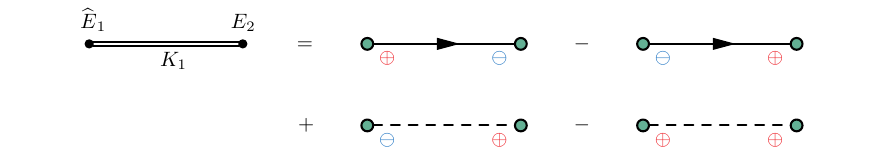}}
\ede
Here the left-hand side is the bulk part of the wavefunction coefficient. On the right-hand side, the four diagrams correspond to the four terms on the right-hand side of (\ref{eq_psi2sol1}). Here, a solid directional line means that the two connected sites are nested and the arrow indicates the time direction. A dashed line means that the two time variables of its two endpoints are factorized. A colored (empty) circle means a positive (negative) energy variable. (In the case of wavefunction, all energy variables are positive, and thus all endpoints are colored.) Finally, the signs of the two momentum variables in a line are explicitly denoted by circled symbols {\color{Red}$\oplus$} and {\color{RoyalBlue}$\ominus$}. 

On the other hand, we can equally choose $\tau_2$ as the earlier time. Then, we should substitute (\ref{eq_Gtau2}) into (\ref{eq_2siteWCint}), and get another expression for the two-site wavefunction coefficient:
\bge
\label{eq_psi2sol2}
  \wt\psi_\text{2-chain}(E_1,\wh E_2,K_1)
  =\ft{2_1 1_{\bar 1}}-\ft{2_{\bar 1}1_1}+\ft{2_{\bar 1}}\ft{1_1}-\ft{2_1}\ft{1_1}.
\ede
The corresponding diagrammatic representation for this decomposition is:
\bge
\parbox{0.8\textwidth}{\includegraphics[width=0.8\textwidth]{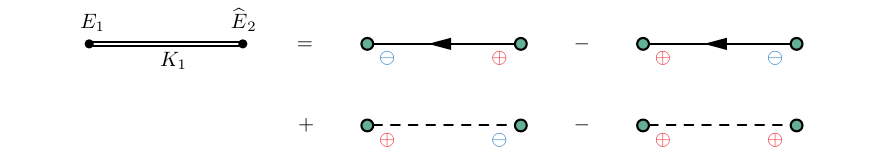}}
\ede

Now, we have obtained two expressions, namely (\ref{eq_psi2sol1}) and (\ref{eq_psi2sol2}), for the same wavefunction coefficient $\wt\psi_\text{2-chain}$. To see the equivalence of the two, we can substitute in the explicit expressions for the family-tree integrals. First, if we use the earliest energy representation (\ref{eq_family}) and finish the summation, then (\ref{eq_psi2sol1}) becomes: 
\begin{align}
\label{eq_2chainWp1}
  &\wt\psi_\text{2-chain}(\wh E_1,E_2,K_1)\n\\
  =&-\FR{1}{[\ii(E_1+K_1)]^{q_{12}}}{~}_2\mathcal{F}_1\left[\bgm q_2,q_{12}\\ q_2+1\edm\middle|\FR{K_1-E_2}{K_1+E_1}\right]
  +\FR{1}{[\ii(E_1-K_1)]^{q_{12}}}{~}_2\mathcal{F}_1\left[\bgm q_2,q_{12}\\ q_2+1\edm\middle|\FR{K_1+E_2}{K_1-E_1}\right]\n\\
  &-\FR{\Gamma[q_1,q_2]}{[\ii(E_1-K_1)]^{q_1}[\ii(E_2+K_1)]^{q_2}}+\FR{\Gamma[q_1,q_2]}{[\ii(E_1+K_1)]^{q_1}[\ii(E_2+K_1)]^{q_2}}.
\end{align} 
Using the same expression (\ref{eq_family}) but choosing $E_2$ as the earliest energy, we can also get the full expression for $\wt\psi_\text{2-chain}(E_1,\wh E_2,K_1)$ as:
\begin{align}
\label{eq_2chainWp2}
&\wt\psi_\text{2-chain}(E_1,\wh E_2,K_1)\n\\
  =&-\FR{1}{[\ii(E_2+K_1)]^{q_{12}}}{~}_2\mathcal{F}_1\left[\bgm q_1,q_{12}\\ q_1+1\edm\middle|\FR{K_1-E_1}{K_1+E_2}\right]
  +\FR{1}{[\ii(E_2-K_1)]^{q_{12}}}{~}_2\mathcal{F}_1\left[\bgm q_1,q_{12}\\ q_1+1\edm\middle|\FR{K_1+E_1}{K_1-E_2}\right]\n\\
  &-\FR{\Gamma[q_1,q_2]}{[\ii(E_2-K_1)]^{q_2}[\ii(E_1+K_1)]^{q_1}}+\FR{\Gamma[q_1,q_2]}{[\ii(E_2+K_1)]^{q_2}[\ii(E_1+K_1)]^{q_1}}.
\end{align} 
On the other hand, we can also use the total-energy series (\ref{eq_family_total}) to express these family trees. The summation can again be done in closed form, and we get another two expressions for the same 2-site chain wavefunction coefficient:
\begin{align}
\label{eq_2chainWt1}
  &\wt\psi_\text{2-chain}(\wh E_1,E_2,K_1)\n\\
  =&-\FR{\Gamma[q_2]}{(\ii E_{12})^{q_{12}}}\bigg\{{}_2\mathcal{F}_1\left[\bgm 1,q_{12}\\ q_2+1\edm\middle|\FR{E_2-K_1}{E_{12}}\right]
  +{}_2\mathcal{F}_1\left[\bgm 1,q_{12}\\ q_2+1\edm\middle|\FR{E_2+K_1}{E_{12}}\right]\bigg\}\n\\
  &-\FR{\Gamma[q_1,q_2]}{[\ii(E_1-K_1)]^{q_1}[\ii(E_2+K_1)]^{q_2}}+\FR{\Gamma[q_1,q_2]}{[\ii(E_1+K_1)]^{q_1}[\ii(E_2+K_1)]^{q_2}},
\end{align}
and,
\begin{align}
\label{eq_2chainWt2}
  &\wt\psi_\text{2-chain}(E_1,\wh E_2,K_1)\n\\
  =&-\FR{\Gamma[q_1]}{(\ii E_{12})^{q_{12}}}\bigg\{{}_2\mathcal{F}_1\left[\bgm 1,q_{12}\\ q_1+1\edm\middle|\FR{E_1-K_1}{E_{12}}\right]
  +{}_2\mathcal{F}_1\left[\bgm 1,q_{12}\\ q_1+1\edm\middle|\FR{E_1+K_1}{E_{12}}\right]\bigg\}\n\\
  &-\FR{\Gamma[q_1,q_2]}{[\ii(E_2-K_1)]^{q_2}[\ii(E_1+K_1)]^{q_1}}+\FR{\Gamma[q_1,q_2]}{[\ii(E_2+K_1)]^{q_2}[\ii(E_1+K_1)]^{q_1}}.
\end{align}
One can check that the four expressions in (\ref{eq_2chainWp1}), (\ref{eq_2chainWp2}), (\ref{eq_2chainWt1}), and (\ref{eq_2chainWt2}) are in fact equivalent. 

\paragraph{Correlation function} Next, let us look at the correlation function $\wt{\mathcal{T}}_\text{2-chain}$ for the two-side chain graph. According to the diagrammatic rule based on the Schwinger-Keldysh formalism, the corresponding integral representation for $\wt{\mathcal{T}}_\text{2-chain}$ is:
\begin{align}
\label{eq_T2chain}
  &\wt{\mathcal{T}}_\text{2-chain}(E_1,E_2,K_1)\n\\
  &=(-\ii)^2\sum_{\aa,\bb=\pm}\aa\bb\int_{-\infty}^0\di\tau_1\di\tau_2(-\tau_1)^{q_1-1}(-\tau_2)^{q_2-1}\wt D_{\aa}(E_1;\tau_1)\wt D_{\bb}(E_2;\tau_2)\wt D_{\aa\bb}(K_1;\tau_1,\tau_2).
\end{align}
This integral can be readily calculated, but once again we want to make a detour. The idea is similar to the previous example of wavefunction coefficient: We rewrite the bulk propagator in partially ordered form. There are again two ways to do this:
\begin{align}
\label{eq_Dtau1}
  &\wt D_{\pm\pm}(K_1;\wh\tau_1,\tau_2)=\Big[e^{\pm\ii K_1(\tau_1-\tau_2)}-e^{\mp\ii K_1(\tau_1-\tau_2)}\Big]\theta(\tau_2-\tau_1)+e^{\mp\ii K_1(\tau_1-\tau_2)};\\
  \label{eq_Dtau2}
  &\wt D_{\pm\pm}(K_1;\tau_1,\wh\tau_2)=\Big[e^{\mp\ii K_1(\tau_1-\tau_2)}-e^{\pm\ii K_1(\tau_1-\tau_2)}\Big]\theta(\tau_1-\tau_2)+e^{\pm\ii K_1(\tau_1-\tau_2)}.
\end{align}
The first choice (\ref{eq_Dtau1}) amounts to choosing $\tau_1$ as the earlier time variable. Substituting (\ref{eq_Dtau1}) into (\ref{eq_T2chain}), we get:
\bge
\label{eq_T2chainTau1}
  \wt{\mathcal{T}}_\text{2-chain}(\wh{E}_1,E_2,K_1)= \ft{1_1 2_{\bar 1}}-\ft{1_{\bar 1}2_1}+\ft{1_{\bar 1}}\ft{2_1}-\ft{1_1}\ft{\bar 2_{\bar 1}} +\text{c.c.}.
\ede
As before, a barred quantity means a sign flip. Thus, $\bar 2_{\bar 1}$ means $-E_2-K_1$.
Here we have used a fact about general scalar tree graphs with parity-even couplings: In general, for a graph with $V$ vertices, there are $2^V$ different SK contours, corresponding to the SK label $\aa=\pm$ for $V$ vertices. However, the two choices with a simultaneous flip of all signs on all vertices are related by complex conjugation. Thus, we can always fix the SK label for one vertex, and sum over the SK indices of the rest $V-1$ vertex. This reduces the total number of graphs to be computed to $2^{V-1}$. We then add up these $2^{V-1}$ diagrams, taking the complex conjugation, and get the final result for the whole correlator. As a rule, in this work, we always fix the SK index of the \emph{earliest} energy variable to be $+1$. 

According to the rule of diagrammatic representation, a plus type vertex is represented by a filled circle, while a minus vertex is represented by an unfilled circle. A circle with shadings means that two choices for the SK label are summed. (See App.\ \ref{app_notations}.) Then, we can rewrite (\ref{eq_T2chainTau1}) with this diagrammatic notation as: 
\bge
\label{eq_T2chainFig}
\parbox{0.8\textwidth}{\includegraphics[width=0.8\textwidth]{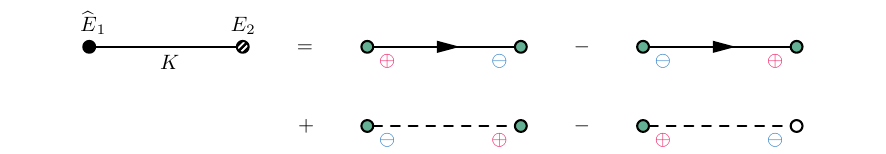}}
\ede
That is, we have fixed the SK label of the earlier site ($\wh{E}_1$ in this example) to be plus (filled), and summed over all SK labels for the rest of vertices. Note that, for a correlator, it is possible to have a white circle in the family tree integral, like the last diagram in (\ref{eq_T2chainFig}). On the other hand, every line must have one {\color{Red}$\oplus$} and one {\color{RoyalBlue}$\ominus$}; The situation that two {\color{Red}$\oplus$} or {\color{RoyalBlue}$\ominus$} are attached to the same line is forbidden.  

Similarly, we can as well choose $\tau_2$ as the earlier time, and thus use (\ref{eq_Dtau2}) when computing the integral (\ref{eq_T2chain}). Then, we get:
\bge
\label{eq_T2chainTau2}
  \wt{\mathcal{T}}_\text{2-chain}(E_1,\wh{E}_2,K_1)= \ft{2_1 1_{\bar 1}}-\ft{2_{\bar 1} 1_1}+\ft{1_1}\ft{2_{\bar 1}}-\ft{\bar 1_{\bar 1}}\ft{2_{1}} +\text{c.c.}.
\ede
We can represent this decomposition diagrammatically as:
\bge
\label{eq_T2chainFig2}
\parbox{0.8\textwidth}{\includegraphics[width=0.8\textwidth]{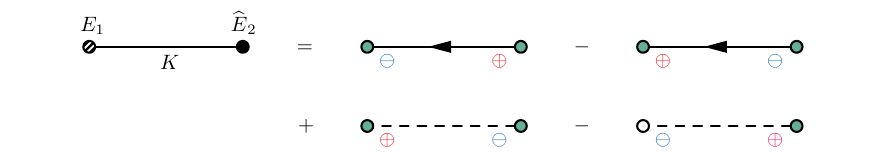}}
\ede
One can check in a similar way that the two expressions (\ref{eq_T2chainTau1}) and (\ref{eq_T2chainTau2}) are actually equivalent. We collect the explicit expressions of these correlators in terms of hypergeometric functions in App.\ \ref{app_examples}.

\subsection{Three-site chain} 
\label{sec_3chain}

Next, let us consider a more complicated example, namely a graph with three vertices. At the tree level, there is only one topology to connect three sites, namely a 3-site chain. This corresponds to an arbitrary tree graph with two bulk exchanges. Below, we will again consider the wavefunction coefficient and the correlation functions separately. 

\paragraph{Wavefunction coefficient} The integral for the rescaled wavefunction coefficient  $\wt\psi_\text{3-chain}$ for a 3-site chain can be written as:
\begin{align}
\label{eq_3siteWCint}
 & \wt\psi_\text{3-chain}(E_1,E_2,E_3,K_1,K_2)
 =\ii^3\int_{-\infty}^0\prod_{i=1}^3\Big[\di\tau_i\,(-\tau_i)^{q_i-1}B(E_i;\tau_i)\Big]\wt G(K_1;\tau_1,\tau_2)\wt G(K_2;\tau_2,\tau_3).
\end{align}

As in the previous example, our strategy is still to assign a partial order to the graph as the first step.
For a 3-site chain, we have 3 different ways to introduce a partial order. Using our notation for family-tree integrals, they correspond to $[123]$, $[2(1)(3)]$, and $[321]$. Let us consider these cases in turn.

For the $[123]$ case, we have $\tau_1<\tau_2<\tau_3$, and we should use $\wt G(K_1;\wh\tau_1,\tau_2)$ and $\wt G(K_2;\wh\tau_2,\tau_3)$ for the two bulk propagators. Following exactly the same procedure as above, we can decompose the wavefunction coefficients as linear combinations of (products) of family tree integrals:
\begin{align}
\label{eq_3chainWC1}
  -\wt\psi_\text{3-chain}(\wh 1)
  =&~\ft{1_12_{\bar 12}3_{\bar 2}}-\ft{1_12_{\bar 1\bar 2}3_{2}}-\ft{1_{\bar 1}2_{12}3_{\bar 2}}+\ft{1_{\bar 1}2_{1\bar 2}3_{2}}\n\\
   &+\ft{1_12_{\bar 1\bar 2}}\ft{3_2}-\ft{1_12_{\bar 12}}\ft{3_2}-\ft{1_{\bar 1}2_{1\bar 2}}\ft{3_2}+\ft{1_{\bar 1}2_{12}}\ft{3_2}\n\\
   &+\ft{1_{\bar 1}}\ft{2_{12}3_{\bar 2}}-\ft{1_{\bar 1}}\ft{2_{1\bar 2}3_{2}}-\ft{1_{1}}\ft{2_{12}3_{\bar 2}}+\ft{1_{1}}\ft{2_{1\bar 2}3_{2}}\n\\
   &+\ft{1_{\bar 1}}\ft{2_{1\bar 2}}\ft{3_2}-\ft{1_{\bar 1}}\ft{2_{12}}\ft{3_2}-\ft{1_{1}}\ft{2_{1\bar 2}}\ft{3_2}+\ft{1_{1}}\ft{2_{12}}\ft{3_2}.
\end{align}
Here we have an additional sign on the left-hand side, because of, on the one hand, the overall factor $\ii^3$ in the integral expression (\ref{eq_3siteWCint}), and, on the other hand, the inclusion of three $-\ii$'s in the definition of family tree integrals. Also, we are using the shorthand notation $\wt{\psi}_\text{3-chain}(\wh{1})\equiv\wt{\psi}_\text{3-chain}(\wh{E}_1,E_2,E_3,K_1,K_2)$ to highlight that we are choosing $E_1$-site as the earliest site.

The expression starts to look complicated, but once again, we can use a diagrammatic representation to make this result easier to read:
\bge
\label{eq_3chainWC1_fig}
\parbox{0.9\textwidth}{\hspace{-6mm}\includegraphics[width=0.98\textwidth]{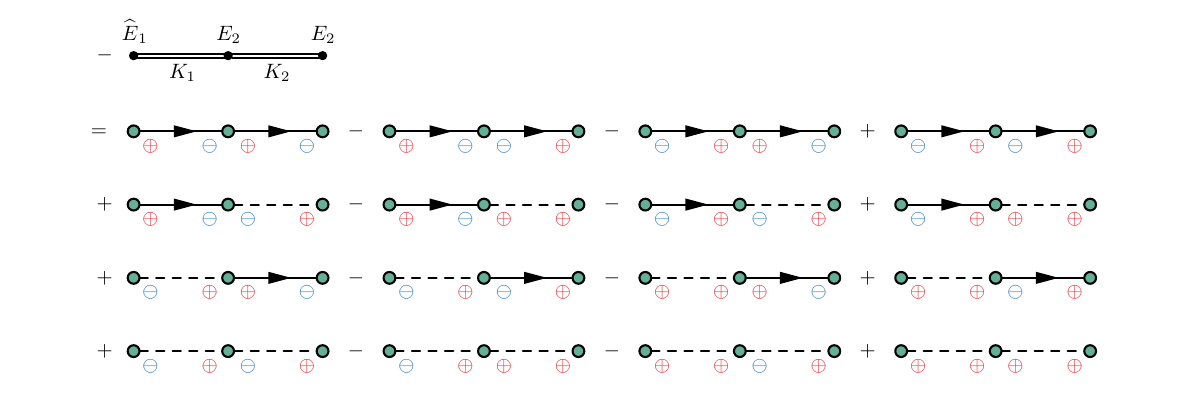}}
\ede
Clearly, we see a pattern emerging here. The 16 diagrams on the right-hand side are organized into four rows according to how we ``cut'' the chain. By cutting a chain, we mean to replace the directional line by a factorized (dashed) line. The four rows from top to bottom correspond to cutting no line, the right line, the left line, and the both lines, respectively. Then, for each type of ``cut'' within a row, we have four terms, corresponding to different assignments of {\color{Red}$\oplus$} and {\color{RoyalBlue}$\ominus$}. To make the pattern of these assignments more transparent, we can rewrite (\ref{eq_3chainWC1}) into a more compact form:
\begin{align}
\label{eq_3chainWC1comp}
  -\wt\psi_\text{3-chain}(\wh 1)
  =\sum_{\aa,\bb
  =\pm}\aa\bb\Big\{
  \ft{1_{1^\aa}2_{\bar 1^\aa2^\bb}3_{\bar 2^\bb}}
  +\ft{1_{1^\aa}2_{\bar 1^\aa\bar 2^\bb}}\ft{3_2}
  +\ft{1_{\bar 1^\aa}}\ft{2_{12^\bb}3_{\bar 2^\bb}}
  +\ft{1_{\bar 1^\aa}}\ft{2_{1\bar 2^\bb}}\ft{3_2}
  \Big\}.
\end{align} 
Here we are introducing yet another set of superscript $\aa,\bb=\pm$ to decorated some entries, with the obvious meaning that, for example, $1_{1^\aa}=E_1+\aa K_1$ and $2_{\bar 1^\aa 2^\bb}=E_2-\aa K_1+\bb K_2$. The expression is simple enough that we can observe a general rule for directly writing down this result. We will summarize this rule later, but the essential points are clear: We assign each internal energy $K_\ell$ a summation variable $\aa_\ell=\pm$, and then execute a kind of ``wavefunction cut'' to the whole partially ordered graph in all possible ways. By a wavefunction cut, we mean to break the family tree into some small family trees, and whenever we make such a cut, we switch the barred and unbarred internal energies of the cut line, and also remove the summation index associated with the cut internal energy at the daughter site. Thus, for example, if we cut the $K_1$ line in $[1_{1^\aa}2_{\bar 1^\aa 2^\bb}3_{\bar 2^\bb}]$, we get $[1_{\bar 1^\aa}][2_{1 2^\bb}3_{\bar 2^\bb}]$. Then, the whole wavefunction coefficient is given by the sum of all possible cuts and the sum over all indices $\aa_\ell$ associated with all bulk lines, together with an overall sign $(-1)^N$ where $N$ is the number of sites. We will restate and prove this rule in a more complete fashion in Sec.~\ref{sec_general_rules}.

At this point, one can insert the explicit results for all family tree integrals in (\ref{eq_3chainWC1comp}) to find an analytical expression for the 3-site chain wavefunction coefficient. In particular, the 3-site family like $[123]$ can be expressed in terms of a Kampé de Fériet function, the 2-site family $[12]$ in terms of Gaussian hypergeometric function, and 1-site family $[1]$ as a simple power. The full expression is quite long, and we collect it in App.\;\ref{app_examples}.  

Above we have found an explicit expression for the 3-site chain in terms of family-tree integrals, which is obtained by assigning a particular partial order $[123]$ to the 3-site chain at the very beginning. Next, let us consider a different choice of partial order for the 3-site chain. The order $[321]$ is not really different, as it can obtained from $[123]$ by the replacement $1\leftrightarrow 3$. So, let us consider the case of $[2(1)(3)]$. To find the family-tree expression for this choice, we substitute $\wt{G}(K_1,\tau_1,\wh\tau_2)$ and $\wt{G}(K_2,\wh\tau_2,\tau_3)$ in (\ref{eq_3siteWCint}) with their explicit expressions given in (\ref{eq_Gtau1}) and (\ref{eq_Gtau2}). Then, following the same procedure as above, we get the following decomposition:
\begin{align}
\label{eq_3chainWC2comp}
  -\wt\psi_\text{3-chain}(\wh 2)
  =\sum_{\aa,\bb
  =\pm}\aa\bb\Big\{
  \ft{2_{1^\aa 2^\bb}(1_{\bar 1^\aa})(3_{\bar 2^\bb})}
  +\ft{2_{1^\aa\bar 2^\bb}1_{\bar 1^\aa}}\ft{3_2}
  +\ft{1_{1}}\ft{2_{\bar 1^\aa2^\bb}3_{\bar 2^\bb}}
  +\ft{1_1}\ft{2_{\bar 1^\aa\bar 2^\bb}}\ft{3_2}
  \Big\}.
\end{align} 
As a simple exercise, one can easily verify that this result can be obtained very easily from the rule summarized below (\ref{eq_3chainWC1comp}). Also, this result can again be represented diagrammatically as:
\bge
\label{eq_3chainWC2_fig}
\parbox{0.9\textwidth}{\hspace{-6mm}\includegraphics[width=0.98\textwidth]{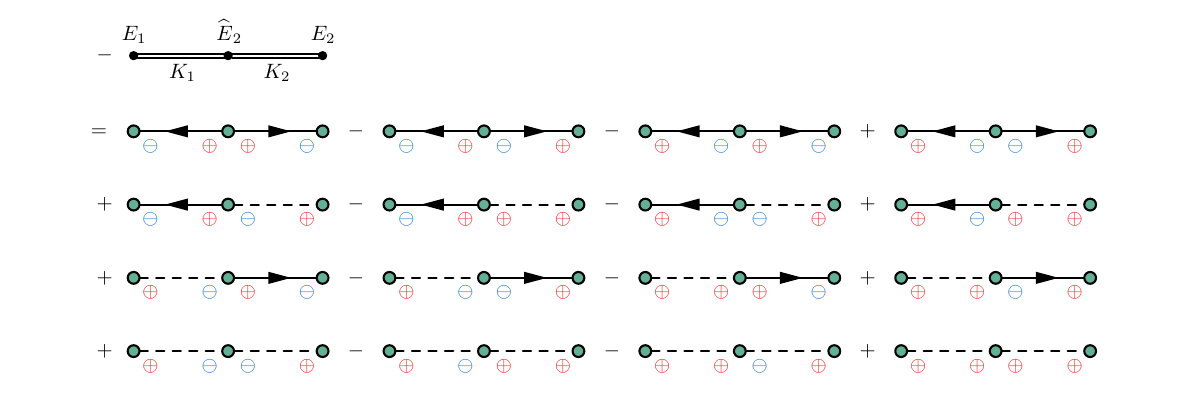}}
\ede
According to our earlier discussion on 2-site chain, the two expressions (\ref{eq_3chainWC1comp}) and (\ref{eq_3chainWC2comp}), although looked different, are in fact equivalent. 

Incidentally, this choice of partial order leads to a close-form expression for the final result in terms of Appell function. We collect explicit expressions in App.\;\ref{app_examples}.

\paragraph{Correlation function} Now let us look at the correlator of 3-site chain. As before, to compute the three-site correlator $\mathcal{T}_\text{3-chain}$ with partial order $[123]$, we can fix the SK label of Site 1 to be $+$, and we only need to compute $\mathcal{T}_{+\aa_2\aa_3}$, with $\aa_2,\aa_3=\pm$. From the diagrammatic rule reviewed in Sec.\;\ref{sec_frw}, it is straightforward to compute these objects. For instance:
\begin{align}
  \wt{\mathcal{T}}_{+++}
  =&~(-\ii)^3\int_{-\infty}^0\prod_{i=1}^3\Big[\di\tau_i\,(-\tau_i)^{q_i-1}e^{+\ii E_i\tau_i}\Big] \Big[\Big(e^{+\ii K_1(\tau_1-\tau_2)}-e^{-\ii K_1(\tau_1-\tau_2)}\Big)\theta(\tau_2-\tau_1)\n\\
   &+e^{-\ii K_1(\tau_1-\tau_2)}\Big]  \Big[\Big(e^{+\ii K_2(\tau_2-\tau_3)}-e^{-\ii K_2(\tau_2-\tau_3)}\Big)\theta(\tau_3-\tau_2) +e^{-\ii K_2(\tau_2-\tau_3)}\Big]\\
  =&~\ft{1_1 2_{\bar 12} 3_{\bar 2}}-\ft{1_1 2_{\bar 1\bar 2} 3_2}-\ft{1_{\bar 1}2_{12} 3_{\bar 2}}+\ft{1_{\bar 1} 2_{1\bar 2} 3_{2}}+\ft{1_1 2_{\bar 1\bar 2}}\ft{3_2}-\ft{1_{\bar 1} 2_{1\bar 2}}\ft{3_2}\n\\
   &+\ft{ 1_{\bar 1}}\ft{ 2_{12} 3_{\bar 2}}-\ft{1_{\bar 1}}\ft{2_{1\bar 2}3_2}+\ft{1_{\bar 1}}\ft{2_{1\bar 2}}\ft{3_2}.
\end{align}
Similarly, we get results for the other three branches:
\begin{align}
  \wt{\mathcal{T}}_{++-}
  =&-\ft{1_12_{\bar 12}}\ft{\bar 3_{\bar 2}}+\ft{1_{\bar 1}2_{12}}\ft{\bar 3_{\bar 2}}-\ft{1_{\bar 1}}\ft{2_{12}}\ft{\bar 3_{\bar 2}},\\
  \wt{\mathcal{T}}_{+--}
  =&+\ft{1_1}\ft{\bar 2_{\bar 1\bar 2}\bar 3_{2}}-\ft{1_1}\ft{\bar 2_{\bar 12}\bar 3_{\bar 2}}+\ft{1_1}\ft{\bar 2_{\bar 12}}\ft{\bar 3_{\bar 2}},\\
  \wt{\mathcal{T}}_{+-+}=&-\ft{1_1}\ft{\bar 2_{\bar 1\bar 2}}\ft{3_2}.
\end{align}
As said, the result for the 3-site chain correlator is simply the sum of the above four expressions together with their complex conjugates, $\wt{\mathcal{T}}_\text{3-chain}=(\wt{\mathcal{T}}_{+++}+\wt{\mathcal{T}}_{++-}+\wt{\mathcal{T}}_{+--}+\wt{\mathcal{T}}_{+-+})+\text{c.c.}$. It is useful to reorganize the final result in the following way:
\begin{align}
  \wt{\mathcal{T}}_\text{3-chain}
  =&~ \Big\{\ft{1_1 2_{\bar 12} 3_{\bar 2}}-\ft{1_1 2_{\bar 1\bar 2} 3_2}-\ft{1_{\bar 1}2_{12} 3_{\bar 2}}+\ft{1_{\bar 1} 2_{1\bar 2} 3_{2}}\n\\
   &+\ft{1_1 2_{\bar 1\bar 2}}\ft{3_2}-\ft{1_12_{\bar 12}}\ft{\bar 3_{\bar 2}}-\ft{1_{\bar 1} 2_{1\bar 2}}\ft{3_2}+\ft{1_{\bar 1}2_{12}}\ft{\bar 3_{\bar 2}}\n\\
   &+\ft{ 1_{\bar 1}}\ft{ 2_{12} 3_{\bar 2}}-\ft{1_{\bar 1}}\ft{2_{1\bar 2}3_2}-\ft{1_1}\ft{\bar 2_{\bar 12}\bar 3_{\bar 2}}+\ft{1_1}\ft{\bar 2_{\bar 1\bar 2}\bar 3_{2}}\n\\
   &+\ft{1_{\bar 1}}\ft{2_{1\bar 2}}\ft{3_2}-\ft{1_{\bar 1}}\ft{2_{12}}\ft{\bar 3_{\bar 2}}-\ft{1_1}\ft{\bar 2_{\bar 1\bar 2}}\ft{3_2}+\ft{1_1}\ft{\bar 2_{\bar 12}}\ft{\bar 3_{\bar 2}}\Big\}+\text{c.c.}.
\end{align}
Again, this is very similar to the structure we got for the wavefunction coefficient, but also with important difference. Let us express this result with the following diagram:
\bge
\parbox{0.9\textwidth}{\hspace{-6mm}\includegraphics[width=0.98\textwidth]{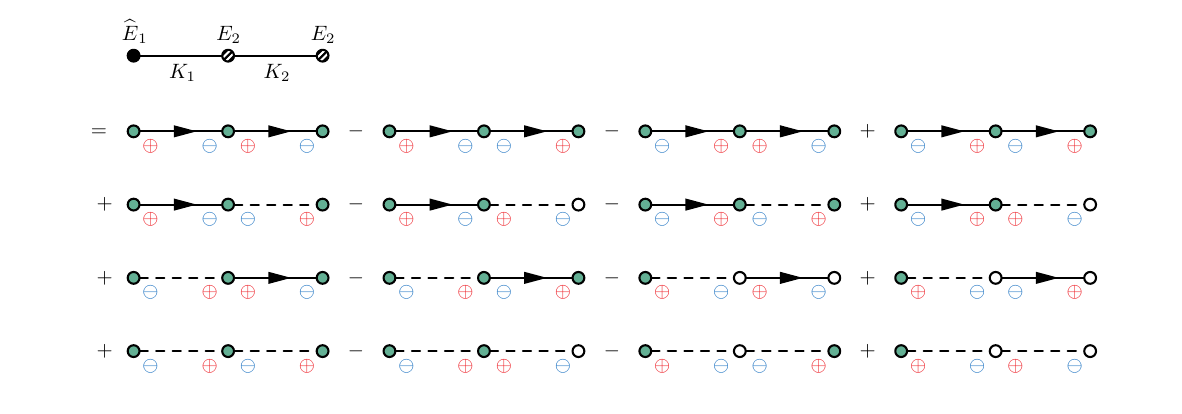}}
\ede
We can also observe a pattern here, which can be made more explicit by further rewriting the above expression as:
\begin{align}
  \wt{\mathcal{T}}_\text{3-chain}
  =&~ \sum_{\aa,\bb=\pm}\aa\bb\Big\{\ft{1_{1^\aa} 2_{\bar1^\aa 2^\bb} 3_{\bar 2^\bb}} 
    +\ft{1_{1^\aa} 2_{\bar 1^\aa\bar 2^\bb}}\ft{3^\bb_{2^\bb}} 
    +\ft{ 1_{\bar 1^\aa}}\ft{ 2^\aa_{1^\aa2^\bb} 3^\aa_{\bar 2^\bb}} 
    +\ft{1_{\bar 1^\aa}}\ft{2^\aa_{1^\aa\bar 2^\bb}}\ft{3^\bb_{2^\bb}}\Big\}+\text{c.c.}.
\end{align}
Here we are using again the decorations $\aa,\bb=\pm$ for some entries, this time not only for subscripts (internal energies), but also for the main entries (external energies). Thus, for example, $3_{\bar 2^\bb}^\aa$ means $\aa E_3-\bb K_2$.

As in the case of wavefunction coefficients, here we can also observe a general rule for directly writing down the result for correlators. The initial steps are identical: We assign properly the external and internal energy variables to a family tree, such as in $[1_1 2_{12} 3_2]$. Then, for every pair of internal energy variable $K_\ell$, we assign a bar to the \emph{later} one, and assign a summation variable $\aa_i$ to both of them. For the three-site chain example, we get $[1_{1^\aa}2_{\bar 1^\aa 2^\bb}3_{\bar 2^\bb}]$. The rule starts to differ from the one for wavefunction coefficients when we taking cuts. For correlators, we should take a different kind of cut which we call ``correlator cut.'' By a correlator cut, we mean to break the family tree by cutting some internal line $K_\ell$. At the same time, we switch the barred and unbarred pair of the cut internal energy $K_\ell$, and also \emph{add} the summation variable $\aa_\ell$ associated to $K_\ell$ to all \emph{external energies} of the one resulting daughter family tree (but not other disconnected family trees resulting from cutting lines other than $K_\ell$). Therefore, for instance, if we take the correlator cut of $K_1$ in $[1_{1^\aa} 2_{\bar1^\aa 2^\bb} 3_{\bar 2^\bb}]$, we get $[ 1_{\bar 1^\aa}][ 2^\aa_{1^\aa2^\bb} 3^\aa_{\bar 2^\bb}]$. Then, the whole correlator is given by the sum over all possible cuts, and over all summation variables $\aa_\ell$ associated to all bulk lines $K_\ell$, together with their complex conjugate. Again, we will restate and prove this rule in a more complete fashion in Sec.~\ref{sec_general_rules}.

Now, similar to our treatment of 3-site chain wavefunction coefficient, we can also consider the 3-site chain correlator with a different partial order $[2(1)(3)]$. One can repeat all the diagrammatic calculations as detailed above, or just apply the above rule to the family tree $[2_{12}(1_{\bar 1})(3_{\bar 2})]$. Either way, we should get the same result:
\begin{align}
  \wt{\mathcal{T}}_\text{3-chain}(\wh 2)
  =&\sum_{\aa,\bb=\pm}\aa\bb\Big\{
  \ft{2_{1^\aa 2^\bb}(1_{\bar 1^\aa})(3_{\bar 2^\bb})}
 +\ft{2_{1^\aa \bar 2^\bb}(1_{\bar 1^\aa})}\ft{3_{2^\bb}^\bb}\n\\
 &+\ft{1_{1^\aa}^\aa}\ft{2_{\bar 1^\aa 2^\bb}3_{\bar 2^\bb}} 
 +\ft{1_{1^\aa}^\aa}\ft{2_{\bar 1^\aa \bar 2^\bb}}\ft{3_{2^\bb}^\bb}
  \Big\}+\text{c.c.}.
\end{align}
We can also illustrate this result with the following diagrammatic expression:
\bge
\parbox{0.9\textwidth}{\hspace{-6mm}\includegraphics[width=0.98\textwidth]{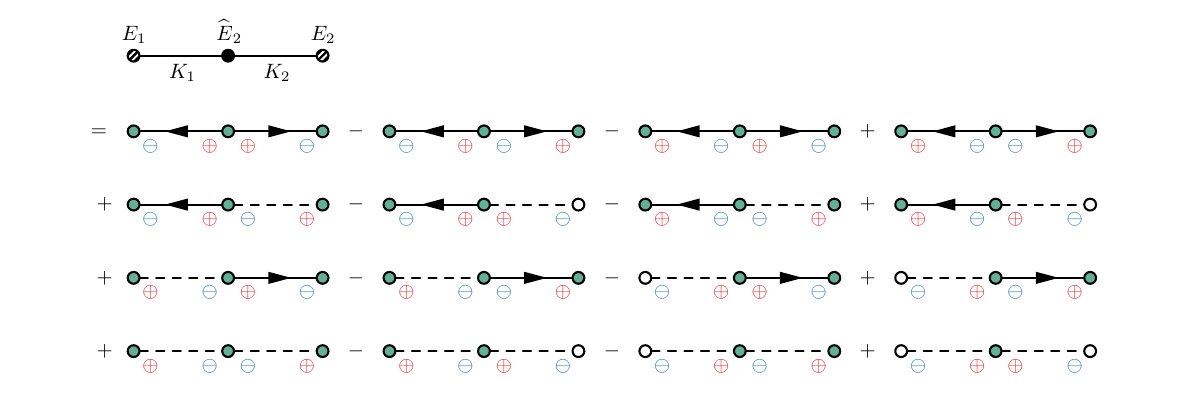}}
\ede 
As we should expect, there is an exchange symmetry $(E_1,K_1)\leftrightarrow(E_3,K_2)$ in the expression. We can also write down explicit expressions of all above decompositions for three-site chain correlators in terms of (multivariate) hypergeometric functions. We collect these results in App.\;\ref{app_examples}.

\subsection{Four-site star}

One can use the rules discovered above to directly write down expressions for any tree-level wavefunction coefficients or correlators of conformal scalars. The final results are expressed in terms of family tree integrals, which can in turn be expressed as power series  of energy variables. The procedure is totally mechanical. For a graph with $N$ sites, we will get an expression with $4^{N-1}$ terms. In the next section, we will provide explicit proofs for these rules. To finish the current section, we apply these rules to write down explicit expressions for the wavefunction coefficient and correlator of a 4-site example with a star structure. We can schematically express this decomposition for the wavefunction as follows:
\bge
\parbox{0.8\textwidth}{\includegraphics[width=0.8\textwidth]{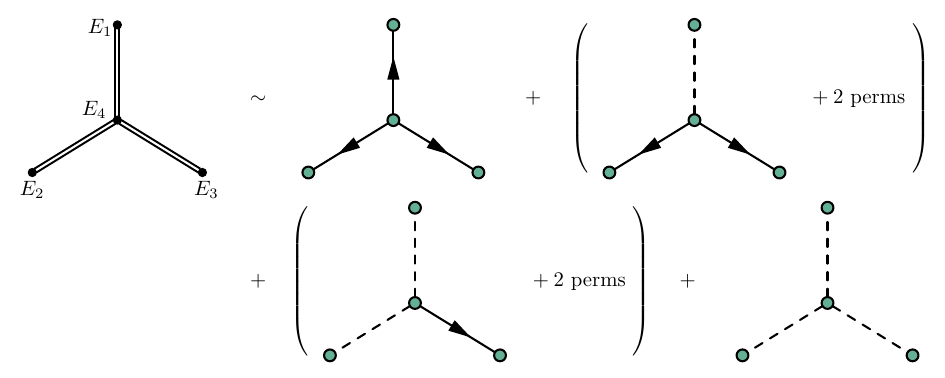}}
\ede 
As shown in this diagram, we are taking the partial order of the graph to be $[4(1)(2)(3)]$. Here we have been schematic and neglected all assignments and summations over internal energy decorations. Using the rules presented before, it is now very easy to write down the expression for the wavefunction coefficient $\wt\psi_\text{4-star}$ as:
\begin{align}
  \wt{\psi}_\text{4-star}
  =&\sum_{\aa,\bb,\cc=\pm}\aa\bb\cc\Big\{\ft{4_{1^{\aa}2^{\bb}3^{\cc}}(1_{\bar 1^{\aa}})(2_{\bar 2^{\bb}})(3_{\bar 3^\cc})}+\Big(\ft{4_{{\bar 1}^{\aa}2^{\bb}3^{\cc}}(2_{\bar 2^{\bb}})(3_{\bar 3^\cc})}\ft{1}+\text{2 perms}\Big)\n\\
  &+\Big(\ft{4_{\bar 1^\aa\bar 2^\bb 3^\cc}3_{\bar 3^\cc}}\ft{1_1}\ft{2_2}+\text{2 perms}\Big)
  +\ft{4_{\bar 1^\aa\bar 2^\bb\bar 3^\cc}}\ft{1_1}\ft{2_2}\ft{3_3}\Big\}.
\end{align}
Similarly, we can find the corresponding correlator $\wt{\mathcal{T}}_\text{4-star}$ as:
\begin{align}
  \wt{\mathcal{T}}_\text{4-star}
  =&\sum_{\aa,\bb,\cc=\pm}\aa\bb\cc\Big\{\ft{4_{1^{\aa}2^{\bb}3^{\cc}}(1_{\bar 1^{\aa}})(2_{\bar 2^{\bb}})(3_{\bar 3^\cc})}+\Big(\ft{4_{{\bar 1}^{\aa}2^{\bb}3^{\cc}}(2_{\bar 2^{\bb}})(3_{\bar 3^\cc})}\ft{1_{1^{\aa}}^\aa}+\text{2 perms}\Big)\n\\
  &+\Big(\ft{4_{\bar 1^\aa\bar 2^\bb 3^\cc}3_{\bar 3^\cc}}\ft{1_{1^\aa}^\aa}\ft{2_{2^\bb}^\bb}+\text{2 perms}\Big)
  +\ft{4_{\bar 1^\aa\bar 2^\bb\bar 3^\cc}}\ft{1_{1^\aa}^\aa}\ft{2_{2^\bb}^\bb}\ft{3_{3^\cc}^\cc}\Big\}+\text{c.c.}.
\end{align}
We note that the four-site family $[4(1)(2)(3)]$ has a closed form expression in terms of Lauricella's $F_A$ function. This is actually true for any $N$-site family tree with two generations. We give explicit expressions for both $\wt\psi_\text{4-star}$ and $\wt{\mathcal{T}}_\text{4-star}$ in App.\;\ref{app_examples}.

\section{General Rules for Arbitrary Tree Graphs}
\label{sec_general_rules}

Now that we have examined a number of examples for both wavefunction coefficients and correlators, we are ready to state the general rules for directly writing down the analytic answer for these objects in terms of family tree integrals. Below, we will summarize these rules in a more systematic way, and then provide direct proofs for them. Once again, we consider wavefunction coefficients and correlators separately.

\paragraph{Notation  for a $N$-site graph} Before we state the rules for writing down the answer, let us summarize our notational rules for family tree, which we have been using in this section.

Our rule for writing down a family tree is the following:
\begin{enumerate}
  \item Given a graph of $N$ sites, we label all of its $N$ external energies as $E_1,\cdots,E_N$, and all $N-1$ internal energies as $K_1,\cdots,K_{N-1}$. 
  \item When writing down a family tree, an external energy $E_A$ will be denoted as $A$, and an internal energy $K_\ell$ attached to Site $A$ will appear in the subscript of $A$ as $A_\ell$. Thus, for example, $A_{k\ell}\equiv E_A+K_k+K_\ell$ 
  \item Whenever we add a bar to an energy variable, we mean to take its opposite. Thus, $\bar A=-A$; $A_{\bar\ell m}\equiv E_A-K_\ell+K_m$.
  \item We also add a summation index $\aa,\bb,\cdots=\pm$ as superscript of some energy variables. Whenever an energy variable carries such a superscript, it means that we should further keep or flip the sign of the energy variable according to the sign of this superscript. Thus, for instance, $A^\aa_{\bar \ell^\aa m^\bb}\equiv \aa E_A-\aa K_\ell+\bb K_m$.  
\end{enumerate}

\paragraph{Rules for wavefunction coefficients} For an arbitrary $N$-site tree-graph contribution to a wavefunction coefficient, we can follow the these rules to write down its analytical answer in terms of family trees.

\begin{enumerate} 
  \item For the given graph, we decide a partial order. The partial order can be chosen arbitrarily, but once we make a choice, we stick to it in all following steps. 
  
  To demonstrate this and following steps, we take a 4-site chain as an example. For a 4-site chain, we have four external energies $E_1,\cdots, E_4$ corresponding to all four sites, and three $K_1,K_2,K_3$ corresponding to all three lines. Now, let us choose a partial order such that $\tau_2<\tau_1$ and $\tau_2<\tau_3<\tau_4$.
 
  \item With the partial order given, we can already write down an expression of the largest family tree, with external energies as main entries and internal energies as subscripts. 
  
  Thus, for the four-site chain with the above partial order, we have $\ft{2_{12}(1_1)(3_{23}4_3)}$.
  
  \item Note that every internal energy $K_\ell$ in the subscripts appears exactly twice, i.e., they come in pairs, one associated to an earlier time and one to the later time. For each internal-energy pair, we assign a bar to the \emph{later} one, and assign a summation variable $\aa_\ell$ to both of them. 
  
  Then, for the above 4-site chain, we get $\ft{2_{1^\aa 2^\bb}(1_{\bar 1^\aa})(3_{\bar 2^\bb3^\cc}4_{\bar 3^\cc})}$.
  
  \item We then take all possible ``wavefunction cut.'' (Here we call it wavefunction cut to distinguish it from another cut for computing correlators to be introduced below.) By a wavefunction cut, we mean: i) to break the family tree into a product of several smaller family trees; 2) to switch the unbarred-barred internal energy pair of the cut line, and then \emph{remove} the summation variable associated with the \emph{later} site. 
  
For our 4-site chain example, we can choose to cut or not to cut either of its three lines, resulting in 8 possibilities in total:
  \begin{align}
  \label{eq_4chain_cuts_WC}
  &\text{No cut:}~\ft{2_{1^\aa 2^\bb}(1_{\bar 1^\aa})(3_{\bar 2^\bb3^\cc}4_{\bar 3^\cc})},
  &&\text{Cutting $K_1,K_2$}:~\ft{2_{\bar 1^\aa \bar 2^\bb}}\ft{1_1}\ft{3_{2 3^\cc}4_{\bar 3^\cc}},\n\\
  &\text{Cutting $K_1$}:~\ft{2_{\bar 1^\aa 2^\bb}3_{\bar 2^\bb3^\cc}4_{\bar 3^\cc}}\ft{1_1},
  &&\text{Cutting $K_2,K_3$}:~\ft{2_{1^\aa \bar 2^\bb}1_{\bar 1^\aa}}\ft{3_{2\bar 3^\cc}}\ft{4_{3}}\n\\
  &\text{Cutting $K_2$}:~\ft{2_{1^\aa \bar 2^\bb}1_{\bar 1^\aa}}\ft{3_{23^\cc}4_{\bar 3^\cc}},
  &&\text{Cutting $K_1,K_3$}:~\ft{2_{\bar 1^\aa 2^\bb}3_{\bar 2^\bb\bar 3^\cc}}\ft{1_1}\ft{4_{3}},\n\\
  &\text{Cutting $K_3$}:~\ft{2_{1^\aa 2^\bb}(1_{\bar 1^\aa})(3_{\bar 2^\bb\bar 3^\cc})}\ft{4_{3}},
  &&\text{Cutting $K_1,K_2,K_3$}:~\ft{2_{\bar 1^\aa \bar 2^\bb}}\ft{1_1}\ft{3_{2\bar 3^\cc}}\ft{4_{3}}.
  \end{align}

  \item We take all possible wavefunction cuts as above, and sum over all of them, and also sum over all superscripts $\aa_i=\pm$. Together with an overall factor $(-1)^N$, this gives  the final expression for the wavefunction coefficient.
    
  Thus, for our 4-site chain example, the final answer is given by:
  \begin{align}
  \wt\psi_\text{4-chain}=&~(-1)^4\sum_{\aa,\bb,\cc=\pm}\aa\bb\cc\Big\{
  \ft{2_{1^\aa 2^\bb}(1_{\bar 1^\aa})(3_{\bar 2^\bb3^\cc}4_{\bar 3^\cc})}
  +\ft{2_{\bar 1^\aa 2^\bb}3_{\bar 2^\bb3^\cc}4_{\bar 3^\cc}}\ft{1_1}
  +\ft{2_{1^\aa \bar 2^\bb}1_{\bar 1^\aa}}\ft{3_{23^\cc}4_{\bar 3^\cc}}\n\\
  &+\ft{2_{1^\aa 2^\bb}(1_{\bar 1^\aa})(3_{\bar 2^\bb\bar 3^\cc})}\ft{4_{3}}
  +\ft{2_{\bar 1^\aa \bar 2^\bb}}\ft{1_1}\ft{3_{2 3^\cc}4_{\bar 3^\cc}}
  +\ft{2_{1^\aa \bar 2^\bb}1_{\bar 1^\aa}}\ft{3_{2\bar 3^\cc}}\ft{4_{3}}\n\\
  &+\ft{2_{\bar 1^\aa 2^\bb}3_{\bar 2^\bb\bar 3^\cc}}\ft{1_1}\ft{4_{3}}
  +\ft{2_{\bar 1^\aa \bar 2^\bb}}\ft{1_1}\ft{3_{2\bar 3^\cc}}\ft{4_{3}} 
   \Big\}.
  \end{align}
\end{enumerate}

\paragraph{Proof} The proof of the above rule is pretty straightforward. The key point is that, for the graph with a given partial order, we can always rewrite the $N-1$ bulk propagators so that the $\theta$ functions appeared in these expressions are all consistent with the prescribed partial order. In this way, every bulk propagator consists of four terms. For instance, consider the rescaled propagator $\wt G(K_\ell;\tau_A,\tau_B)$ of  Line $\ell$, which connects Site $A$ and Site $B$, and suppose that our partial order requires $\tau_A<\tau_B$, namely, Site $A$ is the mother of Site $B$. Then, from (\ref{eq_bulkpropWC}) and (\ref{eq_resG}), we see that the propagator can be written as:
\bge
  \wt G(K_\ell;\tau_A,\tau_B)=\Big[e^{+\ii K_\ell(\tau_A-\tau_B)}-e^{-\ii K_\ell(\tau_A-\tau_B)}\Big]\theta(\tau_B-\tau_A)+e^{-\ii K_\ell(\tau_A-\tau_B)}-e^{+\ii K_\ell(\tau_A+\tau_B)}.
\ede
It is a matter of triviality to check that the above propagator can be rewritten as:
\bge
\label{eq_GtildeReorg}
  \wt G(K_\ell;\tau_A,\tau_B)=\sum_{\aa =\pm}\aa \Big[e^{\ii \aa K_\ell(\tau_A-\tau_B)}\theta(\tau_B-\tau_A)+e^{-\ii \aa K_\ell\tau_A+\ii K_\ell\tau_B}\Big].
\ede
Now we can use expressions as above for all $N-1$ bulk propagators, and thus split the original integral expression for the wavefunction coefficient into $4^{N-1}$ terms. In particular, the two terms in (\ref{eq_GtildeReorg}) give the following structure in the final expression:
\bge
  \sum_{\aa=\pm}\aa\Big\{ \ft{\cdots A_{\ell^\aa }B_{\bar \ell^\aa }\cdots}+ \ft{\cdots A_{\bar\ell^\aa}}\ft{B_{\ell}\cdots}\Big\}.
\ede
From this we can directly see the origin of all our rules: 
\begin{enumerate}
  \item We have two terms in (\ref{eq_GtildeReorg}). The first term with $\theta$ function corresponds to the uncut line and the second term without $\theta$ function corresponds to the cut line. So, for an $N$-site graph, there are $2^{N-1}$ possible cut structures. 
  \item In the uncut term in (\ref{eq_GtildeReorg}), the internal energy subscript $\ell$ always appears in a pair, one barred and the other unbarred. Also, the unbarred ($+K_\ell$) always appears at the earlier site, and the barred one ($-K_\ell$) at the later site. This follows from the fact that we have a factor $e^{\ii\aa K_\ell(\tau_A-\tau_B)}$ in the uncut term.
  \item Compared to the uncut term, the cut term amounts to a replacement $(A_{\ell^\aa},B_{\bar\ell^\aa})\to (A_{\bar \ell^\aa},B_{\ell})$. This is exactly our wavefunction cut.
  \item Finally, due to the summation over $\aa=\pm$, we have $2^{N-1}$ summations in the whole expression for the wavefunction coefficient. Also, due to the different sign conventions between the wavefunction Feynman rule and the family tree integrals, there is another overall factor of $(-1)^N$ to be included. This finishes our proof for the wavefunction rule.
\end{enumerate}
As we can see, our rule produces an expression of $4^{N-1}$ terms, each of which has a distinct cut structure and kinematic combination. We note that the number of distinct terms  is in accordance with what was observed in \cite{Arkani-Hamed:2023kig} from a study of differential equations.

\paragraph{Rules for correlators}

Now we turn to correlators. The rule for writing down the analytical answer for arbitrary tree-graph contribution to a correlator is quite similar to the rule for wavefunction coefficients. The first three steps are actually identical, so we will not repeat ourselves. The difference starts to show up in Step 4 when we take the cut. Below we show the correlator rules starting from Step 4:
\begin{enumerate}\setcounter{enumi}{3}
  \item We take all possible ``correlator cut.'' By a correlator cut, we mean: i) to break the family tree into a product of several smaller family trees; 2) to switch the unbarred-barred internal energy pair of the cut line, and then attach the summation variable of the cut internal energy to all \emph{external energies} of the resulting daughter family tree (but not other disconnected family trees resulting from cutting lines other than $K_\ell$). 
  Thus, for our 4-site chain example, we again get 8 possible cuts for the correlator, but their expressions are different from the wavefunction coefficients except for the uncut one:
\begin{align}
  \label{eq_4chain_cuts_CF}
  &\text{No cut:}~\ft{2_{1^\aa 2^\bb}(1_{\bar 1^\aa})(3_{\bar 2^\bb3^\cc}4_{\bar 3^\cc})},
  &&\text{Cutting $K_1,K_2$}:~\ft{2_{\bar 1^\aa \bar 2^\bb}}\ft{1_{1^\aa}^\aa}\ft{3_{2^\bb 3^\cc}^\bb 4_{\bar 3^\cc}^\bb},\n\\
  &\text{Cutting $K_1$}:~\ft{2_{\bar 1^\aa 2^\bb}3_{\bar 2^\bb3^\cc}4_{\bar 3^\cc}}\ft{1_{1^\aa}^\aa},
  &&\text{Cutting $K_2,K_3$}:~\ft{2_{1^\aa \bar 2^\bb}1_{\bar 1^\aa}}\ft{3_{2^\bb\bar 3^\cc}^\bb}\ft{4_{3^\cc}^\cc}\n\\
  &\text{Cutting $K_2$}:~\ft{2_{1^\aa \bar 2^\bb}1_{\bar 1^\aa}}\ft{3_{2^\bb 3^\cc}^\bb 4_{\bar 3^\cc}^\bb},
  &&\text{Cutting $K_1,K_3$}:~\ft{2_{\bar 1^\aa 2^\bb}3_{\bar 2^\bb\bar 3^\cc}}\ft{1_{1^\aa}^\aa}\ft{4_{3^\cc}^\cc},\n\\
  &\text{Cutting $K_3$}:~\ft{2_{1^\aa 2^\bb}(1_{\bar 1^\aa})(3_{\bar 2^\bb\bar 3^\cc})}\ft{4_{3^\cc}^\cc},
  &&\text{Cutting $K_1,K_2,K_3$}:~\ft{2_{\bar 1^\aa \bar 2^\bb}}\ft{1_{1^\aa}^\aa}\ft{3_{2^\bb\bar 3^\cc}^\bb}\ft{4_{3^\cc}^\cc}.
\end{align}

  \item We take all possible correlator cuts as above, sum over all of them, and also sum over all superscripts $\aa_i=\pm$. Then, we sum the result with its complex conjugate, and get the final expression for the correlator.
  
  For our 4-site chain example, the final expression for the rescaled correlator is:
  \begin{align}
  \wt{\mathcal{T}}_\text{4-chain}=&\sum_{\aa,\bb,\cc=\pm}\aa\bb\cc\Big\{
  \ft{2_{1^\aa 2^\bb}(1_{\bar 1^\aa})(3_{\bar 2^\bb3^\cc}4_{\bar 3^\cc})}
  +\ft{2_{\bar 1^\aa 2^\bb}3_{\bar 2^\bb3^\cc}4_{\bar 3^\cc}}\ft{1_{1^\aa}^\aa}
  +\ft{2_{1^\aa \bar 2^\bb}1_{\bar 1^\aa}}\ft{3_{2^\bb 3^\cc}^\bb 4_{\bar 3^\cc}^\bb}\n\\
  &+ \ft{2_{1^\aa 2^\bb}(1_{\bar 1^\aa})(3_{\bar 2^\bb\bar 3^\cc})}\ft{4_{3^\cc}^\cc}
  + \ft{2_{\bar 1^\aa \bar 2^\bb}}\ft{1_{1^\aa}^\aa}\ft{3_{2^\bb 3^\cc}^\bb 4_{\bar 3^\cc}^\bb}
  +\ft{2_{1^\aa \bar 2^\bb}1_{\bar 1^\aa}}\ft{3_{2^\bb\bar 3^\cc}^\bb}\ft{4_{3^\cc}^\cc}\n\\
  &+\ft{2_{\bar 1^\aa 2^\bb}3_{\bar 2^\bb\bar 3^\cc}}\ft{1_{1^\aa}^\aa}\ft{4_{3^\cc}^\cc}
  +\ft{2_{\bar 1^\aa \bar 2^\bb}}\ft{1_{1^\aa}^\aa}\ft{3_{2^\bb\bar 3^\cc}^\bb}\ft{4_{3^\cc}^\cc}
   \Big\}+\text{c.c.}.
  \end{align}

\end{enumerate}

\paragraph{Proof} The proof is similar to the previous case. We still rewrite all SK propagators such that any $\theta$ functions appearing in these propagators are in accordance with the prescribed partial order. However, a new complication appears here: Since there is a possibility of sign flips for \emph{external} energies, we need to keep track of them. Then, for the (rescaled) bulk propagator $\wt D_{\aa_A\aa_B}(K_\ell;\tau_A,\tau_B)$ that connects $\tau_A$ and $\tau_B$, the corresponding time integrals have the following structure: 
\begin{align}
  \sum_{\aa_A,\aa_B=\pm}\aa_A\aa_B\int_{\tau_A,\tau_B}e^{\ii \aa_A E_A\tau_A+\ii\aa_B E_B\tau_B}\wt D_{\aa_A\aa_B}(K_\ell;\tau_A,\tau_B)\cdots
\end{align}
Here the neglected part ``$\cdots$'' denotes other propagators also attached to $\tau_A$ or $\tau_B$ which are irrelevant to our argument. At this point, it is helpful to expand the summation over the two SK indices $\aa_A,\aa_B=\pm$, and we get:
\begin{align}
\label{eq_DbulkExpand}
  &\int_{\tau_A,\tau_B}e^{+\ii(E_A\tau_A+E_B\tau_B)}\bigg\{\Big[e^{+\ii K_\ell(\tau_A-\tau_B)}-e^{-\ii K_\ell(\tau_A-\tau_B)}\Big]\theta(\tau_B-\tau_A)+e^{-\ii K_\ell(\tau_A-\tau_B)}\bigg\}\n\\
  +&\int_{\tau_A,\tau_B}e^{-\ii(E_A\tau_A+E_B\tau_B)}\bigg\{\Big[e^{-\ii K_\ell(\tau_A-\tau_B)}-e^{+\ii K_\ell(\tau_A-\tau_B)}\Big]\theta(\tau_B-\tau_A)+e^{+\ii K_\ell(\tau_A-\tau_B)}\bigg\}\n\\
  -&\int_{\tau_A,\tau_B}e^{+\ii (E_A\tau_A-E_B\tau_B)}e^{+\ii K_\ell(\tau_A-\tau_B)}-\int_{\tau_A,\tau_B}e^{-\ii (E_A\tau_A-E_B\tau_B)}e^{-\ii K_\ell(\tau_A-\tau_B)}.
\end{align}
Obviously, we can reorganize the above terms in the following way: 
\begin{align}
\label{eq_DbulkRearrange}
  &\sum_{\aa,\bb=\pm}\aa \bb \int_{\tau_A,\tau_B}e^{+\ii\bb E_A\tau_A}\bigg\{e^{+\ii \bb E_B\tau_B}e^{+\ii\aa K_\ell(\tau_A-\tau_B)}\theta(\tau_B-\tau_A)+e^{+\ii \aa  E_B\tau_B} e^{-\ii\aa K_\ell(\tau_A-\tau_B)}\bigg\} .
\end{align}
We emphasize that the two summation variables $\aa,\bb$ are not the original SK indices $\aa_A,\aa_B$. The two new variables $\aa,\bb$ are introduced only to make the full expression (\ref{eq_DbulkExpand}) compact. Indeed, from (\ref{eq_DbulkRearrange}), we can see that the relevant contribution to the final result from this part of the integral reads:
  \bge
  \label{eq_corCut}
  \sum_{\aa=\pm}\aa\Big\{\ft{\cdots A_{\ell^\aa}^\bb B_{\bar\ell^\aa}^\bb\cdots}+\ft{\cdots A_{\bar\ell^\aa}^\bb}\ft{B_{\ell^\aa}^\aa\cdots}\Big\}.
  \ede   
Here we have deliberately neglected the summation over $\bb$ for a reason to be explained below. As in the previous discussion for wavefuncitons, this expression explains our cutting rule for correlators. The new phenomenon here is that, in the factorized term in (\ref{eq_DbulkRearrange}), the sign flip of $K$ induces a sign flip of the external energy $E_B$ of the \emph{daughter} site, since the sign-controlling $\aa$ also appears in $e^{+\ii\aa E_B\tau_B}$ in the factorized term. In particular, the external energy and the internal energy of the daughter site of a cut line always have the same sign, as is clear from the factor $e^{\color{RoyalBlue}{+\ii\aa E_B\tau_B}}e^{-\ii\aa K_\ell{\color{RoyalBlue}+\ii\aa K_\ell\tau_B}}$.

On the other hand, in the nested term (i.e., the uncut term), all external energies should have the same sign, due to the factor $e^{+\ii\bb E_A\tau_A}e^{+\ii\bb E_B\tau_B}$. This fact, together with the previous one for the cut term, shows that the variable $\bb$ is in fact redundant: Once we fix the SK index of the earliest site to be plus, the signs of all external energies in a graph are completely fixed by the cut structure. So, in practice, we can just decide the signs of external energies by how we cut the graph, and discard the variable $\bb$ completely, just as what we stated in the rule above.

As a result, we can once again make two independent choices for each bulk propagator: We choose to cut or not to cut it, and we choose a sign for $\aa_i$. The resulting family trees are completely determined by the ``correlator cutting rule'' in (\ref{eq_corCut}). Thus, we again get a final result with $4^{N-1}$ terms, together with their complex conjugates. In particular, we note that the uncut terms of the correlator are completely identical to the uncut terms in the wavefunction coefficient for the same graph.

\section{Chronological Energy Integrals of a Family Tree}
\label{sec_energyint}

In this section, we discuss an alternative understanding of the family tree integrals, which can be thought of as a boundary dual of doing bulk time integrals. This dual viewpoint has been emphasized in recent years as a useful way to study cosmological amplitudes. In this boundary treatment, the time variables disappear, and instead, one deals with energy integrals rather than time integrals. Although the energy integrals are not nested, they are not necessarily easier to compute than nested time integrals, since the energy integrand typically involves fractions of energy sums. Rather, the interesting aspect of the energy integral is that the integrand is nothing but the corresponding amplitude in flat spacetime. It has a simple structure such that one can derive recursive rules to built up the integrand directly without using diagrammatic rules in the bulk. 

We are not going to repeat the story of building up full energy integrand for wavefunction coefficients, which has been explored in \cite{Arkani-Hamed:2017fdk} in great detail. Instead, we want to address the following question in this section: Since we are decomposing a full amplitude into family trees, what is the corresponding energy-integral representation for any given family tree? In particular, is there a pattern for the energy integrand? Quite remarkably, as we shall see, there is a very simple and intuitive rule for generating the energy integrand for any family tree. In short, the rule is that, we simply write down a \emph{chronological energy integrand} which is a simple fraction and encodes the ``birthdays'' of all family members. Then, the full energy integrand for a family tree is just the sum of chronological integrands with all possible assignment of birthdays consistent with the partial order structure. Technically, as we shall show, this is achieved by merging subfamilies of the same level through a \emph{shuffle product}.

Below we will first explain how to turn a nested bulk time integral into an energy integral, and then explain the birthday rule with a few examples.

\paragraph{From time integral to energy integral}
For family trees with exponent $q\neq n-1$ ($n\in \mathbb{N}$), the energy integral is very conveniently obtained by using the following identity:
\bge
\label{eq_energyrep}
  (-\tau)^{q-1}=\FR{\ii^{1-q}}{\Gamma(1-q)}\int_{0}^\infty \di\ep\, e^{\ii\ep \tau}\ep^{-q}
\ede
The integral is well defined for $q<1$ and can be analytically continued to any value of $q\in \mathbb{C}$ except at the poles of $\Gamma(1-q)$, namely $q=1,2,\cdots$. The convergence in the upper limit $\ep\to \infty$ is guaranteed by the $\ii\ep$ prescription of the time variable. 

Then, for arbitrary nested time integral given in (\ref{eq_NTI}), we have:
\begin{align}
  \mb{T}_{\mathscr{N}}^{(q_1\cdots q_N)}(\omega_1,\cdots,\omega_N)
  =&~(-\ii)^N\int_{-\infty}^0\prod_{i=1}^N\Big[\di\tau_i\,(-\tau_i)^{q_i-1}e^{\ii \omega_i\tau_i}\Big]\prod_{(i,j)\in\mathscr{N}}\theta(\tau_i-\tau_j)\n\\
  =&~\int_0^\infty\prod_{i=1}^N\bigg[\FR{\di\ep_i\,(\ii\ep_i)^{-q_i}}{\Gamma[1-q_i]} \int_{-\infty}^0\di\tau_i\,e^{\ii \mathcal{E}_i\tau_i}\bigg]\prod_{(i,j)\in\mathscr{N}}\theta(\tau_i-\tau_j),
  \label{eq_energy_int_mid}
\end{align}
where we have defined $\E_i\equiv \omega_i+\ep_i$.
That is, we get an integral over $N$ energy variables $\ep_i$ ($i=1,\cdots,N$), and the integrand is obtained by finishing the nested time integral:
\begin{align}
  \int_{-\infty}^0\prod_{\ell=1}^N\Big[\di\tau_\ell\,e^{\ii \mathcal{E}_\ell\tau_\ell}\Big] \prod_{(i,j)\in\mathscr{N}}\theta(\tau_i-\tau_j).
\end{align}
To see how to finish this integral, we turn to a few examples.

\paragraph{$\bm N$-site chain}

We begin with an $N$-site chain $[1\cdots N]$, with no branching structure, namely, every mother has at most one daughter. In this case, the time integral has the following form:
\begin{align}
  \ii^N\prod_{i=1}^N\int_{-\infty}^0\Big[\di\tau_\ell\, e^{\ii \mathcal{E}_\ell\tau_\ell}\Big]\theta_{N,N-1}\theta_{N-1,N-2}\cdots\theta_{21}.
\end{align}
Thanks to the chain structure, we can directly finish this integral starting from the earliest time variable:
\begin{align}
  &\ii^N\int_{-\infty}^0\prod_{\ell=1}^N\Big[\di\tau_\ell\, e^{\ii \mathcal{E}_\ell\tau_\ell}\Big]\theta_{N,N-1}\cdots\theta_{21}=\FR{\ii^{N-1}}{\E_1}\int_{-\infty}^0\prod_{\ell=2}^N\Big[\di\tau_\ell\, e^{\ii \mathcal{E}_\ell\tau_\ell}\Big]e^{\ii\E_1\tau_2}\theta_{N,N-1}\cdots\theta_{32}\n\\
  =&~\FR{\ii^{N-2}}{\E_1\E_{12}}\int_{-\infty}^0\prod_{\ell=3}^N\Big[\di\tau_\ell\, e^{\ii \mathcal{E}_\ell\tau_\ell}\Big]e^{\ii\E_{12}\tau_3}\theta_{N,N-1}\cdots\theta_{43}=\cdots=\FR{1}{\E_1\E_{12}\cdots\E_{1\cdots N}}.
\end{align}

Here we have inserted a factor of $\ii^N$ for convenience. The result is a simple fraction, with the denominator being products of $N$ successive partial sums of (shifted) energy variables $\E_{1\cdots j}\equiv \E_1+\cdots+\E_j$ with $j=1,\cdots, N$. 

Let us introduce another shorthand notation:
\begin{align}
  \label{eq_energy_integrand}
  \ef{a_1a_2\cdots a_N}\equiv &~\FR{1}{\E_{a_1}\E_{a_1a_2}\cdots \E_{a_1\cdots a_N}},\\
  \label{eq_energy_integral}
  \ei{a_1a_2\cdots a_N}\equiv &~(-\ii)^N\int_0^\infty\prod_{i=1}^N\bigg[\FR{\di\ep_i\,(\ii\ep_i)^{-q_i} }{\Gamma[1-q_i]} \bigg]\ef{a_1a_2\cdots a_N}.
\end{align}
Clearly, the energy integral $\{a_1a_2\cdots a_N\}$ has a chain structure, and its $N$ members are ordered according to the time ordering. So we call $\{a_1a_2\cdots a_N\}$ a \emph{chronological energy integral}.
Thus, we just discovered that a family tree of a chain structure is completely equivalent to a chronological energy integral: 
\begin{keyeqn}
\begin{align}
\label{eq_chainenergyint}
  \ft{1\cdots N} =\ei{1\cdots N}.
\end{align}\vspace{-7mm}
\end{keyeqn}
 
We can already use this result to build wavefunction coefficients of chain diagrams. For instance, from the family-tree decomposition of the two-site chain $\wt\psi_\text{2-chain}$ in (\ref{eq_psi2sol1}), we see the corresponding energy integrand is given by:
\begin{align}
   &\ef{1_1 2_{\bar 1}}-\ef{1_{\bar 1}2_{1}}+\ef{1_{\bar 1}}\ef{2_1}-\ef{1_1}\ef{2_1}\n\\
  =&~ \FR{1}{(\E_1+K_1)\E_{12}}-\FR{1}{(\E_1-K_1)\E_{12}}+\FR{1}{(\E_1-K_1)(\E_2+K_1)}-\FR{1}{(\E_1+K_1)(\E_2+K_1)} \n\\
  =&~\FR{2K_1}{(\E_1+K_1)(\E_2+K_1)\E_{12}},
\end{align} 
in agreement with the result obtained with other methods such as the recursive rule given in \cite{Arkani-Hamed:2017fdk}. 

Similarly, we can consider the three-site chain $\wt\psi_\text{3-chain}(\wh 1)$. According to (\ref{eq_3chainWC1comp}), the corresponding energy integrand is:
\begin{align} 
  &\sum_{\aa,\bb=\pm}\aa\bb\Big\{
  \ef{1_{1^\aa}2_{\bar 1^\aa2^\bb}3_{\bar 2^\bb}}
  +\ef{1_{1^\aa}2_{\bar 1^\aa\bar 2^\bb}}\ef{3_2}
  +\ef{1_{\bar 1^\aa}}\ef{2_{12^\bb}3_{\bar 2^\bb}}
  +\ef{1_{\bar 1^\aa}}\ef{2_{1\bar 2^\bb}}\ef{3_2}
  \Big\}\n\\
  =&\sum_{\aa,\bb=\pm}\aa\bb\bigg\{\FR{1}{(\E_1+\aa K_1)(\E_{12}+\bb K_2)\E_{123}}+\FR{1}{(\E_1+\aa K_1)(\E_{12}-\bb K_2)(\E_3+K_2)}\n\\
  &+\FR{1}{(\E_1-\aa K_1)(\E_2+K_1+\bb K_2)(\E_{23}+K_1)}+\FR{1}{(\E_1-\aa K_1)(\E_2+K_1-\bb K_2)(\E_3+K_2)}\bigg\}\n\\
  =&~\FR{4K_1K_2}{\E_{123}(\E_1+K_1)(\E_2+K_{12})(\E_3+K_2)}\bigg[\FR{1}{\E_{12}+K_2}+\FR{1}{\E_{23}+K_1}\bigg],
\end{align} 
also in agreement with the result obtained from the recursive rule.

\paragraph{Branching tree } Things start to get complicated once there are branchings in a given family tree. Take the simplest three-site tree $[2(1)(3)]$ as an example, the relevant time integral is:
\begin{align}
  \ii^3\int_{-\infty}^0\di\tau_1\di\tau_2\di\tau_3\, e^{\ii \mathcal{E}_1\tau_1+\ii \mathcal{E}_2\tau_2+\ii \mathcal{E}_3\tau_3}\theta_{32}\theta_{12}.
\end{align}
Clearly, the complication is that the two time variables $\tau_1$ and $\tau_3$ do not come with a definite order, and we can no longer finish the integral recursively as in the previous chain example. To solve this problem, we use the following simple relation:
\bge
  \theta_{32}\theta_{12}=(\theta_{31}+\theta_{13})\theta_{32}\theta_{12}=\theta_{31}\theta_{12}+\theta_{13}\theta_{32}.
\ede
Then, on the right-hand side, every term has a chain structure which allows us to finish the time integral as before. So, we get:
\begin{align}
  \ft{2(1)(3)}=\ei{213}+\ei{231}.
\end{align}
This is a general solution: Whenever we encounter time variables, say $\tau_i$ and $\tau_j$, which are undecided by the partial order structure in the family tree, we insert the identity $1=\theta_{ij}+\theta_{ji}$ to break the integrand into two separate terms, each of which has a fixed order for $\tau_i$ and $\tau_j$. We can do this for all non-ordered pairs of time variables, so that, in the end, we break the energy integrand into many terms. Each of these terms has a chain structure, ordered chronologically. Clearly, we are simply enumerating all possible orders of ``birthday'' for all family members, and each possible order appears once and only once. So we call this a ``birthday rule.'' 

The birthday rule is better illustrated with examples. First, consider a 4-site family with a cubic vertex, namely $[4(1)(2)(3)]$. Using the above method, we can find:
\begin{align}
\label{eq_4123eint}
   \ft{4(1)(2)(3)} 
  =&~\ei{4123}+\ei{4132}+\ei{4231}+\ei{4213}+\ei{4312}+\ei{4321}.
\end{align}
This can be represented diagrammatically as:
\bge
\parbox{0.7\textwidth}{\includegraphics[width=0.7\textwidth]{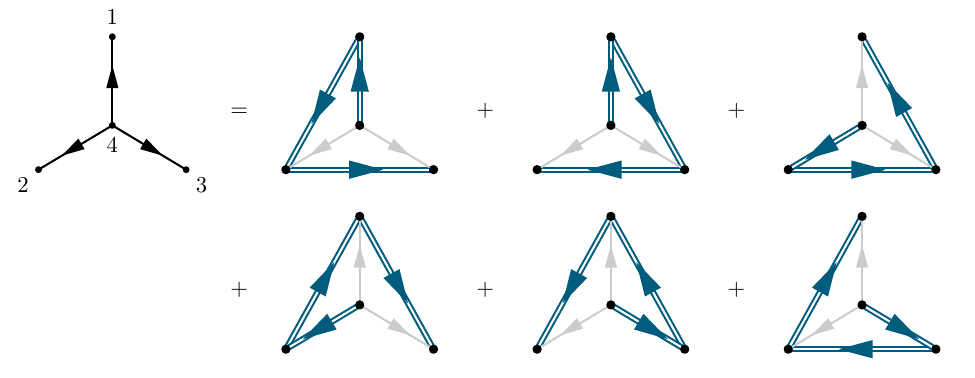}}
\ede
Here, each diagram on the right-hand side represents a chronological energy integral, with the directional double line indicating the birthday order.

At this point, it is useful to introduce a more compact way to collect the six terms on the right-hand side of (\ref{eq_4123eint}), using the notation of shuffle product. For any two ordered strings of letters, namely ``words,'' the shuffle product $\shuffle$ is defined to be the sum of all words composed of letters in the original two words, without changing the order of letters within each of the two original words. For example,
\bge
  (12)\shuffle(34)=1234+1324+1342+3124+3142+3412.
\ede 
By construction, the shuffle product is associative and commutative. For convenience, we also require that the shuffle product used here is distributive with respect to normal addition. For example, $1\shuffle(2+3)=1\shuffle 2+1\shuffle 3$.

Then it is clear that the birthday rule is nothing but taking shuffle products among subfamilies of several sisters. So, we have:
\begin{align}
  \ft{1(2)(3)(4)}
  =&~\ei{1(2)\shuffle(3)\shuffle(4)}.
\end{align}

Let us look at two more family trees with 5 sites. The first one is $[1(24)(35)]$. According to the birthday rule, we find that it can be written as a sum of six chronological energy integrals, as follows:
\bge
\parbox{0.7\textwidth}{\includegraphics[width=0.7\textwidth]{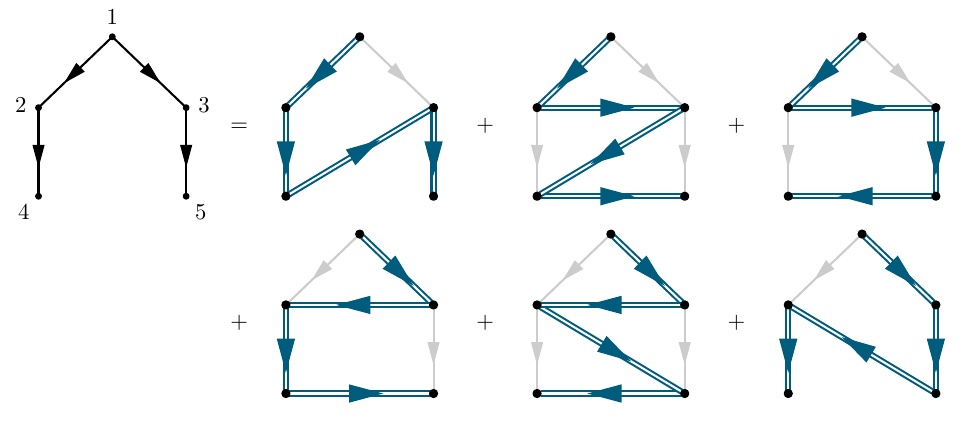}}
\ede
This decomposition is again generated by a shuffle product between two subfamilies $(24)$ and $(13)$. So, we have:
\begin{align}
  \ft{1(24)(35)} 
  = &~\ei{1(24)\shuffle(35)}\n\\
  = &~\ei{12435}+\ei{12345}+\ei{12354}+\ei{13245}+\ei{13254}+\ei{13524}.
\end{align}
Our last example is a 5-site family tree with two branchings, namely $[1(2)(3(4)(5))]$. Diagrammatically, we can decompose this family tree into chronological energy integrals as follows:
\bge
\parbox{0.88\textwidth}{\includegraphics[width=0.88\textwidth]{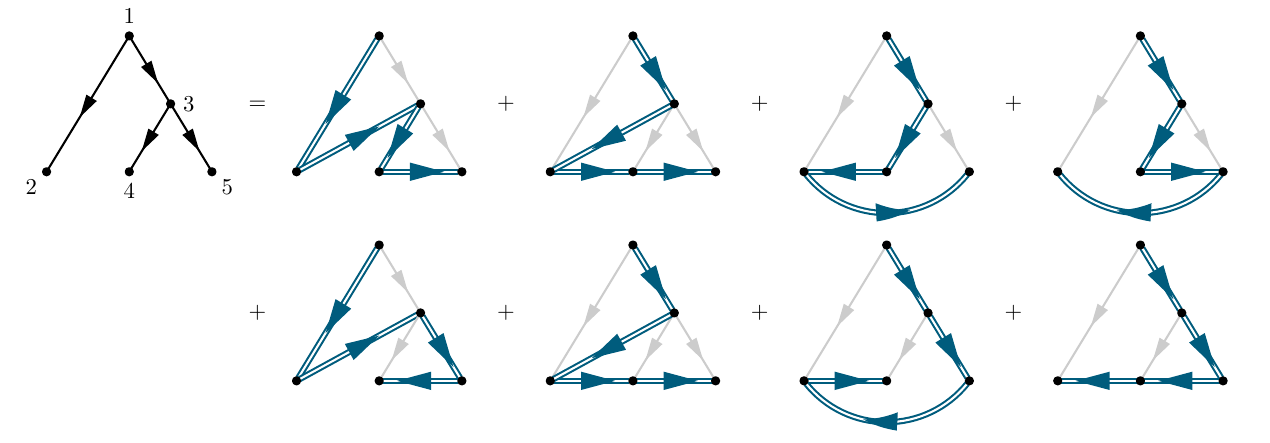}}
\ede
To get this result, we need to take shuffle product starting from the inner group of subfamilies of $[1(2)(3(4)(5))]$, namely $(4)$ and $(5)$. This shuffle product has the effect of merging the two subfamilies with their common ancestor $3$ to get a larger subfamily, namely $(345)$ or $(354)$. We then take shuffle product between the two subfamilies of outer layer, namely $(2)$ and $(3(4)\shuffle(5))$, to get the final answer, where all parentheses are removed:
\begin{align}
  \ft{1(2)\big(3(4)(5)\big)}
  =&~\ei{1(2)\shuffle\big(3(4)\shuffle(5)\big)}\n\\
  =&~\ei{1(2)\shuffle(345)}+\ei{1(2)\shuffle(354)}\n\\
  =&~\ei{12345}+\ei{13245}+\ei{13425}+\ei{13452}\n\\
   &+\ei{12354}+\ei{13245}+\ei{13524}+\ei{13542}.
\end{align}
This is a general phenomenon of our birthday rule: We always take shuffle products from the innermost layer of subfamilies, and use this product to remove the parentheses. We do this layer by layer in the outgoing direction, until all parentheses are removed.

\paragraph{Birthday rule} Now we are ready to state our birthday rule for writing down the energy integral of an arbitrary family tree: The energy integral representation of an arbitrary family tree $[\mathscr{P}(1\cdots N)]$ is the sum of chronological energy integrals corresponding to all possible orders of birthdays of all family members. Mathematically, this amounts to take the shuffle products of all subfamilies of the same layer, from the innermost layer to the outermost layer. This rule can be summarized in the following equation:
\begin{keyeqn}
\begin{align}
 \ft{\mathscr{P}(1\cdots N)}=\big\{\bigshuffle\mathscr{P}(1\cdots N)\big\},
\end{align}\vspace{-7mm}
\end{keyeqn}
where the shuffle product of the family $\bigshuffle \mathscr{P}(1\cdots N)$ is as described above.

From our previous discussion with specific example, one can already see how the shuffle product emerges when we try to finish the nested time integral. As we showed, the time integral is more tractable if it has a linear structure without any branching. Thus, to organize any family tree into a sum of chain integrals, we need to enumerate all possible permutations of time variables whose orders are undecided in the family tree. This is essentially the definition of shuffle product. 

To conclude this section, we want to emphasize that, although the above birthday rule is introduced for convenient conversion from a time integral to energy integrals, the way to do layer-by-layer shuffle product for a family tree is completely general. We can use this strategy to reduce any family tree into a sum of family chains. From the perspective of performing bulk time integrals alone, this reduction means that we can fully characterize any nested bulk time integrals by a class of chain integrals, which are completely specified by the number of sites in the chain. We shall pursue this reduction and discuss the properties of chain integrals in a future work.

\section{Singularity Structure of Conformal Amplitudes}
\label{sec_singular}

In this section, we briefly examine the singularity structure of conformal amplitudes in general FRW background. The singularity structure of arbitrary tree diagrams is complicated by the free parameters $q_i$ at all vertices. Here we will only consider some universal features. In particular, the familiar total-energy singularity and partial-energy singularities are generic singular features of a tree graph in any FRW background. Also, generic values of $q_i$ typically turn these singularities into branch points, generating further branch cuts on the complex planes of energy variables. We leave a complete characterization of these singularities to a future work. Here, we shall focus on another question: Given that a tree graph can always be decomposed into a sum of products of family trees, it would be useful to know, in a specific limit (total or partial energy limits), which terms in this decomposition contain singularities. Meanwhile, as we shall see, many terms in the family tree decomposition may contain fake singularities, which are canceled out in the final result. Below, we address these points, first with a few by-far familiar examples, and then discuss the general patterns we can learn from these case studies.

As mentioned above, typically, a tree graph (for either a wavefunction coefficient or a correlation function) possesses a singularity when the magnitude sum of all momenta flowing into any of its subgraph goes to zero. There are two complementary viewpoints about this singularity: From the bulk viewpoint, the singularity arises because the vanishing energy sum signals a failure of $\ii\ep$ prescription in the time integral. Thus, such a singularity must be produced as a divergence of the time integral in the early time limit. From the boundary viewpoint, the energy sum appears in the denominator of the energy integrand. Then, one can use a standard Landau analysis to diagnose the singular behavior \cite{Eden:1966dnq}. In this language, the energy sum approaching to zero corresponds to an endpoint singularity. Given that the two treatments should yield identical results, below we shall mainly focus on the energy integrand, which turns out to be an easier approach. 

\paragraph{Two-site chain} We again begin with the two-site chain. In the notation of energy integrals, we have, for the two-site chain wavefunction coefficient:\begin{align}
  \wt\psi_\text{2-chain}(\wh E_1,E_2,K)= \ei{1_1 2_{\bar 1}}-\ei{1_{\bar 1}2_{1}}+\ei{1_{\bar 1}}\ei{2_1}-\ei{1_1}\ei{2_1}.
\end{align} 
The energy integrand corresponding to these four terms can be written as:
\begin{align}
  &\FR{1}{(E_1+K_1+\ep_1)(E_1+E_2+\ep_{12})}-\FR{1}{(E_1-K_1+\ep_1)(E_1+E_2+\ep_{12})}\n\\
  +&\FR{1}{(E_1-K_1+\ep_1)(E_2+K_1+\ep_2)}-\FR{1}{(E_1+K_1+\ep_1)(E_2+K_1+\ep_2)}
\end{align}
Consider, for example, the first term. The fraction possesses two poles at $\ep_1=-E_1-K_1$ and $\ep_{12}=-E_1-E_2$. Thus, the full energy integral may become singular when either $E_1+K_1\to 0$ or $E_1+E_2\to 0$, since in each of the two limits, a pole of the integrand would hit the endpoint of the integration $\ep_1=0$ or $\ep_{12}=0$, resulting in an endpoint singularity. Clearly, this analysis can be applied to all four terms, and we conclude that the full result may be singular in the following several situations:
\begin{enumerate}
  \item $E_1+E_2\to 0$: In this limit, the first and second terms generate endpoint singularities. This is conventionally called a \emph{total-energy singularity} since $E_1+E_2$ is the total energy of the graph.
  \item $E_1+K_1\to 0$: The first and the fourth terms contribute to the singularity. This is conventionally called a \emph{partial-energy singularity} since $E_1+K_1$ is the total energy of the subgraph containing Site 1, after the Line $K_1$ is cut open. 
  \item $E_2+K_1\to 0$: The third and the fourth terms contribute to the singularity. This is another partial-energy singularity which corresponds to the total energy of the subgraph containing Site 2.
\end{enumerate}
In addition, there is a spurious singularity: When $E_1-K_1\to 0$, both the second and the third terms become singular, but one can readily check that the singularities cancel out between the two terms:
\begin{align}
  &\lim_{E_1-K_1\to 0}\bigg[-\FR{1}{(E_1-K_1+\ep_1)(E_1+E_2+\ep_{12})}+\FR{1}{(E_1-K_1+\ep_1)(E_2+K_1+\ep_2)}\bigg]\n\\
  =&~\FR{1}{(E_{12}+\ep_{12})(E_2+K_1+\ep_2)}.
\end{align}
The phenomenon we observe here for the two-site chain is in fact quite general: Whenever we send the total energy of the whole graph or a connected subgraph to zero, we may encounter a genuine albeit unphysical singularity. Also, whenever we try to split the whole wavefunction or correlator into pieces with fixed partial order of sites, we get spurious singularities in the ``colinear'' limit such as $E_{1}-K_1\to 0$, which must be canceled among themselves\footnote{This is called a colinear limit, since the external energy $E_1=|\bm k_1|+\cdots|\bm k_n|$ is itself a sum of magnitudes of external momenta of all external legs attached to Site 1. Thus, the 3-momentum conservation $\bm k_1+\cdots+\bm k_n-\bm K_1=0$ implies that the limit $E_1-K_1\to 0$ is reached by physical configurations only when all $\bm k_i$'s and $\bm K_1$ are colinear.}.

The physical reason for the total-energy  singularity (of whole graph or subgraphs) is energy conservation: The wavefunction or correlator are quantities defined at a fixed time slice, and the energy of all external legs must be nonnegative in the physical region. This construction explicitly breaks the time translation symmetry. Consequently, the external energies never sum to zero in the physically reachable configurations unless they are all vanishing. Thus, a total-energy or partial-energy limit is necessarily unphysical if we require that all external energies are strictly positive. However, if we extrapolate external energies to unphysical regions and reach the total-energy limit, we are reintroducing the energy conservation, and thus recover the time translation symmetry. Therefore, the singularity we have discovered is nothing but integrating a constant quantity over indefinitely long time duration. It is well known that under such situations, the correct quantity to look for is the rate, which is nothing but the scattering matrix. Thus, we expect that the residue of the total-energy pole to be related to the flat-space scattering amplitude. Indeed, it is easy to check the total-energy limit of the 2-site chain energy integrand to be $(\ep_{12})^{-1}(E_1^2-K_1^2)^{-1}$. We recognize that the residue $-1/(E_1^2-K^2)$ is nothing but the flat-space scattering amplitude of a single exchange of a massless particle in the $s$-channel.  

On the other hand, the cancellation of spurious colinear poles are physically guaranteed by our choice of conformal vacuum. If we choose the initial state other than the conformal vacuum and allow for negative frequency terms in the mode function (\ref{eq_modefunction}), we would in general expect to see colinear singularities as well. 

The above singularity structure revealed from energy integrand can be easily confirmed by directly taking the corresponding limits of explicit results such as (\ref{eq_2chainWp1})-(\ref{eq_2chainWt2}).

\paragraph{Three-site chain}

Having understood the singularity structure of the two-site chain example, now we proceed to the case of three-site chain, which provides more information for an intuitive understanding of the distribution of singularities among different terms in the family tree decomposition.

\begin{figure}[t]
\centering
\includegraphics[width=0.7\textwidth]{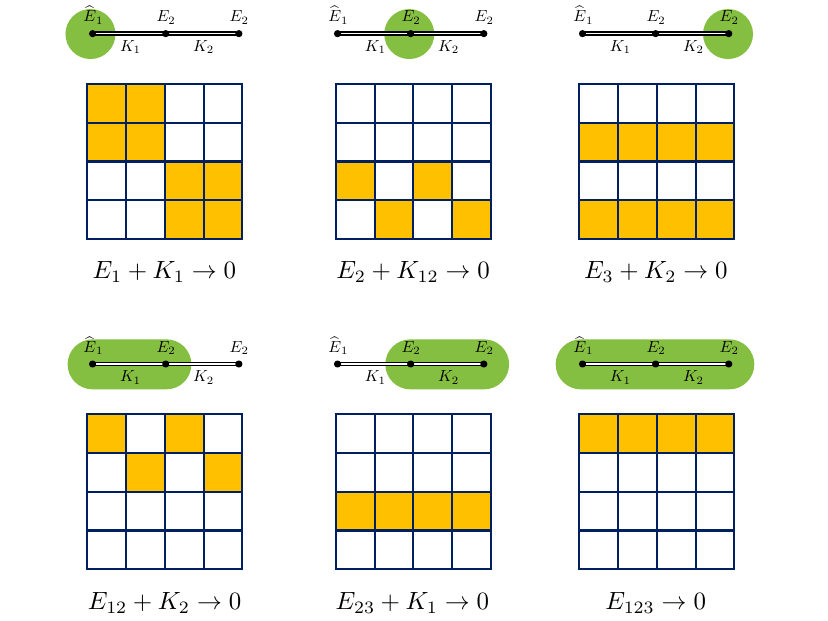} 
\caption{The pole structure of family trees contributing to the three-site chain wavefunction coefficients $\wt\psi_\text{3-chain}$ in various subgraph total-energy limits. In each panel, we consider a specific limit where the total energy of a subgraph goes to zero, as indicated by the expression below the $4\times 4$ box, and the corresponding subgraph is enclosed by the green shading above the box. In the box, the $4\times 4$ cells correspond to the 16 terms on the right-hand side of Eq.\ (\ref{eq_3chainWC1_fig}). That is, we consider the family tree decomposition with partial order $[123]$. The divergent terms in the indicated limit are shaded in yellow.}
\label{fig_3site_pole1}
\end{figure}

As shown in Sec.~\ref{sec_3chain}, the family-tree decomposition of a three-site chain amplitude has 16 terms, and it would be cumbersome to list all the poles of the 16 terms in various kinematic limits. Thus we resort to a diagrammatic presentation. For definiteness, we start from decomposition based on the partial $[123]$, whose final result is given in (\ref{eq_3chainWC1}), or diagrammatically, in (\ref{eq_3chainWC1_fig}). In either of these two equations, the 16 terms on the right-hand side are organized in a $4\times 4$ block, whose four rows correspond to 4 different cut structures, and the four columns correspond to 4 choices of summation indices for the two bulk lines. Then, applying the ``birthday rule'' which is trivial in the current case, one can directly write down the energy integrands for all these 16 terms and analyze their pole structure. We show the result in Fig.\;\ref{fig_3site_pole1}. As is clear, there are six connected subgraphs which can give rise to singularities when their total energy goes to zero. In each case, the terms that become divergent are shaded in yellow. 

\begin{figure}[t]
\centering
\includegraphics[width=0.7\textwidth]{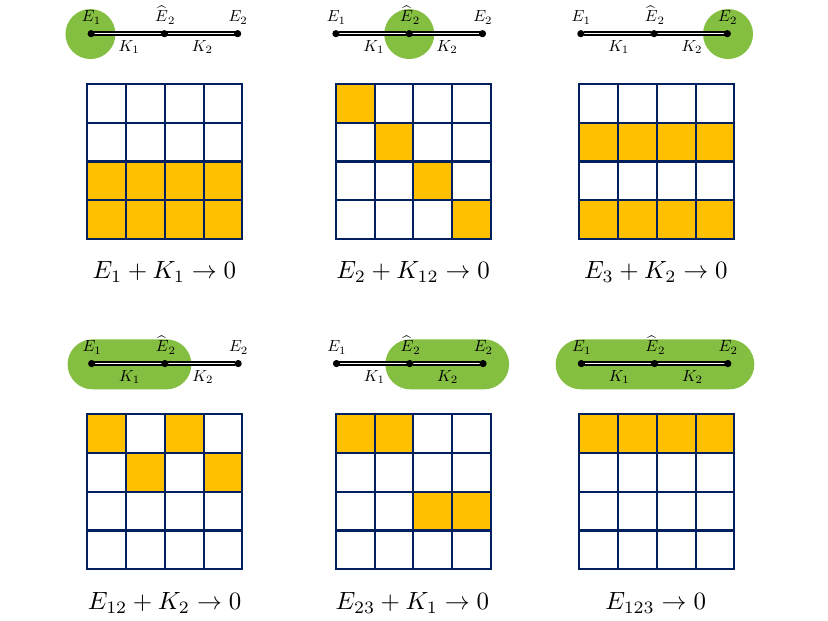} 
\caption{The same as Fig.\;\ref{fig_3site_pole1}, except that we choose the partial order of the family-tree decomposition to be $[2(1)(3)]$. The 16 cells in each box correspond to the 16 terms on the right-hand side of Eq.\ (\ref{eq_3chainWC2_fig}).}
\label{fig_3site_pole2}
\end{figure}

Likewise, one can consider the same three-site chain amplitude, but with a different choice of family-tree decomposition. In Fig.\;\ref{fig_3site_pole2}, we show the singular terms among the total of 16 terms in the family-tree decomposition of $\wt\psi_\text{3-chain}$ with partial order $[2(1)(3)]$. The $4\times 4$ grids here are in one-to-one correspondence with the $4\times 4$ terms on the right-hand side of (\ref{eq_3chainWC2_fig}). Comparing Fig.\ \ref{fig_3site_pole1} with Fig.\ \ref{fig_3site_pole2}, we see that, although the singularities of the final answer must be independent of how we decompose it, the distribution of singularities among all terms obviously depends the decomposition. 

\paragraph{General rule} To conclude this section, we briefly comment on the general case. In particular, we want to ask the question: Given a family-tree decomposition of a tree-level conformal amplitude, which terms could become singular in a given subgraph zero-energy limit? (Here we treat the whole graph as its own subgraph so that the total-energy and partial-energy limits can be discussed in a unified way.) From the above examples, it is not hard to observe an answer to this question. That is, for a term to be singular in a particular subgraph zero-energy limit, it is necessary and sufficient to satisfy the following two conditions:
\begin{enumerate}
  \item The term is fully nested within the subgraph. 
  \item On all lines connecting the subgraph to the rest (if there is any), there are no arrows pointing towards the subgraph, and the sign mark [namely {\color{Red}$\oplus$} and {\color{RoyalBlue}$\ominus$} introduced below (\ref{eq_T2chainFig})] on the side of the subgraph should be {\color{Red}$\oplus$}. 
\end{enumerate}
We do not spell out a formal proof to this rule as it is almost self-evident.

\section{Inflationary Limit: From Family Trees to Polylogarithms}
\label{sec_inflation}

In previous sections, we remained very generic about the choices of $q$ parameters and allow the value of $q$ to vary from vertex to vertex. However, it is also interesting to look at a few particular examples with fixed values of $q$, either because such values may correspond to physically relevant scenarios, or because there may emerge new interesting mathematical structure. In this section, we consider a few of such particular examples. We will only consider wavefunction coefficients for these examples, and the treatment for correlators is completely the same.

For definiteness, we will focus on the $\phi^3$ theory in $(3+1)$-dimensional spacetime, namely, a conformal scalar with the cubic self-interaction. This model, albeit incomplete as a field theory on its own, is of considerable interest for perturbative studies of field-theoretic amplitudes. It also serves as a very useful toy example for studying cosmological correlators in general. From (\ref{eq_actionflat}), we see that, in this special case, the parameter $q_\ell$ at all vertices equal to $\wt p$ given in (\ref{eq_wtp}).

The four cases listed in (\ref{eq_wtp}) are obviously of immediate interest, among which the flat-space one with $q=\wt p+1=1$ is relatively simple: The amplitudes of conformal scalars in flat space is nothing but the energy integrand discussed in Sec.\;\ref{sec_energyint}, as has been well studied in previous works. On the other hand, the cases of radiation domination $q=2$ and matter domination $q=3$ can be obtained from the flat-space result by acting differential operators, owing to the following simple relation for the nested time integral $\mb{T}^{(q_1\cdots q_\ell \cdots q_N)}_{\mathscr{N}}(\omega_1,\cdots,\omega_N)$ defined in (\ref{eq_NTI}):
\begin{align}
  \ii \frac{\partial}{\partial \omega_\ell}\mb{T}^{(q_1\cdots q_\ell \cdots q_N)}_{\mathscr{N}}(\omega_1,\cdots,\omega_N) =\mb{T}^{(q_1 \cdots (q_{\ell}+1) \cdots q_N)}_{\mathscr{N}}(\omega_1,\cdots,\omega_N) \ .~~~~(\ell=1,\cdots,N)
\end{align}
This relation can be readily checked with either the original definition (\ref{eq_NTI}) or the series representation (\ref{eq_family}).

Thus, the only nontrivial example is the inflationary case with $q=0$. This case is of particular interest for several reasons. First, it corresponds to amplitudes in inflation background, which is a most prominent scenario for primordial cosmology. Thus the study of conformal amplitudes in this case may help us to better understand realistic inflation correlation functions. Second, the amplitudes in $q\to 0$ limit has an enhanced conformal symmetry, which is a consequence of the enlarged bulk isometries of dS spacetime relative to a generic FRW background. Third, as we shall see, the $\phi^3$ amplitudes in inflation naturally have polylogarithmic representations, and the highest transcendental weight of a graph is simply given by the number of sites $N$. Moreover, there is a simple relationship between the transcendental weight of a term in the final expression with the cut structure.

Technically, the inflationary limit, or $q\to 0$ limit, is a bit subtle in that each family tree integral is typically divergent in this limit, but the sums of all the family trees in wavefunction coefficients or in correlators remain finite. As mentioned above, the finite part is typically represented by polylogarithmic functions. We will see that our family tree decomposition method makes it easy to find the polylogarithmic expression for generic tree graphs. In particular, it is also straightforward to find the symbol of the final amplitudes. Below we elucidate these points with several explicit examples with increasing complexity.

\paragraph{Two-site chain}
Once again we begin with the simplest nontrivial case, namely the two-site chain. Previously, we have worked out the 2-site chain wavefunction coefficient in terms of family-tree integrals as:
\bge
\label{eq_psi2chainInf}
  \wt\psi_\text{2-chain}= \ft{1_1 2_{\bar 1}}-\ft{1_{\bar 1}2_1}+\ft{1_{\bar 1}}\ft{2_1}-\ft{1_1}\ft{2_1}.
\ede
In the case of a $\phi^3$ theory in an inflation background with exact exponential expansion, we need to take $q_1,q_2\to 0$. In this limit, each individual term in the above expression diverges, which can be traced back to an infrared (late-time) divergence of the bulk time integral. However, the divergences cancel when taking the sum. To see this point more explicitly, we take $q_1=q_2=q$ and let $q\to 0$. It is convenient to work with the series representation for the family tree as in (\ref{eq_family}). For example, the first term in (\ref{eq_psi2chainInf}) can be evaluated as:
\begin{align}
  \ft{1_1 2_{\bar 1}}_{q_1=q_2=q}
  =&~\FR{-1}{[\ii(E_1+K)]^{2q}}\sum_{n=0}^\infty\FR{(-1)^n}{n!}\FR{\Gamma[n+2q]}{n+q}\Big(\FR{E_2-K}{E_1+K}\Big)^n.
\end{align}
Then we take the limit $q\to 0$. Clearly, the $n=0$ term in the summation is divergent, and thus we isolate it from the rest, and set $q=0$ directly in $n\neq 0$ terms:
\begin{align}
 \lim_{q\to 0}\ft{1_12_{\bar 1}} 
 =&-\Big\{1-2q\log[\ii(E_1+K)]+2q^2\log^2[\ii(E_1+K)]\Big\}\n\\
  &\times\bigg\{\FR{\Gamma(2q)}{q}+\sum_{n=1}^\infty\FR{1}{n^2}\Big(\FR{K-E_2}{K+E_1}\Big)^n\bigg\}+\order{q}\n\\
  =&-\FR{1}{2q^2}+\FR{\ga_E+\log[\ii(E_1+K)]}{q}\n\\
  &-\text{Li}_2\FR{K-E_2}{K+E_1}-\Big(\log[\ii(E_1+K)]+\ga_E\Big)^2-\FR{\pi^2}{6}+\order{q}.
\end{align}
Thus, we get divergent terms up to $\order{1/q^2}$. In the finite terms $\sim \order{q^0}$, we recognize that the summation is nothing but the dilogarithmic function $\text{Li}_2(z)$. Already at this point, we see the appearance of polylogarithmic functions that is frequently encountered when studying conformal amplitudes in inflation. 

We can repeat the above exercise for all terms in (\ref{eq_psi2chainInf}). Each of these terms contain divergent terms up to $\order{1/q^2}$. However, all such terms cancel completely. The cancelation is guaranteed since we know that the final answer should be finite. A very similar phenomenon appears if we directly do the bulk time integral directly in inflation background. In that case, the time integral for each individual family tree is divergent at the late-time limit. However, this divergence cancels out in the final answer. In fact, our $q\to 0$ procedure here can be viewed as a regularized version of direct bulk integration in inflation background. Here, we list the divergent terms in the $q\to 0$ limit for all individual terms in (\ref{eq_psi2chainInf}) to explicitly show their mutual cancelation:
\begin{align}
  \label{eq_2site_finite}
  \text{Div}\Big\{\ft{1_1 2_{\bar 1}}\Big\}
  =&-\FR{1}{2q^2}+\FR{\ga_E+\log[\ii(E_1+K)]}{q},\\
  \text{Div}\Big\{-\ft{1_{\bar 1} 2_{1}}\Big\}
  =&+\FR{1}{2q^2}-\FR{\ga_E+\log[\ii(E_1-K)]}{q},\\
  \text{Div}\Big\{\ft{1_{\bar 1}}\ft{2_1}\Big\}
  =&-\FR{1}{q^2}+\FR{\log[\ii(E_1-K)]+\log[\ii(E_2+K)]+2\ga_E}{q}\\
  \text{Div}\Big\{-\ft{1_1}\ft{2_1}\Big\}
  =&+\FR{1}{q^2}-\FR{\log[\ii(E_1+K)]+\log[\ii(E_2+K)]+2\ga_E}{q}
\end{align}
Here $\text{Div}\{\cdots\}$ collects all terms of $\order{q^n}$ in the $q\to 0$ limit with $n<0$. 

After the cancelation of divergent terms, we are left with finite terms of $\order{q^0}$, which we denote by $\text{Fin}\{\cdots\}$. For completeness, we also list the parts from all terms in (\ref{eq_psi2chainInf}):
\begin{align}
  \label{eq_2site_divergent}
  \text{Fin}\Big\{\ft{1_1 2_{\bar 1}}\Big\}
  =&-\text{Li}_2\FR{K-E_2}{K+E_1}-\Big(\log[\ii(E_1+K)]+\ga_E\Big)^2-\FR{\pi^2}{6}\\
  \text{Fin}\Big\{-\ft{1_{\bar 1} 2_{1}}\Big\}
  =&+\text{Li}_2\FR{K+E_2}{K-E_1}+\Big(\log[\ii(E_1-K)]+\ga_E\Big)^2+\FR{\pi^2}{6}\\
  \text{Fin}\Big\{\ft{1_{\bar 1}}\ft{2_1}\Big\}
  =&+\FR{1}{2}\Big(\log[\ii(E_1-K)]+\log[\ii(E_2+K)]+4\ga_E\Big)\n\\
  &\times \Big(\log[\ii(E_1-K)]+\log[\ii(E_2+K)]\Big)-2\ga_E^2-\FR{\pi^2}6,\\
  \text{Fin}\Big\{-\ft{1_{1}}\ft{2_1}\Big\}
  =&-\FR{1}{2}\Big(\log[\ii(E_1+K)]+\log[\ii(E_2+K)]+4\ga_E\Big)\n\\
  &\times \Big(\log[\ii(E_1+K)]+\log[\ii(E_2+K)]\Big)+2\ga_E^2+\FR{\pi^2}6,
\end{align}
Now we can sum over all four terms in (\ref{eq_psi2chainInf}) to get the following finite expression with $q=0$:
\begin{align}
 \wt\psi_\text{2-chain}
 =&~ \text{Li}_2\FR{K+E_2}{K-E_1}-\text{Li}_2\FR{K-E_2}{K+E_1}+\FR{1}{2}\log\FR{E_1-K}{E_1+K}\log\FR{(E_1-K)(E_1+K)}{(E_2+K)^2}.
\end{align}
It is possible to use various identities involving dilogrithmic functions (see App.\ \ref{app_specialfunctions}) to recast the above result to the following more symmetric expression:
\begin{align}
  \wt\psi_\text{2-chain}
 =&~\text{Li}_2\FR{E_2-K}{E_{12}}+\text{Li}_2\FR{E_1-K}{E_{12}}+\log\FR{E_1+K}{E_{12}}\log\FR{E_2+K}{E_{12}}-\FR{\pi^2}6.
\end{align}
This expression agrees with existing result in the literature \cite{Arkani-Hamed:2015bza,Baumann:2021fxj}. In particular, it is straightforward to find the symbol of the two-site chain as:
\begin{align}
  \mathcal{S}(\wt\psi_\text{2-chain})
  =+\FR{E_2+K}{E_{12}}\otimes\FR{E_1+K}{E_1-K}+\FR{E_1+K}{E_{12}}\otimes\FR{E_2+K}{E_2-K}.
\end{align}

\paragraph{Three-site chain}

Next let us look at the wavefunction coefficient $\wt\psi_\text{3-chain}$ of the three-site chain. We can repeat the above analysis for the two-site chain, starting from the family-tree decomposed expressions such as (\ref{eq_3chainWC1comp}) or (\ref{eq_3chainWC2comp}), setting $q_1=q_2=q_3=q$, and then send $q\to 0$. This exercise is straightforward but quite tedious. Thus we will not show the full result. In particular, we will not consider the terms with cuts, since these terms only involves the one-site and two-site families, whose $q\to 0$ limits have been explicitly taken in the previous example. Thus, we will only concentrate on the uncut part of the result, namely the terms involving three-site family trees. 

Technically, we have two choices: We can either work with 3-site chain $[123]$ as in (\ref{eq_3chainWC1comp}), or work with the two-generation 3-site family $[2(1)(3)]$ as in (\ref{eq_3chainWC2comp}). Let us  consider the latter case first. 

To avoid notational clutter, we will drop the subscripts for internal energies, and restore the original notation for the family trees as in (\ref{eq_family}). It is again convenient to work with the series representation directly:  
\begin{align}
\ft{2(1)(3)}
=&~\FR{\ii}{(\ii \omega_2)^{q_{123}}}\sum_{n_1,n_3=0}^\infty\FR{(-1)^{n_{13}}\Gamma[n_{13}+q_{123}]}{(n_1+q_1)(n_3+q_3)}\FR{u_1^{n_1}}{n_1!}\FR{u_3^{n_3}}{n_3!},
\end{align}
where we have defined $u_1\equiv \omega_1/\omega_2$ and $u_3\equiv \omega_3/\omega_2$. 
Now, we can set $q_1=q_2=q_3=q$, and expand the above expression around $q=0$. Once again, in the summation, the terms with $n_1=0$ or $n_3=0$ diverge in this limit. Thus we isolate them, and set $q\to 0$ directly in the rest of terms, as follows:
\begin{align}
\label{eq_ft213qto0}
\ft{2(1)(3)}
=&~\FR{\ii}{(\ii \omega_2)^{3q}} 
 \bigg\{\FR{\Gamma[3q]}{q^2}+\FR{1}{q}\sum_{n_1=1}^\infty\FR{(-1)^{n_1}\Gamma[n_1]}{n_1}\FR{u_1^{n_1}}{n_1!}+\FR{1}{q}\sum_{n_3=1}^\infty\FR{(-1)^{n_3}\Gamma[n_3]}{n_3}\FR{u_3^{n_3}}{n_3!}\n\\
&+\sum_{n_1,n_3=1}^\infty\FR{(-1)^{n_{13}}\Gamma[n_{13}]}{n_1n_3}\FR{u_1^{n_1}}{n_1!}\FR{u_3^{n_3}}{n_3!}\bigg\}+\order{q}.
\end{align}
The single-layer summations in the first line are again the familiar dilogarithmic function $\text{Li}_2$. To see what the double-layer summation in the second line gives, we can take its derivatives with respect to $\log u_1$ and $\log u_3$:
\begin{align}
  &~\FR{\pd}{\pd\log u_1}\FR{\pd}{\pd\log u_3}\sum_{n_1,n_3=1}^\infty\FR{(-1)^{n_{13}}\Gamma[n_{13}]}{n_1n_3}\FR{u_1^{n_1}}{n_1!}\FR{u_3^{n_3}}{n_3!}= \sum_{n_1,n_3=1}^\infty\FR{(-1)^{n_{13}}\Gamma[n_{13}]}{n_1!n_3!}u_1^{n_1}u_3^{n_3}.
\end{align} 
It is easy to see that the right-hand side sums into a linear combination of logarithmic functions:
\begin{align}
\sum_{n_1,n_3=1}^\infty\FR{(-1)^{n_{13}}\Gamma[n_{13}]}{n_1!n_3!}u_1^{n_1}u_3^{n_3}
  =&~\log(1+u_1)+\log(1+u_3)-\log(1+u_{13}).
\end{align}
Therefore, the original double-layer summation in (\ref{eq_ft213qto0}) corresponds to the following function:
\bge
  \mb{L}_3(u_1,u_3)\equiv \int_0^{u_1}\FR{\di u_1'}{u_1'}\int_0^{u_3}\FR{\di u_3'}{u_3'}\Big[\log(1+u_1')+\log(1+u_3')-\log(1+u_{13}')\Big].
\ede
The lower limits of the integrals here are chosen by the boundary condition $\mb{L}_3(0,u_3)=0$ and $\mb{L}_3(u_2,0)=0$, which can be easily seen from the original summation expression in (\ref{eq_ft213qto0}). The above integral can actually be done, and the result is given by complicated combinations of polylogarithmic functions, whose highest transcendental weight is $3$. We will not bother to show the full expression here, which can be rather long. Instead, we note that it is a simple exercise to read the symbol of $\mb{L}_3$ function from its definition, as follows:
\begin{align}
  \mathcal{S}(\mb{L}_3)=&~\FR{(1+u_1)(1+u_3)}{1+u_{13}}\otimes u_1\otimes u_3+\FR{(1+u_1)(1+u_3)}{1+u_{13}}\otimes u_3\otimes u_1\n\\
  &~+\FR{1+u_{13}}{1+u_{1}}\otimes(1+u_1)\otimes u_1+\FR{1+u_{13}}{1+u_{3}}\otimes(1+u_3)\otimes u_3.
\end{align}
Thus, together with the boundary condition $\mb{L}_3(0,u_3)=0$ and $\mb{L}_3(u_2,0)=0$, this symbol uniquely determines the function $\mb{L}_3$.

With the piece of highest transcendental weight obtained, now it is straightforward to find an expanded expression for the family tree $[2(1)(3)]$:
\begin{align}
\ft{2(1)(3)}
=&~\ii\,\Big[1-3q\log(\ii \omega_2)+\FR{9}{2}q^2\log^2(\ii \omega_2)-\FR{9}2q^3\log^3(\ii \omega_2)\Big]\n\\
 &\times\bigg\{\FR{1}{3q^3}-\FR{\ga_E}{q^2}+\FR{\text{Li}_2(-u_1)+\text{Li}_2(-u_3)+\fr32\ga_E^2+\fr14\pi^2}{q} +\mb{L}_3(u_1,u_3)\bigg\}+\order{q}.
\end{align}
Once again, we get divergent terms, but this time the most divergent term is of $\order{1/q^3}$. On the other hand, we know that these divergent terms must cancel themselves when we combine all the family trees into the wavefunction coefficient. Thus, so far as the final result for the wavefunction is concerned, we can throw away all divergent terms, and keep the $\order{q^0}$ terms only:
\begin{align}
  \text{Fin}\Big\{ \ft{2(1)(3)} \Big\}
 =&~\ii\,\Big\{\mb{L}_3(u_1,u_3)-3\log(\ii \omega_2)\Big[\text{Li}_2(-u_1)+\text{Li}_2(-u_3)+\fr32\ga_E^2+\fr14\pi^2\Big]\n\\
  &-\FR{9}2\ga_E\log^2(\ii \omega_2)-\FR{3}2\log^3(\ii \omega_2)\Big\}.
\end{align}
Now that we have obtained the finite parts of all kinds of family trees appeared in the 3-site chain wavefunction coefficients $\wt\psi_\text{3-chain}$, we can just sum them together to get the final result. The message we want to convey through this example is that, although the $q\to 0$ is a bit subtle due to the appearance of divergent terms in each family tree, the mutual cancelation of those divergences in the final result is guaranteed. Consequently, we can expand each term in the family tree decomposition in the $q\to 0$ limit and keep the $\order{q^0}$ terms only. As we have seen, the result is typically expressed as combinations of polylogarithmic functions. 

Now, let us look at the same 3-site chain wavefunction, but with different choice of partial order, as in (\ref{eq_3chainWC1comp}). Thus, the new 3-site family here is the 3-site chain $[123]$. Similar to the above case, we again take $q_1=q_2=q_3=q$, and then send $q\to 0$:
\begin{align}
  \ft{123}
  =&~\FR{\ii}{(\ii\omega_1)^{3q}}\sum_{n_2,n_3=0}^\infty\FR{(-1)^{n_{23}}\Gamma[n_{23}+q_{123}]}{(n_{23}+q_{23})(n_3+q_3)}\FR{v_{2}^{n_2}}{n_2!}\FR{v_3^{n_3}}{n_3!}\n\\
  =&~\FR{\ii}{(\ii\omega_1)^{3q}}\bigg\{\FR{\Gamma[3q]}{2q^2}+\FR{1}{q}\sum_{n_2=1}^{\infty}\FR{(-v_2)^{n_{2}}}{n_2^2}  
   +\sum_{n_2=0}^\infty\sum_{n_3=1}^\infty\FR{(-1)^{n_{23}}\Gamma[n_{23}]}{n_{23}n_3}\FR{v_{2}^{n_2}}{n_2!}\FR{v_3^{n_3}}{n_3!}\bigg\}
\end{align}
Here we have defined $v_2\equiv\omega_2/\omega_1$ and $v_3\equiv\omega_3/\omega_1$.
In the $q\to 0$ limit, all terms with $n_3=0$ are divergent. We can further separate them into two parts, with $n_2=n_3=0$ and $n_2\geq1$, $n_3=0$, respectively. The single-layer summation is easily recognized to be the dilogarithmic function. To see what the double summation gives, we again take its derivative with respect to $\log v_3$, and get:
\begin{align}
&\FR{\pd}{\pd\log v_3}\sum_{n_2=0}^\infty\sum_{n_3=1}^\infty\FR{(-1)^{n_{23}}\Gamma[n_{23}]}{n_{23}n_3}\FR{v_{2}^{n_2}}{n_2!}\FR{v_3^{n_3}}{n_3!}
=\sum_{n_2=0}^\infty\sum_{n_3=1}^\infty\FR{(-1)^{n_{23}}n_{23}!}{n_{23}^2}\FR{v_{2}^{n_2}}{n_2!}\FR{v_3^{n_3}}{n_3!}\n\\
=&~\text{Li}_2(-v_{23})-\text{Li}_2(-v_2).
\end{align}
So we can define another function:
\bge
  \wt{\mb{L}}_3(v_2,v_3)\equiv\int_{0}^{v_3}\FR{\di v_3'}{v_3'}\Big[\text{Li}_2(-v_2-v_3')-\text{Li}_2(-v_2)\Big].
\ede
This integral can again be carried out explicitly in terms of weight-3 polylogarithmic functions, although the expression is rather long. However, it is again easy to find the symbol of $\wt{\mb{L}}_3$, which can be used to simplify the expression:
\begin{align}
  \mathcal{S}(\wt{\mb L}_3)=(1+v_{23})\otimes v_{23}\otimes \FR{v_2}{v_3}-(1+v_2)\otimes v_2\otimes \FR{v_2}{v_3}-\FR{1+v_{23}}{1+v_2}\otimes \FR{v_3}{1+v_2}\otimes v_2.
\end{align}
Then, the finite terms of the family tree $[123]$ in the $q\to 0$ limit can be expressed as:
\begin{align}
  \text{Fin}\Big\{\ft{123}\Big\}
  =&~\ii\Big\{\wt{\mb L}_3(v_2,v_3)-3\zeta_3-\fr32\ga_E^3-\fr34\pi^2\ga_E-\Big[3\text{Li}_2(-v_2)+\fr92\ga_E^2+\fr34\pi^2\Big]\log(\ii\omega_1)\n\\
   &-\fr92\ga_E\log^2(\ii\omega_1)-\fr32\log^3(\ii\omega_1)\Big\}.
\end{align}

\paragraph{Arbitrary two-generation families} 

From the above examples, it is already clear how to take $q\to 0$ limit of an arbitrary wavefunction coefficient. All we need to do is to keep the finite part of the limit, since all divergent terms should cancel out in the final result. In the finite part, our series representation for the family trees typically lead to polylogarithmic functions. A curious question is that, using the above procedures of taking derivatives and then trying to resum, can we find a rule to directly write down the answer of an arbitrary family tree in terms of polylogarithmic functions, starting from \emph{any} of its series representation? This is an interesting question on its own right but is a bit beyond the main scope of this work. However, it is likely that our procedure works in most general situation, for which a more complete analysis will be presented in a future work. Here, we give one more example to show how our method works for more general situations. 

Our example is an arbitrary family tree with $N$ sites, but with only two generations. One can call it a $(N-1)$-star graph. Using our notation, this family tree can be denoted as $[1(2)\cdots (N)]$. Using our series representation (\ref{eq_family}), we have:
\begin{align}
  \ft{1(2)\cdots (N)}= \FR{(-\ii)^N}{(\ii\omega_1)^{q_{1\cdots N}}}\sum_{n_2,\cdots ,n_N=0}^\infty\FR{\Gamma[q_{1\cdots N}+n_{2\cdots N}]}{(q_2+n_2)\cdots(q_N+n_N)}\FR{v_2^{n_2}}{n_2!}\cdots\FR{v_N^{n_N}}{n_N!},
\end{align}
where $v_i\equiv\omega_i/\omega_1$ ($i=2,\cdots, N$). Incidentally, this summation can be represented in terms of Lauricella's $F_A$ function, although we will not use this fact here. 

Then, taking all $q_i=q$ ($i=1,\cdots, N$) and set $q\to 0$, we see that the terms with any $n_i=0$ $(i=2,\cdots,N)$ will be divergent. On the other hand, setting some $n_i$'s to zero reduces the layer of summation, which results in polylogarithmic functions of smaller weight. Thus, let us focus on the part of the result that has the highest transcendental weight (weight $N$) and cannot be written as products of lower-weight functions. This corresponds to the $(N-1)$-fold summation with all $n_i\geq 1$. To see what this summation gives, once again, we take its derivatives with respect to $\log v_i$ with $i=2,\cdots,N$. Then:
\begin{align}
  &\FR{\pd}{\pd\log v_2}\cdots\FR{\pd}{\pd \log v_N}\sum_{n_2,\cdots,n_N=1}^\infty\FR{(-1)^{n_{2\cdots N}}\Gamma[n_{2\cdots N}]}{n_2\cdots n_N}\FR{v_2^{n_2}}{n_2!}\cdots\FR{v_N^{n_N}}{n_N!}\n\\
  =&~\sum_{n_2,\cdots,n_N=1}^\infty\FR{\Gamma[n_{2\cdots N}](-v_2)^{n_2}\cdots(-v_N)^{n_N}}{n_2!\cdots n_N!} 
\end{align}
With a bit of algebra one can show that this summation is nothing but a linear combination of normal logarithmic functions:
\begin{align}
&\sum_{n_2,\cdots,n_N=1}^\infty\FR{\Gamma[n_{2\cdots N}](-v_2)^{n_2}\cdots(-v_N)^{n_N}}{n_2!\cdots n_N!}\n\\
=&-\log(1+v_{2\cdots N})+\sum_{j=2}^N\log(1+v_{2\cdots N}-v_j)-\sum_{j,k=2,j\neq k}^N\log(1+v_{2\cdots N}-v_{jk})\n\\
  &+\cdots+(-1)^{N-1}\sum_{j=2}^N\log(1+v_j).
\end{align}  
So we see that the summation is nothing but an $(N-1)$-fold iterative integral of normal logarithmic functions with respect to $\log v_i$, and this is just a polylogarithmic function $\mb L_N$ of weight $N$, of which an explicit integral representation can be given as:
\begin{align}
  \mb L_N(v_2,\cdots, v_N) 
  =&-\int_{0}^{v_2}\FR{\di v_2'}{v_2'}\cdots\int_0^{v_N}\FR{\di v_N'}{v_N'}\bigg[\log(1+v_{2\cdots N}')-\sum_{j=2}^N\log(1+v_{2\cdots N}'-v_j')\n\\
  &+\sum_{j,k=2,j\neq k}^N\log(1+v_{2\cdots N}'-v_{jk}') 
   +\cdots+(-1)^{N}\sum_{j=2}^N\log(1+v_j')\bigg].
\end{align}
Then, in terms of this function and its lower-weight counterparts, we can write down the analytical expression for any 2-generation graph in the $q\to 0$ limit. 

\section{Conclusion and Outlook}
\label{sec_conclusion}

The cosmological correlators encode rich physical information about both fundamental physics and the primordial universe, and are under active measurements by current and future cosmological experiments. Theory-wise, these objects represent the equal-time correlations of quantum fields living in a time-dependent cosmological background and show interesting mathematical structure. Cosmological correlators in inflationary background (dS) have been actively studied in past years, while correlators in general power-law FRW universes start to receive more attentions in recent studies. 

In this work, we presented simple algorithms for directly writing down the analytical answers for arbitrary tree-level amplitudes for a self-interacting conformal scalar in a horizon-exiting and power-law FRW universe, including both the wavefunction coefficients and the correlators. These analytical results are expressed in terms of the family-tree integrals, which can in turn be expressed as Taylor series of energy variables of any site or in the total energy. Our results are made possible by the recently proposed family-tree decomposition for arbitrary nested time integrals over exponential and power functions. We also found interesting relation between the family-tree integrals and the energy integral representation. Furthermore, in the special case of $\phi^3$ theory in inflation, we find that our general results reduce to functions of polylogarithmic type, and the series representation for the family-tree integrals provides a convenient way to identify these polylogarithmic functions and their symbols.

There are a number of interesting follow-ups which we mention below and leave for future investigations.

First, we have only consider the amplitudes for conformal scalar fields in this work. While this is a very interesting toy model, we are ultimately interested in the more realistic amplitudes of fields of arbitrary masses and arbitrary (nonzero) spins. Including arbitrary mass could be quite nontrivial as the mode function in general takes much complicated form. In dS space (exponential expansion), the mode function for a general massive scalar is represented by the Hankel function, while the mode function for a general massive scalar in general power-law FRW universe does not have a known special function representation. However, we do expect that, at least for some special cases, the massive amplitudes can be much more simplified with the partial Mellin-Barnes repesentation. Thus, it would be interesting to explore this technique in general FRW universe, to generalize our result to massive amplitudes.

Second, it would be interesting to explore the loop amplitudes for conformal scalars. The loop amplitudes are obviously complicated by the loop (spatial) momentum integrals. Unlike full energy-momentum integrals for covariant Feynman diagrams for flat-space scattering amplitudes, the spatial momentum integrals for cosmological amplitudes are typically elliptical integrals even at the 1-loop level, which makes a full analytical understanding rather challenging. On the other hand, the nested time integrals in loop integrals do not have any additional complication than tree graphs, as time orderings never form a loop. Therefore, we expect that the techniques employed in this work can at least help to sort out the time integrals for general loop amplitudes.  Also, it has been observed that, once we go beyond the tree level, the wavefunction coefficient and the correlator for the same graph do not necessarily share the same ``transcendental weight.'' Given that we are able to compute the arbitrary time integrals with the family-tree decomposition, it would be interesting to further explore the relationship between these two objects at loop level.

Third, we have shown that the family-tree integrals reduce to functions of polylogarithmic type in a special limit, namely the cubic self-interaction in inflation. While we only show how to make this reduction for a class of simple examples, we believe that our procedure is completely general. It would be interesting to complete this reduction for arbitrary tree graphs. In particular, it would be interesting to look for a (hopefully simple) algorithm to find the symbol for arbitrary tree graphs in this limit.

Fourth, it would be interesting to study the relation between our family-tree decomposition and the recently developed differential equations for these amplitudes based on canonical forms \cite{Arkani-Hamed:2023kig,Arkani-Hamed:2023bsv}. In particular, by a transformation of basis, one should be able to show explicitly that our results satisfy those differential equations, although the basis transformation may be complicated with increasing number of sites. 

Last, while we do have series representations for arbitrary family tree integrals,  we have also seen that the (partially ordered) family tree integrals can be further reduced to (totally ordered) ``family chain'' integrals by using the birthday rule. This reduction has the obvious advantage that it reduces the class of integrals we need to consider for evaluating arbitrary nested time integrals. The resulting family chains have the standard form of iterated integrals and thus enjoy all mathematical structures of such integrals. Exploiting these structures has turned out to be extremely fruitful for polylogarithmic functions, which is, as mentioned above, a special limit of our family-tree or family-chain integrals. Thus, it would be very interesting to study more systematically the mathematical properties of family chain integrals, including their algebraic structure, the analytical properties, as well as practical strategy for numerical implementations. We leave these interesting mathematical questions for future explorations. 

\paragraph{Acknowledgments.} We thank Jiaju Zang for useful discussions. We also thank Xingang Chen, Hayden Lee, and Guilherme Pimentel for helpful comments on a draft version of this work. This work is supported by NSFC under Grant No.\ 12275146, the National Key R\&D Program of China (2021YFC2203100), and the Dushi Program of Tsinghua University.

\newpage
\begin{appendix}

\section{Notations}
\label{app_notations}

For readers' convenience, in this appendix, we tabulate some frequently used notations, both symbolic and graphic. First, in Table \ref{tab_notations}, we list some symbolic notations together with the numbers of equations where they are defined or first appear.  
\begin{table}[tbph] 
  \centering
   \caption{List of symbolic notations}
   \vspace{2mm}
  \begin{tabular}{lll}
   \toprule[1.5pt]
    Notation 
    &\multicolumn{1}{c}{Description} & Equation \\ \hline 
    $p$ & Power of physical time in the FRW scale factor  & (\ref{eq_SF_PT}) \\
    $\wt{p}$ & Power of comformal time in the FRW scale factor & (\ref{eq_SF_CT}) \\
    $P_n$ or $q_\ell-1$ & Power of conformal time in the effective Minkowski coupling & (\ref{eq_actionflat}) \\
    $\bm{k}_i$ & Spatial momentum of external line & (\ref{eq_ctowf}) \\
    $\bm{K}_i$ & Spatial momentum of internal line & Above (\ref{eq_reswfcoef}) \\
    $E_i$ & External energy & (\ref{eq_1site_CF}) \\
    $K_i$ & Internal energy & (\ref{eq_5ptWF}) \\
    $\wh{E}_i$ & The maximal external energy & (\ref{eq_psi2sol1}) \\
    $\mathcal{E}_i$ & Sum of energy $\omega_i$ and integral variable $\epsilon_i$ & (\ref{eq_energy_int_mid}) \\
    $G(k;\tau_1,\tau_2)$ & Bulk propagator of wavefunction coefficients & (\ref{eq_bulkpropWC}) \\
    $B(k;\tau)$ & Bulk-to-boundary propagator of wavefunction coeffients & (\ref{eq_btobpropWC}) \\
    $D_\mathsf{ab}(k;\tau_1,\tau_2)$ & Bulk propagator of correlation functions & (\ref{eq_Dpmpm},\ref{eq_Dpmmp}) \\
    $D_\aa(k;\tau)$ & Bulk-to-boundary propagator of correlation functions & (\ref{eq_btobpropSK}) \\
    $\wt{G}(k;\tau_1,\tau_2)$ & Rescaled bulk propagator of wavefunction coefficients & (\ref{eq_resG}) \\
    $\wt{D}_{\aa\bb}(k;\tau_1,\tau_2)$ & Rescaled bulk propagator of correlation functions & (\ref{eq_resD}) \\
    $\wt{D}_\aa(k;\tau)$ & Rescaled bulk-to-boundary propagator of correlation functions & (\ref{eq_resDbtob}) \\
    $\aa,\bb,\cdots$ & SK or SK-like indices take value from $\pm 1$ & (\ref{eq_5ptC}) \\
    $\psi_n$ & Wavefunction coefficient & (\ref{eq_wfcoefdef}) \\
    $\mathcal{T}_n$ & Correlation function & (\ref{eq_crfuncdef}) \\
    $\wt{\psi}_n$ & Rescaled wavefunction coefficient & (\ref{eq_reswfcoef}) \\
    $\wt{\mathcal{T}}_n$ & Rescaled correlation function & (\ref{eq_rescrfunc}) \\ 
    $\mb{T}^{(q_1\cdots q_N)}_{\mathscr{N}}$ & General $N$-site nested time integral with tree structure $\mathscr{N}$& (\ref{eq_NTI}) \\
    $\ft{123\cdots}$ & Family tree integral & (\ref{eq_familytree_123def}) \\
    $\ei{123\cdots}$ & Energy integral & (\ref{eq_energy_integral}) \\
    $\ef{123\cdots}$ & Energy integrand & (\ref{eq_energy_integrand}) \\
    $\text{Div}\{\cdots\}$ & Divergent part of a function at $q\to 0$ & (\ref{eq_2site_finite}) \\
    $\text{Fin}\{\cdots\}$ & Finite part of a function at $q\to 0$ & (\ref{eq_2site_divergent}) \\    
    \bottomrule[1.5pt] 
  \end{tabular}
  \label{tab_notations}
\end{table}

In the meantime, we also make use of a number of graphic notations for various types of amplitudes. First, for rescaled wavefunction coefficients $\wt\psi$, we use \emph{double lines} to represent the propagator. We use an empty box to denote a boundary point, and a small black dot to denote a bulk vertex. Thus, the bulk-to-boundary propagator $B(k;\tau)$ with momentum $k$, and the \emph{rescaled} bulk propagator $G(K;\tau_1,\tau_2)$ with momentum $K$ are respectively represented as:
\begin{align}
&\parbox{28mm}{\includegraphics[width=28mm]{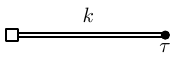}}~= B(k;\tau) ,
&&\parbox{28mm}{\includegraphics[width=28mm]{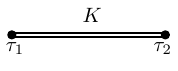}}~= \wt{G}(K;\tau_1,\tau_2)   .
\end{align}
On the other hand, the propagators for correlators in the SK formalism come in various types. This has to do with the two possible SK branches that a bulk time variable can take. We will use a single solid line to denote the propagator for the correlator, in order to distinguish it from the wavefunction propagator. Also, we use a filled/empty dot to denote a bulk vertex of $+$/$-$ type. The boundary point is still denoted by an empty box. Thus, the two boundary-to-bulk propagators $D_{\pm}(k;\tau)$ and the four bulk propagators $D_{\aa\bb}$~($\aa,\bb=\pm$) are respectively represented as:
\begin{align}
&\parbox{28mm}{\includegraphics[width=28mm]{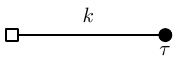}}~= \wt{D}_+(k;\tau), 
&&\parbox{28mm}{\includegraphics[width=28mm]{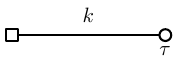}}~= \wt{D}_-(k;\tau),\\
&\parbox{28mm}{\includegraphics[width=28mm]{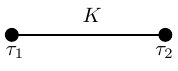}}~= \wt{D}_{++}(K;\tau_1,\tau_2), 
&&\parbox{28mm}{\includegraphics[width=28mm]{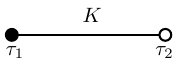}}~= \wt{D}_{+-}(K;\tau_1,\tau_2),\\
&\parbox{28mm}{\includegraphics[width=28mm]{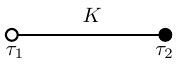}}~= \wt{D}_{-+}(K;\tau_1,\tau_2), 
&&\parbox{28mm}{\includegraphics[width=28mm]{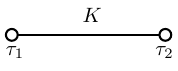}}~= \wt{D}_{--}(K;\tau_1,\tau_2).
\end{align}
Sometimes we will use ``half filled'' vertices to denote a bulk vertex whose SK index is undetermined:
\begin{align}
&\parbox{28mm}{\includegraphics[width=28mm]{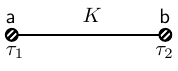}}~= \wt{D}_{\aa\bb}(K;\tau_1,\tau_2)   .
\end{align}
Whenever we use such a half filled vertex with an undetermined SK index $\aa$, it should be understood that one should sum the result over both values of $\aa=\pm$.

We stress that our graphic rules here is slightly different from the one used in previous works from our group in that the diagrammatic lines always represent the \emph{rescaled} propagators.

To graphically represent the \emph{final} (integrated) result for wavefunction coefficients or correlators, we employ additional graphic notations for family tree integrals and energy integrals, as detailed in the main text. For a family-tree integrals, we use an arrowed solid line to denote an internal line that contains a time-ordering $\theta$ function whose direction denotes the direction of time. We use a dash line to represent a line without such $\theta$ function factors so that the two sites connected by a dashed line are factorized:
\begin{align}
&\parbox{31mm}{\includegraphics[width=31mm]{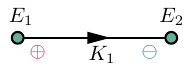}\vspace{1mm}}~= \ft{1_12_{\bar 1}} ,
&&\parbox{31mm}{\includegraphics[width=31mm]{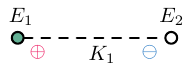}\vspace{1mm}}~= \ft{1_1}\ft{\bar 2_{\bar 1}}.
\end{align}
In these examples, we also use several decorations: For the vertices, a filled circle denotes an external energy variable (say $E_2$) which appears in the family-tree integral as a positive entry $+E_2$ while an unfilled circle means $-E_2$. Also, the notation {\color{Red}$\oplus$} and {\color{RoyalBlue}$\ominus$} are used to mark the sign of the internal energy (say $K_1$) associated with the site.

Finally, we use blue double arrowed lines to denote energy integral. Since we always use the birthday rule to break a family tree into chain diagrams when writing the energy integral, our energy integrals always have a chain structure, whose direction denotes the arrow of time of the corresponding time integral:
\begin{align}
&\parbox{31mm}{\includegraphics[width=31mm]{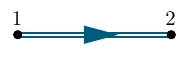}}~= \ei{12}. 
\end{align}

\section{Special Functions and Useful Formulae}
\label{app_specialfunctions}
In this appendix we present the definitions of the special functions involved in the main text and a few useful formulae. 

\paragraph{Gamma functions}
As in previous works of our group along this direction, we use a compact notation for products and fractions of Euler $\Gamma$ functions:
\begin{align}
  \Gamma\big[z_1,\cdots,z_p\big]&\equiv \Gamma(z_1)\cdots\Gamma(z_p) \ , \\
  \Gamma\left[
  \begin{matrix}
    z_1,\cdots,z_p \\
    w_1,\cdots,w_q
  \end{matrix}
  \right]
  &\equiv \frac{\Gamma(z_1)\cdots\Gamma(z_1)}{\Gamma(w_1)\cdots\Gamma(w_q)} \ .
\end{align}
We also use one of the Pochhammer symbols:
\bge
  (z)_n\equiv\Gamma\bgb z+n \\ z \edb.
\ede

\paragraph{Hypergeometric series and functions}
Some of our results in the main text involve a few well-studied hypergeometric series. We present both the standard version and the ``dressed'' version of these functions. In the main text, it is always more convinient to use the dressed version of these functions. A detailed exposition of hypergeometric functions can be found in \cite{Slater:1966}.

The Gauss hypergeometric$\ _2\text{F}_1(z)$ function is defined by the following series in the disk $|z|<1$, and by analytical continuation when $z$ is outside the disk:
\begin{align}
  \ _2\text{F}_1
  \left[
  \begin{matrix}
    a,b \\
    c
  \end{matrix}
  \bigg|
  z
  \right]
  \equiv \sum_{n=0}^\infty \frac{(a)_n (b)_n}{(c)_n}\frac{z^n}{n!} \ .
\end{align}
The dressed version of Gauss hypergeometric function is 
\begin{align}
  \ _2 \mathcal{F}_1
  \left[
  \begin{matrix}
    a,b \\
    c
  \end{matrix}
  \bigg|
  z
  \right]
  &\equiv\Gamma 
  \left[
  \begin{matrix}
    a,b \\
    c
  \end{matrix}
  \right]
  \ _2F_1
  \left[
  \begin{matrix}
    a,b \\
    c
  \end{matrix}
  \bigg|
  z
  \right] \notag \\
  &= \sum_{n=0}^\infty 
  \Gamma 
  \left[
  \begin{matrix}
    a+n,b+n \\
    c+n
  \end{matrix}
  \right]
  \frac{z^n}{n!} \ .
\end{align}
There are many well-known variable transformation formulae of the Gauss hypergeometric functions. For example,
\begin{align}
  \ _2F_1 
  \left[
  \begin{matrix}
    a,b \\
    c
  \end{matrix}
  \bigg|z
  \right]
  =(1-z)^{-b} \ _2F_1 
  \left[
  \begin{matrix}
    c-a,b \\
    c
  \end{matrix}
  \bigg|\frac{z}{z-1}
  \right] \ .
  \label{2F1ToVExp}
\end{align}
More of such transformation formulae can be found in \cite{nist:dlmf}.

For any types of hypergeometric functions, the relations between the standard version and the dressed version are all similar. The dressed versions are obtained from the standard version by removing the gamma functions in the denominators of the Pochhammer symbols. We will only present the dressed version of the multi-variable hypergeometric functions below.

Now we come to the bivariate hypergeometric functions. These are much less understood than univariate functions.   A relatively better studied class of functions are the Appell functions, which come in 4 types, $F_1,\cdots,F_4$. In this work we only need the second type. The dressed version of Appell $F_2$ function is defined by the following series when it is convergent and by analytical continuation otherwise:
\begin{align}
  \mathcal{F}_2 
  \left[
  a \bigg|
  \begin{matrix}
    b_1,b_2 \\
    c_1,c_2
  \end{matrix}
  \bigg| z,w
  \right]
  =\sum_{m,n=0}^\infty
  \Gamma 
  \left[
  \begin{matrix}
    a+m+n,b_1+m,b_2+n \\
    c_1+m, c_2+n
  \end{matrix}
  \right]
  \frac{z^m w^n}{m! n!} \ .
\end{align}

We also need more general version of two-variable hypergeometric functions. The dressed version of Kampé de Fériet functions are defined as
\begin{align}
  &\ ^{p+q}\mathcal{F}_{r+s}
  \left[
  \begin{matrix}
    a_1,\cdots,a_p \\
    c_1,\cdots,c_r
  \end{matrix}
  \bigg|
  \begin{matrix}
    b_1,b_1';\cdots;b_q,b_q' \\
    d_1,d_1';\cdots;d_s,d_s'
  \end{matrix}
  \bigg|
  z,w
  \right] \notag \\
  =&\sum_{n,m=0}^\infty 
  \Gamma 
  \left[
  \begin{matrix}
    a_1+n+m,\cdots,a_p+n+m \\
    c_1+n+m,\cdots,c_r+n+m
  \end{matrix}
  \right]
  \Gamma 
  \left[
  \begin{matrix}
    b_1+n,\cdots,b_q+n \\
    d_1+n,\cdots,d_s+n
  \end{matrix}
  \right]
  \Gamma 
  \left[
  \begin{matrix}
    b_1'+m,\cdots,b_q'+m \\
    d_1'+m,\cdots,d_s'+m
  \end{matrix}
  \right]
  \frac{z^{n} w^{m}}{n! m!} \ .
\end{align}
Sometimes we may omit some of the $b,d$ parameters by replacing them with a dash ``-''. This means that we simply drop the corresponding $\Gamma$ function in the definition. 

The Appell $F_2$ function can actually be expressed as
\begin{align}
  \mathcal{F}_2 
  \left[
  a \bigg|
  \begin{matrix}
    b_1,b_2 \\
    c_1,c_2
  \end{matrix}
  \bigg| z,w
  \right]
  =\ ^{1+1}\mathcal{F}_{0+1}
  \left[
  \begin{matrix}
    a \\
    \text{\ -\ }
  \end{matrix}
  \bigg|
  \begin{matrix}
    b_1,b_2\\
    c_1,c_2
  \end{matrix}
  \bigg|
  z,w
  \right] \ .
\end{align}

Finally, some of our results lead to a type of $N$-variable hypergeometric functions. The dressed version of the $N$-variable Lauricella's $F_A$ function is 
\begin{align}
  &\mathcal{F}_A
  \left[
  a \bigg|
  \begin{matrix}
    b_1,\cdots,b_N \\
    c_1,\cdots,c_N
  \end{matrix}
  \bigg| z_1,\cdots,z_N
  \right] \notag \\
  =&\sum_{n_1,\cdots,n_N=0}^\infty
  \Gamma \left[
  \begin{matrix}
    a+n_1+\cdots+n_N,b_1+n_1,\cdots,b_N+n_N \\
    c_1+n_N,\cdots,c_N+n_N
  \end{matrix}
  \right]
  \frac{z_1^{n_1}\cdots z_N^{n_N}}{n_1! \cdots n_N!} \ .
\end{align}
The Appell $F_2$ function is actually the 2-variable situation of Lauricella's $F_A$ function.

\paragraph{Polylogarithmic functions and symbols}

Polylogarithmic functions are very extensively studied in both mathematical and physical literatures. We will only make slight use of some of these results in Sec.\ \ref{sec_inflation}. More comprehensive introductions to this topic can be found in \cite{Duhr:2012fh}.

Polylogarithmic functions are in a sense generalizations of the original logarithmic function $\log z$. The simplest generalization is the dilogarithm function $\text{Li}_2$ function, which can be defined by the following series:
\begin{align}
  \text{Li}_2(z)=\sum_{n=1}^\infty \frac{z^n}{n^2} \ .
\end{align}
Another useful definition is the following integral:
\begin{align}
  \text{Li}_2(z)=-\int_0^z \di w\, \frac{\log(1-w)}{w} \ .
\end{align}
There are a number of useful functional identities associated with dilogarithmic function. In the main text, we made use of the following two identities:
\bge
  \text{Li}_2(z)+\text{Li}_2\Big(\FR{z}{z-1}\Big)=-\FR{1}{2}\log^2(1-z),
\ede
\bge
  \text{Li}_2(z)+\text{Li}_2(1-z)=\FR{\pi^2}{6}-(\log z)\log(1-z).~~~~(0<z<1)
\ede
We can generalize the logarithm and dilogarithm to a more general class of so-called classical polylogarithmic functions $\text{Li}_n(z)$, which can be defined by an iterated integral:
\begin{align}
  \text{Li}_n(z)=\int_0^z\di\log z'\,\text{Li}_{n-1}(z'),~~~~~~\text{Li}_1(z)\equiv-\log(1-z).
\end{align}
One can also consider more general polylogarithmic functions but we do not make explicit use of them in this work. It is worth noting that polylogarithmic functions have a number of interesting mathematical properties, and a very useful tool to study them is the symbol. Consider a class of iterated integrals $f^{(n)}_i(z)$ defined recursively in the following way:
\bge
  \di f^{(n)}(z)=\sum_i f_i^{(n-1)}\di\log R_i(z),
\ede
where $R_i$ are rational functions of $z$ with rational coefficients. Then, one can define the \emph{symbol} of these functions, denoted by $\mathcal{S}$, as:
\bge
  \mathcal{S}(f^{(n)})=\sum_i\mathcal{S}(f_i^{(n-1)})\otimes R_i.
\ede
Note that it is conventional to remove the ``log'' when writing the symbol. In the main text, we use this definition to get the symbols for family tree integrals for $\phi^3$ theory in inflation. 

\section{Explicit Expressions for Conformal Amplitudes}
\label{app_examples}
In this appendix is to collect explicit expressions for wavefunction coefficients and correlators discussed in Sec.\ \ref{sec_example} in terms of named special functions reviewed in App.\ \ref{app_specialfunctions}. To complete our discussions, here we list explicit expressions for all the graph discussed in the main text.

Before starting, we present some of the family tree series (in the maximal energy representation) in their special function forms. They can also be found in 
\cite{Xianyu:2023ytd}:%
{\allowdisplaybreaks%
\begin{align}
  \ft{1}=&~\frac{-\ii}{(\ii \omega_1)^{q_1}}\Gamma[q_1] \ ,\\
  \ft{12}=&~\frac{-1}{(\ii \omega_1)^{q_{12}}} \ _2\mathcal{F}_1
  \left[
  \begin{matrix}
    q_2,q_{12} \\
    q_2+1
  \end{matrix}
  \bigg| -\frac{\omega_2}{\omega_1}
  \right] \ ,\\
  \ft{123}=&~\frac{\ii}{(\ii \omega_1)^{q_{123}}} \ ^{2+1}\mathcal{F}_{1+1}
  \left[
  \begin{matrix}
    q_{123},q_{23} \\
    q_{23}+1
  \end{matrix}
  \bigg|
  \begin{matrix}
    \text{\ -\ },q_3 \\
    \text{\ -\ },q_3+1
  \end{matrix}
  \bigg|
  -\frac{\omega_2}{\omega_1},-\frac{\omega_3}{\omega_1}
  \right] \ ,\\
  \ft{2(1)(3)}=&~\frac{\ii}{(\ii \omega_2)^{q_{123}}} \mathcal{F}_2
  \left[
  q_{123} \bigg|
  \begin{matrix}
    q_1,q_3 \\
    q_1+1,q_3+1
  \end{matrix}
  \bigg| -\frac{\omega_1}{\omega_2},-\frac{\omega_3}{\omega_2}
  \right] \ ,\\
  \ft{4(1)(2)(3)}=&~\frac{1}{(\ii \omega_4)^{q_{1234}}}\mathcal{F}_A
  \left[
  q_{1234}\middle|
  \begin{matrix}
    q_1,q_2,q_3 \\
    q_1+1,q_2+1,q_3+1
  \end{matrix}
  \bigg| -\frac{\omega_1}{\omega_4},-\frac{\omega_2}{\omega_4},-\frac{\omega_3}{\omega_4}
  \right] \ .
\end{align}
}

The 2-site chain correlators involve the hypergeometric functions just like the wavefunction:
\begin{align}
  &\ \wt{\mathcal{T}}_\text{2-chain}(\wh{E_1},E_2,K_1) \notag \\
  =&-\FR{1}{[\ii(E_1+K_1)]^{q_{12}}}{~}_2\mathcal{F}_1\left[\bgm q_2,q_{12}\\ q_2+1\edm\middle|-\FR{E_2-K_1}{E_1+K_1}\right]
  +\FR{1}{[\ii(E_1-K_1)]^{q_{12}}}{~}_2\mathcal{F}_1\left[\bgm q_2,q_{12}\\ q_2+1\edm\middle|-\FR{E_2+K_1}{E_1-K_1}\right] \notag \\
  &-\FR{\Gamma[q_1,q_2]}{[\ii(E_1-K_1)]^{q_1}[\ii(E_2+K_1)]^{q_2}}+\FR{\Gamma[q_1,q_2]}{[\ii(E_1+K_1)]^{q_1}[\ii(-E_2-K_1)]^{q_2}}+\text{c.c.} \ .
\end{align}

\begin{align}
  &\ \wt{\mathcal{T}}_\text{2-chain}(E_1,\wh{E_2},K_1) \notag \\
  =&-\FR{1}{[\ii(E_2+K_1)]^{q_{12}}}{~}_2\mathcal{F}_1\left[\bgm q_1,q_{12}\\ q_1+1\edm\middle|-\FR{E_1-K_1}{E_2+K_1}\right]
  +\FR{1}{[\ii(E_2-K_1)]^{q_{12}}}{~}_2\mathcal{F}_1\left[\bgm q_1,q_{12}\\ q_1+1\edm\middle|-\FR{E_1+K_1}{E_2-K_1}\right]\n\\
  &-\FR{\Gamma[q_1,q_2]}{[\ii(E_2-K_1)]^{q_2}[\ii(E_1+K_1)]^{q_1}}+\FR{\Gamma[q_1,q_2]}{[\ii(E_2+K_1)]^{q_2}[\ii(-E_1-K_1)]^{q_1}}+\text{c.c.} \ .
\end{align}

It is clear that these two expressions can be converted to each other by simply switching $E_1$ and $E_2$.

The 3-site chain graph involves Kampé de Fériet functions for $[123]$ order and Appell $F_2$ functions for $[2(1)(3)]$ order. We omit the $[321]$ order since it can be obtained from $[123]$ in a similar way to the 2-site chain. The expressions are: 
\begin{align}
  &\ \wt{\psi}_\text{3-chain}(\wh{E_1},E_2,E_3,K_1,K_2) \notag \\
  =&-\ii \sum_{{\aa , \bb}=\pm} {\aa \bb}\bigg\{
  \frac{1}{[\ii(E_1+{\aa}K_1)]^{q_{123}}}\ ^{2+1}\mathcal{F}_{1+1}
  \left[
  \begin{matrix}
    q_{123},q_{23} \\
    q_{23}+1
  \end{matrix}
  \left|
  \begin{matrix}
    \text{\ -\ },q_3 \\
    \text{\ -\ },q_3+1
  \end{matrix}
  \right|
  -\frac{E_2-{\aa}K_1+{\bb}K_2}{E_1+{\aa}K_1},-\frac{E_3-{\bb}K_2}{E_1+{\aa}K_1}
  \right] \notag \\
  &+\frac{\Gamma[q_3]}{[\ii(E_3+K_2)]^{q_3}[\ii(E_1+{\aa}K_1)]^{q_{12}}}
  \ _2\mathcal{F}_1 
  \left[
  \begin{matrix}
    q_2,q_{12} \\
    q_2+1
  \end{matrix}
  \bigg|
  -\frac{E_2-{\aa}K_1-{\bb}K_2}{K_1+{\aa}E_1}
  \right]
   \notag \\
  &+\frac{\Gamma[q_1]}{[\ii(E_1-{\aa}K_1)]^{q_1}[\ii(E_2+K_1+{\bb}K_2)]^{q_{23}}}
  \ _2\mathcal{F}_1 
  \left[
  \begin{matrix}
    q_3,q_{23} \\
    q_3+1
  \end{matrix}
  \bigg|
  -\frac{E_3-{\bb}K_2}{E_2+K_1+{\bb}K_2}
  \right] \notag \\
  &+\frac{\Gamma[q_1,q_2,q_3]}{[\ii(E_1-{\aa}K_1)]^{q_1}[\ii(E_2+K_1-{\bb}K_2)]^{q_2}[\ii(E_3+K_2)]^{q_3}}
  \bigg\} \ .
\end{align}

\begin{align}
  &\ \wt{\psi}_\text{3-chain}(E_1,\wh{E_2},E_3,K_1,K_2) \notag \\
  =&-\ii \sum_{{\aa , \bb}=\pm} {\aa \bb} \notag \\
  &\ \bigg\{\frac{1}{[\ii(E_2+{\aa}K_1+{\bb}K_2)]^{q_{123}}}\mathcal{F}_2
  \left[q_{123} \bigg|
  \begin{matrix}
    q_1,q_3 \\
    q_1+1,q_3+1
  \end{matrix}
  \bigg|
  -\frac{E_1-{\aa}K_1}{E_2+{\aa}K_1+{\bb}K_2},-\frac{E_3-{\bb}K_2}{E_3+{\aa}K_1+{\bb}K_2}
  \right] \notag \\
  &+\frac{\Gamma[q_3]}{[\ii(E_3+K_2)]^{q_3}[\ii(E_2+{\aa}K_1-{\bb}K_2)]^{q_{21}}}
  \ _2\mathcal{F}_1 
  \left[
  \begin{matrix}
    q_1,q_{21} \\
    q_1+1
  \end{matrix}
  \bigg|
  -\frac{E_1-{\aa}K_1}{E_2+{\aa}K_1-{\bb}K_2}
  \right]
   \notag \\
  &+\frac{\Gamma[q_1]}{[\ii(E_1+K_1)]^{q_1}[\ii(E_2-{\aa}K_1+{\bb}K_2)]^{q_{23}}}
  \ _2\mathcal{F}_1 
  \left[
  \begin{matrix}
    q_3,q_{23} \\
    q_3+1
  \end{matrix}
  \bigg|
  -\frac{E_3-{\bb}K_2}{E_2-{\aa}K_1+{\bb}K_2}
  \right]
   \notag \\
  &+\frac{\Gamma[q_1,q_2,q_3]}{[\ii(E_1+K_1)]^{q_1}[\ii(E_2-{\aa}K_1-{\bb}K_2)]^{q_2}[\ii(E_3+K_2)]^{q_3}}
  \bigg\} \ .
\end{align}

\begin{align}
  &\ \wt{\mathcal{T}}_\text{3-chain}(\wh{E_1},E_2,E_3,K_1,K_2) \notag \\
  =&\ \ii \sum_{{\aa , \bb}=\pm} {\aa \bb}\bigg\{
  \frac{1}{[\ii(E_1+{\aa}K_1)]^{q_{123}}}\ ^{2+1}\mathcal{F}_{1+1}
  \left[
  \begin{matrix}
    q_{123},q_{23} \\
    q_{23}+1
  \end{matrix}
  \left|
  \begin{matrix}
    \text{\ -\ },q_3 \\
    \text{\ -\ },q_3+1
  \end{matrix}
  \right|
  -\frac{E_2-{\aa}K_1+{\bb}K_2}{E_1+{\aa}K_1},-\frac{E_3-{\bb}K_2}{E_1+{\aa}K_1}
  \right] \notag \\
  &+\frac{\Gamma[q_3]}{[\ii({\bb}E_3+{\bb}K_2)]^{q_3}[\ii(E_1+{\aa}K_1)]^{q_{12}}}
  \ _2\mathcal{F}_1 
  \left[
  \begin{matrix}
    q_2,q_{12} \\
    q_2+1
  \end{matrix}
  \bigg|
  -\frac{E_2-{\aa}K_1-{\bb}K_2}{K_1+{\aa}E_1}
  \right]
   \notag \\
  &+\frac{\Gamma[q_1]}{[\ii(E_1-{\aa}K_1)]^{q_1}[\ii({\aa}E_2+{\aa}K_1+{\bb}K_2)]^{q_{23}}}
  \ _2\mathcal{F}_1 
  \left[
  \begin{matrix}
    q_3,q_{23} \\
    q_3+1
  \end{matrix}
  \bigg|
  -\frac{{\aa}E_3-{\bb}K_2}{{\aa}E_2+{\aa}K_1+{\bb}K_2}
  \right] \notag \\
  &+\frac{\Gamma[q_1,q_2,q_3]}{[\ii(E_1-{\aa}K_1)]^{q_1}[\ii({\aa}E_2+{\aa}K_1-{\bb}K_2)]^{q_2}[\ii({\bb}E_3+{\bb}K_2)]^{q_3}}
  \bigg\} +\text{c.c.} \ .
\end{align}

\begin{align}
  &\ \wt{\mathcal{T}}_\text{3-chain}(E_1,\wh{E_2},E_3,K_1,K_2) \notag \\
  =&\ \ii \sum_{{\aa , \bb}=\pm} {\aa \bb} \notag \\
  &\ \bigg\{\frac{1}{[\ii(E_2+{\aa}K_1+{\bb}K_2)]^{q_{123}}}\mathcal{F}_2
  \left[q_{123} \bigg|
  \begin{matrix}
    q_1,q_3 \\
    q_1+1,q_3+1
  \end{matrix}
  \bigg|
  -\frac{E_1-{\aa}K_1}{E_2+{\aa}K_1+{\bb}K_2},-\frac{E_3-{\bb}K_2}{E_3+{\aa}K_1+{\bb}K_2}
  \right] \notag \\
  &+\frac{\Gamma[q_3]}{[\ii({\bb}E_3+{\bb}K_2)]^{q_3}[\ii(E_2+{\aa}K_1-{\bb}K_2)]^{q_{21}}}
  \ _2\mathcal{F}_1 
  \left[
  \begin{matrix}
    q_1,q_{21} \\
    q_1+1
  \end{matrix}
  \bigg|
  -\frac{E_1-{\aa}K_1}{E_2+{\aa}K_1-{\bb}K_2}
  \right]
   \notag \\
  &+\frac{\Gamma[q_1]}{[\ii({\aa}E_1+{\aa}K_1)]^{q_1}[\ii(E_2-{\aa}K_1+{\bb}K_2)]^{q_{23}}}
  \ _2\mathcal{F}_1 
  \left[
  \begin{matrix}
    q_3,q_{23} \\
    q_3+1
  \end{matrix}
  \bigg|
  -\frac{E_3-{\bb}K_2}{E_2-{\aa}K_1+{\bb}K_2}
  \right]
   \notag \\
  &+\frac{\Gamma[q_1,q_2,q_3]}{[\ii({\aa}E_1+{\aa}K_1)]^{q_1}[\ii(E_2-{\aa}K_1-{\bb}K_2)]^{q_2}[\ii({\bb}E_3+{\bb}K_2)]^{q_3}}
  \bigg\}+\text{c.c.} \ .
\end{align}

For the 4-site star graph with order $[4(1)(2)(3)]$, we need Lauricella’s $F_A$ functions for the fully nested term. The expression for the wavefunction coefficient is:
\begin{align}
  &\ \wt{\psi}_\text{4-star}(E_1,E_2,E_3,\wh{E_4},K_1,K_2,K_3) \notag \\
  =&\sum_{{\aa , \bb , \cc}=\pm} {\aa \bb \cc} \Bigg(\frac{1}{[\ii(E_4+{\aa}K_1+{\bb}K_2+{\cc}K_3)]^{q_{1234}}} 
  \mathcal{F}_A 
  \bigg[
  q_{1234}
  \left|
  \begin{matrix}
    q_1,q_2,q_3 \\
    q_1+1,q_2+1,q_3+1
  \end{matrix}
  \right| \notag \\
  &\qquad\left(
  -\frac{E_1-{\aa}K_1}{E_4+{\aa}K_1+{\bb}K_2+{\cc}K_3},-\frac{E_2-{\bb}K_2}{E_4+{\aa}K_1+{\bb}K_2+{\cc}K_3},-\frac{E_3-{\cc}K_3}{E_4+{\aa}K_1+{\bb}K_2+{\cc}K_3}
  \right)
  \bigg] \notag \\
  &+\bigg\{\frac{\Gamma[q_3]}{[\ii(E_3+K_3)]^{q_3}[\ii(E_4+{\aa}K_1+{\bb}K_2-{\cc}K_3)]^{q_{412}}}\mathcal{F}_2
  \bigg[q_{412} \bigg|
  \begin{matrix}
    q_1,q_2 \\
    q_1+1,q_2+1
  \end{matrix}
  \bigg| \notag \\
  &\qquad\left(
  -\frac{E_1-{\aa}K_1}{E_4+{\aa}K_1+{\bb}K_2-{\cc}K_3},-\frac{E_2-{\bb}K_2}{E_4+{\aa}K_1+{\bb}K_2-{\cc}K_3}
  \right)
  \bigg]+\text{2 perms}\bigg\} \notag \\
  &+\bigg\{\frac{\Gamma[q_1,q_2]}{[\ii(E_1+K_1)]^{q_1}[\ii(E_2+K_2)]^{q_2}[\ii(E_4-{\aa}K_1-{\bb}K_2+{\cc}K_3)]^{q_{43}}} \notag \\
  &\quad\times \ _2\mathcal{F}_1 
  \left[
  \begin{matrix}
    q_3,q_{43} \\
    q_3+1
  \end{matrix}
  \bigg|
  -\frac{E_3-{\cc}K_3}{E_4-{\aa}K_1-{\bb}K_2+{\cc}K_3}
  \right]
  +\text{2 perms}\bigg\} \notag \\
  &+\frac{\Gamma[q_1,q_2,q_3,q_4]}{[\ii(E_1+K_1)]^{q_1}[\ii(E_2+K_2)]^{q_2}[\ii(E_3+K_3)]^{q_3}[\ii(E_4-{\aa}K_1-{\bb}K_2-{\cc}K_3)]^{q_4}}
  \Bigg),
\end{align}
while the expression for the correlator is:
\begin{align}
  &\ \wt{\mathcal{T}}_\text{4-star}(E_1,E_2,E_3,\wh{E_4},K_1,K_2,K_3) \notag \\
  =&\sum_{{\aa , \bb , \cc}=\pm} {\aa \bb \cc} \Bigg(\frac{1}{[\ii(E_4+{\aa}K_1+{\bb}K_2+{\cc}K_3)]^{q_{1234}}} 
  \mathcal{F}_A 
  \bigg[
  q_{1234}
  \left|
  \begin{matrix}
    q_1,q_2,q_3 \\
    q_1+1,q_2+1,q_3+1
  \end{matrix}
  \right| \notag \\
  &\qquad\left(
  -\frac{E_1-{\aa}K_1}{E_4+{\aa}K_1+{\bb}K_2+{\cc}K_3},-\frac{E_2-{\bb}K_2}{E_4+{\aa}K_1+{\bb}K_2+{\cc}K_3},-\frac{E_3-{\cc}K_3}{E_4+{\aa}K_1+{\bb}K_2+{\cc}K_3}
  \right)
  \bigg] \notag \\
  &+\bigg\{\frac{\Gamma[q_3]}{[\ii({\cc}E_3+{\cc}K_3)]^{q_3}[\ii(E_4+{\aa}K_1+{\bb}K_2-{\cc}K_3)]^{q_{412}}}\mathcal{F}_2
  \bigg[q_{412} \bigg|
  \begin{matrix}
    q_1,q_2 \\
    q_1+1,q_2+1
  \end{matrix}
  \bigg| \notag \\
  &\qquad\left(
  -\frac{E_1-{\aa}K_1}{E_4+{\aa}K_1+{\bb}K_2-{\cc}K_3},-\frac{E_2-{\bb}K_2}{E_4+{\aa}K_1+{\bb}K_2-{\cc}K_3}
  \right)
  \bigg]+\text{2 perms}\bigg\} \notag \\
  &+\bigg\{\frac{\Gamma[q_1,q_2]}{[\ii({\aa}E_1+{\aa}K_1)]^{q_1}[\ii({\bb}E_2+{\bb}K_2)]^{q_2}[\ii(E_4-{\aa}K_1-{\bb}K_2+{\cc}K_3)]^{q_{43}}} \notag \\
  &\quad\times \ _2\mathcal{F}_1 
  \left[
  \begin{matrix}
    q_3,q_{43} \\
    q_3+1
  \end{matrix}
  \bigg|
  -\frac{E_3-{\cc}K_3}{E_4-{\aa}K_1-{\bb}K_2+{\cc}K_3}
  \right]
  +\text{2 perms}\bigg\} \notag \\
  &+\frac{\Gamma[q_1,q_2,q_3,q_4]}{[\ii({\aa}E_1+{\aa}K_1)]^{q_1}[\ii({\bb}E_2+{\bb}K_2)]^{q_2}[\ii({\cc}E_3+{\cc}K_3)]^{q_3}[\ii(E_4-{\aa}K_1-{\bb}K_2-{\cc}K_3)]^{q_4}}
  \Bigg)+\text{c.c.} \ .
\end{align}
Here the ``2 perms'' in the equation stands for the other two permutations of the functions inside the curly braces with respect to indices 1,2,3 together with summation signs $\aa,\bb,\cc$ (notice that these functions are already symmetric with respect to 1,2 together with $\aa,\bb$).

\section{Examples of Transformation-of-Variable Formulae}
\label{app_transf}

In the main text we mentioned that some transformation-of-variable formulae can be obtained from the non-uniqueness of the family tree decomposition. Here we collect some examples of these, both in the ``square bracket'' notation for family trees and in terms of named special functions. We will drop the overall phase factors on both sides of the equations.

We start from the most simple case. Different decompositions of a 2-site graph give us
\begin{align}
  \ft{12}+\ft{21}=\ft{1}\ft{2} \ ,
\end{align}
Writing this equation explicitly in terms of special functions, we have:
\begin{align}
  \frac{1}{\omega_1^{q_{12}}} \ _2\mathcal{F}_1
  \left[
  \begin{matrix}
    q_2,q_{12} \\
    q_2+1
  \end{matrix}
  \bigg|
  -\frac{\omega_2}{\omega_1}
  \right]
  +\frac{1}{\omega_2^{q_{12}}} \ _2\mathcal{F}_1
  \left[
  \begin{matrix}
    q_1,q_{12} \\
    q_1+1
  \end{matrix}
  \bigg|
  -\frac{\omega_1}{\omega_2}
  \right]
  =\frac{\Gamma[q_1,q_2]}{\omega_1^{q_1}\omega_2^{q_2}} \ .
\end{align}
This is actually a mixture of several well-known variable transformation formulae of the Gauss hypergeometric functions.

A more complicated example is from the 3-site graph:
\begin{align}
  \ft{123}+\ft{2(1)(3)}=\ft{1}\ft{23} \ ,
  \label{3SiteToV}
\end{align}
In terms of named special functions, this equation gives:
\begin{align}
  &\frac{1}{\omega_1^{q_{123}}} \ ^{2+1}\mathcal{F}_{1+1}
  \left[
  \begin{matrix}
    q_{123},q_{23} \\
    q_{23}+1
  \end{matrix}
  \left|
  \begin{matrix}
    \text{\ -\ },q_3 \\
    \text{\ -\ },q_3+1
  \end{matrix}
  \right|
  -\frac{\omega_2}{\omega_1},-\frac{\omega_3}{\omega_1}
  \right]
  +\frac{1}{\omega_2^{q_{123}}} \mathcal{F}_2
  \left[
  q_{123}
  \left|
  \begin{matrix}
    q_1,q_3 \\
    q_1+1,q_3+1
  \end{matrix}
  \right|
  -\frac{\omega_1}{\omega_2},-\frac{\omega_3}{\omega_2}
  \right] \notag \\
  =&\ \frac{\Gamma[q_1]}{\omega_1^{q_1} \omega_2^{q_{23}}} \ _2\mathcal{F}_1
  \left[
  \begin{matrix}
    q_3,q_{32} \\
    q_3+1
  \end{matrix}
  \bigg|
  -\frac{\omega_3}{\omega_2}
  \right] \ .
\end{align}

We can obtain more formulae from this. Switching $1,3$ in (\ref{3SiteToV}) and combining the result with the original form, we have
\begin{align}
  \ft{123}-\ft{321}=\ft{1}\ft{23}-\ft{21}\ft{3} .
\end{align}
Therefore, we have:
\begin{align}
  &\frac{1}{\omega_1^{q_{123}}} \ ^{2+1}\mathcal{F}_{1+1}
  \left[
  \begin{matrix}
    q_{123},q_{23} \\
    q_{23}+1
  \end{matrix}
  \left|
  \begin{matrix}
    \text{\ -\ },q_3 \\
    \text{\ -\ },q_3+1
  \end{matrix}
  \right|
  -\frac{\omega_2}{\omega_1},-\frac{\omega_3}{\omega_1}
  \right]
  -\frac{1}{\omega_3^{q_{123}}} \ ^{2+1}\mathcal{F}_{1+1}
  \left[
  \begin{matrix}
    q_{123},q_{21} \\
    q_{21}+1
  \end{matrix}
  \left|
  \begin{matrix}
    \text{\ -\ },q_1 \\
    \text{\ -\ },q_1+1
  \end{matrix}
  \right|
  -\frac{\omega_2}{\omega_3},-\frac{\omega_1}{\omega_3}
  \right] \notag \\
  =&\ \frac{\Gamma[q_1]}{\omega_1^{q_1} \omega_2^{q_{23}}} \ _2\mathcal{F}_1
  \left[
  \begin{matrix}
    q_3,q_{32} \\
    q_3+1
  \end{matrix}
  \bigg|
  -\frac{\omega_3}{\omega_2}
  \right]
  -\frac{\Gamma[q_3]}{\omega_3^{q_3} \omega_2^{q_{21}}} \ _2\mathcal{F}_1
  \left[
  \begin{matrix}
    q_1,q_{12} \\
    q_1+1
  \end{matrix}
  \bigg|
  -\frac{\omega_1}{\omega_2}
  \right] \ .
\end{align}

Another way to obtain transformation-of-variable formulae is comparing the maximal energy representation (\ref{eq_family}) and the total energy representation (\ref{eq_family_total}) with the same time order structure. To make use of the total energy representation, we notive that from $\Gamma[z]\Gamma[1-z]=\pi \csc(\pi z)$, for integer $n_j$ the Pochhammer symbol becomes 
\begin{align}
  \frac{1}{(-\wt{q}_j-\wt{n}_j)_{1+n_j}}=-(-1)^{n_j} \Gamma 
  \left[
  \begin{matrix}
    \wt{q}_j+\wt{n}_j-n_j \\
    \wt{q}_j+\wt{n}_j+1
  \end{matrix}
  \right] \ .
\end{align}

Now the series in (\ref{eq_family_total}) can be readily summed into familiar special functions for several simple cases. From order $[12]$ we get a familiar variable transformation formula of the hypergeometric $_2F_1$ functions:
\begin{align}
  \frac{1}{\omega_1^{q_{12}}}
  \ _2\mathcal{F}_1
  \left[
  \begin{matrix}
    q_2,q_{12} \\
    q_2+1
  \end{matrix} 
  \bigg| -\frac{\omega_2}{\omega_1}
  \right]
  =\frac{\Gamma[q_2]}{\omega_{12}^{q_{12}}} 
  \ _2\mathcal{F}_1
  \left[
  \begin{matrix}
    1,q_{12} \\
    q_2+1
  \end{matrix} 
  \bigg| \ \frac{\omega_2}{\omega_{12}}
  \right] \ .
\end{align}
This formula can also be obtained directly from (\ref{2F1ToVExp}).

The total energy representation of the order $[123]$ does not sum into familiar special functions, but it is in some general form of hypergeometric series. For example, we have a nontrivial series representation of the Kampé de Fériet functions:
\begin{align}
  &\frac{1}{\omega_1^{q_{123}}} \ ^{2+1}\mathcal{F}_{1+1}
  \left[
  \begin{matrix}
    q_{123},q_{23} \\
    q_{23}+1
  \end{matrix}
  \left|
  \begin{matrix}
    \text{\ -\ },q_3 \\
    \text{\ -\ },q_3+1
  \end{matrix}
  \right|
  -\frac{\omega_2}{\omega_1},-\frac{\omega_3}{\omega_1}
  \right] \notag \\
  =\ &\frac{\Gamma[q_3]}{\omega_{123}^{q_{123}}}\sum_{n_2,n_3=0}^\infty 
  \Gamma 
  \left[
  \begin{matrix}
    n_{23}+q_{123}, n_3+q_{23},n_2+1,n_3+1 \\
    n_{23}+q_{23}+1,n_3+q_3+1
  \end{matrix}
  \right]
  \frac{1}{n_2! n_3!}
  \left(\frac{\omega_{23}}{\omega_{123}}\right)^{n_2} \left(\frac{\omega_{3}}{\omega_{123}}\right)^{n_3} \ .
\end{align}

For the order $[2(1)(3)]$, we get a formula for Appell $F_2$ functions.
\begin{align}
  &\frac{1}{\omega_2^{q_{123}}} \mathcal{F}_2
  \left[
  q_{123}
  \left|
  \begin{matrix}
    q_1,q_3 \\
    q_1+1,q_3+1
  \end{matrix}
  \right|
  -\frac{\omega_1}{\omega_2},-\frac{\omega_3}{\omega_2}
  \right]
  =\frac{\Gamma[q_1,q_3]}{\omega_{123}^{q_{123}}} \mathcal{F}_2
  \left[
  q_{123}
  \left|
  \begin{matrix}
    1,1 \\
    q_1+1,q_3+1
  \end{matrix}
  \right|
  \frac{\omega_1}{\omega_{123}},\frac{\omega_3}{\omega_{123}}
  \right] \ .
\end{align}

Very similarly, for the 4-site star graph in the $[4(1)(2)(3)]$ order gives a variable transformation formula of Lauricella's $F_A$ functions.
\begin{align}
  &\frac{1}{\omega_4^{q_{1234}}} 
  \mathcal{F}_A 
  \bigg[
  q_{1234}
  \left|
  \begin{matrix}
    q_1,q_2,q_3 \\
    q_1+1,q_2+1,q_3+1
  \end{matrix}
  \right| 
  -\frac{\omega_1}{\omega_4},-\frac{\omega_2}{\omega_4},-\frac{\omega_3}{\omega_4}
  \bigg] \notag \\
  =&\ \frac{\Gamma[q_1,q_2,q_3]}{\omega_{1234}^{q_{1234}}} 
  \mathcal{F}_A 
  \bigg[
  q_{1234}
  \left|
  \begin{matrix}
    1,1,1 \\
    q_1+1,q_2+1,q_3+1
  \end{matrix}
  \right| 
  \frac{\omega_1}{\omega_{1234}},\frac{\omega_2}{\omega_{1234}},\frac{\omega_3}{\omega_{1234}}
  \bigg] \ .
\end{align}

\end{appendix}

\newpage
\bibliography{CosmoCollider} 
\bibliographystyle{utphys}

\end{document}